%% file: ms_RR_revised_round2_arxiv.tex
\documentclass[useAMS,usenatbib]{mn2e}

\usepackage{epsfig,graphicx}
\usepackage{amssymb,subeqnarray}
\usepackage[letterpaper,margin=1in]{geometry}
\usepackage{array}
\usepackage{rotating}

\begin{document}

\input{macros_RM.tex}
\newcommand{\fillincite}{\textbf{!!CITE!!}}

\newcommand{\hMsun}{h^{-1}M_{\odot}}
\newcommand{\hkpc}{h^{-1}{\rm kpc}}
\newcommand{\hmpc}{h^{-1}{\rm Mpc}}
\newcommand{\kms}{{\rm km\ s}^{-1}}
\newcommand{\ds}{\Delta\Sigma}
\newcommand{\cvir}{c_{\rm 200c}}
\newcommand{\rvir}{r_{\rm 200c}}
\newcommand{\mvir}{M_{\rm 200c}}
\newcommand{\mvirb}{M_{\rm 200b}}
\newcommand{\betac}{\beta_{\rm c}}
\newcommand{\mstr}{M_*}
\newcommand{\vopt}{V_{\rm opt}}
\newcommand{\vvir}{V_{\rm 200c}}
\newcommand{\rd}{R_{\rm d}}
\newcommand{\vrot}{V_{\rm rot}}
\newcommand{\vmaxh}{V_{\rm max, h}}
\newcommand{\rmaxh}{r_{\rm max, h}}
\newcommand{\mbar}{M_{\rm bar}}

\newcommand{\beq}{\begin{equation}}
\newcommand{\eeq}{\end{equation}}
\newcommand{\bef}{\begin{figure}}
\newcommand{\eef}{\end{figure}}
\newcommand{\bec}{\begin{center}}
\newcommand{\eec}{\end{center}}
\newcommand{\beqa}{\begin{eqnarray}}
\newcommand{\eeqa}{\end{eqnarray}}

\newcommand{\aaps}{{A\&AS}}
\newcommand{\araa}{{ARA\&A}}
\newcommand{\aap}{{A\&A}}
\newcommand{\apj}{{ApJ}}
\newcommand{\apjl}{{ApJL}}
\newcommand{\apjs}{{ApJS}}
\newcommand{\aj}{{AJ}}
\newcommand{\prd}{{PRD}}
\newcommand{\pasp}{{PASP}}
\newcommand{\mnras}{{MNRAS}}
\newcommand{\jcap}{{JCAP}}
\newcommand{\nat}{{Nature}}
\newcommand{\physrep}{Phys. Rep.}

\def\imagetop#1{\vtop{\null\hbox{#1}}}

\title[Disk galaxy optical-to-virial velocity ratios]{Optical-to-virial velocity ratios of local disk galaxies from combined kinematics and galaxy-galaxy lensing}
\author[Reyes et al.]{
R. Reyes$^{1,2}$\thanks{\tt rreyes@kicp.uchicago.edu}, 
R. Mandelbaum$^{1,3}$,
J. E. Gunn$^1$,
R. Nakajima$^{4,5,6,7}$, U. Seljak$^{5,6,7,8}$,
\newauthor
C. M. Hirata$^9$
\\$^1$Peyton Hall Observatory, Princeton University, 
Peyton Hall, Princeton, NJ 08544, USA
\\$^2$Kavli Institute for Cosmological Physics and Enrico Fermi Institute, The University of Chicago, Chicago, \\ IL 60637, USA
\\$^3$Department of Physics, Carnegie Mellon University, Pittsburgh, PA 15213, USA 
\\$^4$Argelander-Institut f\"{u}r Astronomie, Universit\"{a}t Bonn, 53121 Bonn, Germany
\\$^5$Space Sciences Lab, Department of Physics and Department of Astronomy, University of California, Berkeley, CA 94720 USA
\\$^6$Lawrence Berkeley National Lab, University of California, Berkeley, CA  94720, USA
\\$^7$Institute of the Early Universe, Ewha Womans University, Seoul, Korea
\\$^8$Institute for Theoretical Physics, University of Zurich, Zurich, Switzerland
\\$^9$Department of Astronomy, Caltech M/C 350-17, Pasadena, CA 91125, USA
}

\bibliographystyle{mn2e}
\maketitle

\label{firstpage} 
\begin{abstract}
In this paper, we measure the optical-to-virial velocity ratios $\vopt/\vvir$ of disk galaxies in the Sloan Digital Sky Survey (SDSS) at a mean redshift of $\langle z\rangle = 0.07$ and with stellar masses $10^9 M_\odot < M_* < 10^{11} M_\odot$. $\vopt/\vvir$, the ratio of the circular velocity measured at the virial radius of the dark matter halo ($\sim$150 kpc) to that at the optical radius of the disk ($\sim$10 kpc), is a powerful observational constraint on disk galaxy formation. It links galaxies to their dark matter haloes dynamically and constrains the total mass profile of disk galaxies over an order of magnitude in length scale. 
For this measurement, we combine $\vopt$ derived from the Tully-Fisher relation (TFR) from Reyes et~al. with $\vvir$ derived from halo masses measured with galaxy-galaxy lensing. In anticipation of this combination, we use similarly-selected galaxy samples for both the TFR and lensing analysis. For three $M_*$ bins with lensing-weighted mean stellar masses of 0.6, 2.7, and $6.5\times 10^{10} M_\odot$, we find halo-to-stellar mass ratios $\mvir/M_* = 41$, 23, and 26, with $1\sigma$ statistical uncertainties of around 0.1 dex, and $\vopt/\vvir=1.27\pm 0.08, 1.39\pm 0.06, 1.27\pm 0.08 \,(1\sigma)$, respectively.
Our results suggest that the dark matter and baryonic contributions to the mass within the optical radius are comparable, if the dark matter halo profile has not been significantly modified by baryons. The results obtained in this work will serve as inputs to and constraints on disk galaxy formation models, which will be explored in future work. Finally, we note that this paper presents a new and improved galaxy shape catalogue for weak lensing that covers the full SDSS DR7 footprint.
\end{abstract}


\section{Introduction}
\label{sec:intro}

The basic picture of disk galaxy formation has long been established: both gas and dark matter acquire angular momentum through tidal torques in the early Universe \citep{1969ApJ...155..393P}; the gas then cools and collapses into a rotationally-supported disk at the center of a cold dark matter halo \citep{1978MNRAS.183..341W,1980MNRAS.193..189F,1984Natur.311..517B,1997ApJ...482..659D,1998MNRAS.295..319M}. Today, many aspects of the theory are still not completely understood, such as star formation, feedback from supernovae and active galactic nuclei (AGNs), angular momentum transfer, mergers, and the response of the dark matter halo to the infall of baryons. State-of-the-art hydrodynamic simulations, which incorporate effective prescriptions for some of these physical processes, are starting to produce individual disk galaxies with luminosities, sizes, and rotation curves comparable with observed ones \citep{2010Natur.463..203G,2011ApJ...742...76G,2011MNRAS.410.1391A}. The ultimate test for these simulations is their ability to satisfy a broad set of observational constraints. In addition, observations can serve as a guide to improve the effective models used in the simulations. Careful comparison between the theoretical models and observations will become increasingly important as the simulations continue to improve, and eventually produce cosmological ensembles of galaxies. 

A powerful observational constraint on disk galaxy formation is the dynamical link between disk galaxies and their dark matter haloes, $\vopt/\vvir$, the ratio of the circular velocity measured at the optical radius ($r_{\rm opt} \sim 10$ kpc, a few optical disk scale lengths) to that at the virial radius of the dark matter halo ($r_{\rm vir} \sim 150$ kpc). Because the optical-to-virial velocity ratio is a dynamical quantity, it does not suffer from uncertainties in stellar mass estimates from photometry (or spectroscopy), which are subject to uncertainties in dust extinction, stellar populations, and the stellar initial mass function (IMF) \citep[see, e.g.,][]{2009ApJ...699..486C}. 

The optical-to-virial velocity ratio is sensitive to the total mass profile within the halo virial radius. For example, an isothermal profile (with a mass density proportional to $r^{-2}$) corresponds to $\vopt/\vvir=1$. 
The shape of the halo mass profile, without any modification from their interaction with baryons, is reasonably well-understood from $N$-body simulations \citep{1996ApJ...462..563N,1997ApJ...490..493N,2008MNRAS.387..536G}. 
On the other hand, the amount of response of the dark matter halo to the infall of baryons during their collapse, as well as to the blowing out of baryons due to feedback effects, is poorly-understood and is the subject of active debate in the literature, both from the theoretical \citep{2004ApJ...616...16G,2006PhRvD..74l3522G,2008ApJ...685L.105R,2009ApJ...697L..38J,2010MNRAS.405.2161D,2010MNRAS.407..435A,2010MNRAS.402..776P,2010MNRAS.406..922T,2010Natur.463..203G,2011ApJ...742...76G,2011arXiv1108.5736G} and observational sides \citep{2002ApJ...574L.129S,2006ApJ...646..899H,2006ApJ...650..777Z,2007ApJ...654...27D,2010MNRAS.408.1463S,2010ApJ...721L.163A,2011MNRAS.416..322D}. 
Different formation scenarios will generally predict different values for $\vopt/\vvir$, which can be directly tested against observations. 

The optical-to-virial velocity ratio can be measured by combining two measurements. First, $V_{\rm opt}$ can be measured directly for individual galaxies from H$\alpha$ (or HI) rotation curves. Moreover, there is a well-established tight relation between $\vopt$ and stellar mass $M_*$, referred to hereafter as the Tully-Fisher relation or TFR \citep{1977A&A....54..661T}. Second, the average $V_{\rm vir}$ can be measured for large galaxy samples with galaxy-galaxy lensing or satellite kinematics. 

Previously, \citet{2002MNRAS.334..797S} combined early TFR and galaxy-galaxy lensing measurements and inferred $\vopt/\vvir =1.8$ with a 2$\sigma$ lower limit of 1.4, for $L^*$ late-type galaxies (here, $V_{\rm vir}=\vvir$ is the circular velocity of the dark matter halo at the virial radius $r_{\rm vir}=\rvir$, the radius within which the mean density is 200 times the critical density of the Universe today). They found this result to be consistent with the prediction of the standard model of adiabatic contraction of the dark matter halo due to baryonic infall \citep{1986ApJ...301...27B}. More recently, \citet{2010MNRAS.407....2D} combined a TFR based on data from \citet{2007AJ....134..945P} with published halo-to-stellar mass ratios in the literature from galaxy-galaxy lensing, satellite kinematics, and halo abundance matching, and found $\vopt/\vvir \simeq 1$, for disk galaxies with $M_* \sim 10^{10}$--$10^{11} M_\odot$. This result is lower than the prediction from the standard adiabatic contraction model and suggests that the effect of adiabatic contraction is weaker or that the opposite effect occurs (i.e., the dark matter halo density within the optical radius decreases instead of increases). 

In this work, we perform a new measurement of $\vopt/\vvir$. We use stacked weak lensing measurements of $\sim$$10^5$ disk galaxies from the Sloan Digital Sky Survey \citep[SDSS;][]{2000AJ....120.1579Y}, with well-defined photometry and available fibre spectroscopy, to measure average halo virial masses $\mvir$. We combine these measurements with the TFR derived in \citet[][hereafter R11]{2011MNRAS.417.2347R} from a sample of 189 disk galaxies with measured H$\alpha$ rotation curves that is, by construction, a fair subsample of the lens sample used in this work. Our results constrain the relation between the optical-to-virial velocity ratio $\vopt/\vvir$ and stellar mass $M_*$, of disk galaxies with a mean redshift of 0.07 and stellar masses $M_* \sim 10^{9}$--$10^{11} M_\odot$. Unlike in previous analyses, we use similarly-selected galaxy samples and consistent definitions in both the lensing and TFR measurements to enable a fair combination of the two. 


This paper introduces a new source galaxy catalogue for lensing, which is demonstrably an improvement, both in area coverage and quality, over the one introduced in  \citet[][hereafter M05]{2005MNRAS.361.1287M}. We describe the catalogue properties, various tests of systematics, and calibration of the lensing signal in Sec.~\ref{sec:data_source_shape}; the reader who is not interested in the technical details may skip this section. The generation procedure for the catalogue is described in Appendix A and the differences from the catalogue in M05 are enumerated and discussed in Appendix B.

The organization of the rest of the paper is as follows: in Sec.~\ref{sec:overview}, we discuss the method of this work and details of the lensing calculations; in Sec.~\ref{sec:data_lens}, we describe SDSS data and the selection of our lens sample; in Sec.~\ref{sec:data_source_shape}, we extensively characterize the source catalogue and shape measurements used in this work; in Sec.~\ref{sec:lens_sig}, we present the measured lensing signals, and the results of various systematics tests; in Sec.~\ref{sec:lens_fits}, we describe fits to the lensing signal to derive average halo masses; in Sec.~\ref{sec:results}, we present our main results and their interpretation; finally, in Sec.~\ref{sec:summ}, we present a summary and discuss plans for future work.

For the calculations performed in this work, we adopt the cosmology: $\Omega_{\rm m}=0.27$, $\Omega_\Lambda=0.73$, $\sigma_8=0.8$, $h=H_0/(100 \kms {\rm Mpc}^{-1})=0.7$. Unless otherwise stated, quantities already include the appropriate factors of $h$: stellar masses $M_*$ scale as $h^{-2}$, halo masses $\mvir$ scale as $h^{-1}$, and distances $R$ scale as $h^{-1}$. All distances are expressed in comoving units, unless otherwise noted.

\section{Method}
\label{sec:overview}

We begin with an overview of the methodology of this work (Sec.~\ref{subsec:overview_comb}). In the rest of the section, we describe galaxy-galaxy lensing theory (Sec.~\ref{subsec:lens_theory}), the calculation of the galaxy-galaxy lensing signal (Sec.~\ref{subsec:lens_calc}), and the derivation of the TFR from R11 used in this work (Sec.~\ref{subsec:tfr_deriv}). 
 
\subsection{Overview}
\label{subsec:overview_comb}

In this work, we constrain two relations: (i) the relation between the halo-to-stellar mass ratio $\mvir/M_*$ and stellar mass $M_*$ (or HSMR), and (ii) the relation between the optical-to-virial velocity ratio $\vopt/\vvir$ and $M_*$ (or OVVR). Note that the virial quantities are less tightly-constrained than the corresponding ``optical'' quantities. 

To determine the HSMR, we measure the galaxy-galaxy lensing signal around stacked galaxies in bins of stellar mass $M_*$. We perform fits to the observed lensing signals to determine the best-fitting halo-to-stellar mass ratio for each stellar mass bin, as described in Sec.~\ref{sec:lens_fits}. We also adopt a functional form for the HSMR (Eq.~\ref{eq:hsmratio} of Sec.~\ref{subsubsec:hsmr_fits}) and fit the lensing signal for all three bins simultaneously to determine the best-fitting parameters in the relation. 

To determine the OVVR, we convert the halo virial masses $\mvir$ in the HSMR into halo virial velocities $\vvir = \sqrt{G\mvir/\rvir}$ (where $G$ is the Newtonian gravitational constant), then take its ratio with the relation between $\vopt$ and $M_*$ from the TFR in R11. Details of these derivations, as well as the results, will be presented in Secs.~\ref{subsubsec:hsmr_fits} \& \ref{sec:results}.
 
\subsection{Lensing theory}
\label{subsec:lens_theory}

Galaxy-galaxy lensing is the deflection of light from sources by the mass in intervening lenses, which shows up as a coherent tangential shearing effect. It provides a simple way to probe the connection between galaxies and matter via their cross-correlation function 
\beq
\xi_{\rm gm}(\vec{r}) = \langle \delta_{\rm g} (\vec{x})
\delta_{\rm m}(\vec{x}+\vec{r})\rangle
\eeq
where $\delta_{\rm g}$ and $\delta_{\rm m}$ are overdensities of galaxies and matter, respectively. This cross-correlation can be related to the projected surface density
\beq\label{eq:sigma}
\Sigma(R) = \overline{\rho} \int \left[1+\xi_{\rm gm}\left(\sqrt{R^2 + \chi^2}\right)\right] d\chi
\eeq
(where $r^2=R^2+\chi^2$). For our purposes we neglect the radial window function, which is of order 100 Mpc broad, 
well beyond the scales that are important in this work.  The surface density $\Sigma(R)$ is then related to the observable quantity for lensing,
\beq \label{eq:ds}
\ds(R) = \gamma_{\rm t}(R) \Sigma_c= \overline{\Sigma}(<R) - \Sigma(R),
\eeq
where $\gamma_{\rm t}$ is the tangential shear. In practice, we truncate the integral in Eq.~\ref{eq:sigma} at 1$\hmpc$, 
which is well beyond the halo virial radii (defined in Eq.~\ref{eq:mvir} below) of the galaxies we study. 

The second relation in Eq.~\ref{eq:ds} is true only in the weak lensing limit, for a matter distribution that is axisymmetric 
along the line of sight. This symmetry is naturally achieved by our procedure of 
stacking many galaxies and determining their average lensing signal.
This observable quantity can be expressed as the product of the tangential 
shear $\gamma_t$ and a geometric factor
\beq\label{eq:sigmacrit}
\Sigma_c^{(ls)} = \frac{c^2}{4\pi G} \frac{D_{\rm s}}{D_{\rm l} D_{\rm ls}(1+z_{\rm l})^2},
\eeq
where $D_{\rm l}$ and $D_{\rm s}$ are angular diameter distances to the lens and source, $D_{\rm ls}$ is the angular diameter distance between the lens and source, and the factor of $(1+z_{\rm l})^{-2}$ arises due to our use of comoving coordinates. For a given lens redshift, $\Sigma_c^{-1}$ rises from zero at $z_{\rm s} = z_{\rm l}$ to an asymptotic value at $z_{\rm s} \gg z_{\rm l}$; that asymptotic value is an increasing function of lens redshift.

\subsection{Lensing signal calculation}
\label{subsec:lens_calc}

Calculation of the galaxy-galaxy lensing signal requires us to identify pairs of lens and source galaxies within some physical separation on the sky.  Since we have spectroscopic redshifts for our lens galaxies, we can work with physical tangential separations rather than in angular coordinates (which mixes physical scales for lens galaxies at different redshifts).
To compute the average lensing signal $\ds(R)$, lens-source pairs are first assigned weights according to the error on the shape measurement via
\beq \label{eq:wls}
w_{ls} = \frac{(\Sigma_c^{(ls)})^{-2}}{\sigma_e^2 + \sigma_{SN}^2}
\eeq
where $\sigma_e$ is the estimated shape error per component and $\sigma_{SN}$ is the intrinsic shape noise per component, which was determined as a
function of magnitude in M05, figure 3 (in Sec.~\ref{subsubsec:erms} of this paper we will 
reassess the accuracy of that shape noise estimate). The factor of $(\Sigma_c^{(ls)})^{-2}$
converts the shape noise in the denominator to a noise in $\Delta\Sigma$; it downweights pairs that are close in redshift.

Once we have computed these weights, we calculate the lensing signal in
23 logarithmic radial bins from 0.02 to 2$\hmpc$ as a summation over lens-source pairs via:
\beq \label{eq:dsestimator}
\ds(R) = \frac{\sum_{ls} w_{ls} \gamma_t^{(ls)} \Sigma_c^{(ls)}}{2 {\cal 
R}\sum_{ls} w_{ls}},
\eeq
where the factor of 2 arises due to our definition of ellipticity and ${\cal 
R}$ is the shear responsivity, which describes how the ellipticity estimator 
used in this paper (Eq.~\ref{eq:shapedef}) responds to a shear 
(\citealt{2002AJ....123..583B} and Sec.~\ref{subsubsec:erms} of this work).

There are several additional procedures that must be done when
computing the signal (see M05 for details). First, the signal
computed around random points must be subtracted from the signal
around real lenses to eliminate contributions from systematic
shear. The measured signal around random points is consistent with zero 
over the range of length scales used in this work (c.f. Sec.~\ref{subsubsec:sys_random}). 

Second, the signal must be boosted, i.e., multiplied by $B(R) =
n(R)/n_{\rm rand}(R)$, ratio of the weighted number density of sources around real lenses, 
relative to the weighted number density of sources around random points, 
in order to account for
the dilution of the lensing signal due to sources that are physically
associated with a lens, and 
therefore not lensed. 
The multiplication by the boost factor means that our shear estimator is essentially 
identical to the globally normalized estimator in \cite{2011ApJ...735..118R}.

To determine errors on the lensing signal and boost factors, we divide the
survey area into 200 bootstrap subregions,\footnote{Ideally we would like contiguous, equal-area subregions.  
Given the SDSS survey geometry, we are forced to compromise slightly, and go in the direction of requiring 
strictly equal-area regions while allowing a small fraction to be non-contiguous.}
 and generate 500 bootstrap-resampled datasets. For illustration, we rebin the signal into 7
radial bins and plot the re-binned signal. The computed lensing signals, 
for the real and random galaxies, as well as the boost factors, 
will be presented in Sec.~\ref{sec:lens_sig}.

\subsection{Derivation of the TFR}
\label{subsec:tfr_deriv}


By construction, the galaxy sample used to derive the TFR in R11 is a fair subsample of the lens sample used in this work
(defined in Sec.~\ref{subsec:data_disk} below and sec.~3 of R11). 
There is only one difference in the selection criteria used: an axis ratio cut ($b/a<0.6$) has been applied to the TFR sample, but not to the lens sample.
We found that applying such a cut only modestly changes 
the distributions of basic galaxy properties, such as stellar mass and galaxy colour (c.f. fig. 1 of R11), 
so we do not expect it to introduce a significant bias between the two samples. 
Applying the axis ratio cut to the lens sample would have decreased the sample size by
almost half, so we have chosen not to do so. 

In this work, we follow the recommendation of R11 and use the TFR between stellar mass $M_*$ and optical velocity $\vopt=V_{80}$, the disk rotation velocity at the radius enclosing 80 per cent of the $i$-band galaxy light, $R_{80}$.\footnote{R11 showed that $M_*$ yields a TFR with smaller scatter than single-band optical luminosities (in $ugriz$), and similarly, $V_{80}$ performs better than alternative definitions of $\vopt$, such as $V_{2.2}$--- evaluated at 2.2 times the disk scale length--- or $V_{\rm c}$--- the asymptotic circular velocity. The common choice for $\vopt$ is $V_{2.2}$, the rotation velocity at 2.2 times the disk scale length $\rd$. We refer the reader to secs.~4 and 5 of R11 for the derivation of  radii $\rd$ and $R_{80}$ (from bulge-disk decomposition fits) and of rotation velocities $V_{80}$ and $V_{2.2}$ (from arctangent fits to the measured rotation curves). R11 showed that $R_{80}$ is more likely to sample the flat part of the disk rotation curve, and as a consequence, $V_{80}$ yields a tighter TFR than $V_{2.2}$ (also see Pizagno~et al. 2007). Moreover, $R_{80}$ is not affected by degeneracies involved in bulge-disk decomposition fits, unlike $\rd$. The choice of rotation velocity definition is important when comparing results from different works, as we do later in Sec.~\ref{subsec:prev_work}.} 

Here (and throughout this work), stellar masses $M_*$ correspond to the Kroupa (\citeyear{2002Sci...295...82K}) IMF.\footnote{The normalization of the Kroupa IMF is 0.3 dex lower than that for a Salpeter IMF with a lower mass cut-off of 0.1 $M_\odot$ and 0.05 dex higher than that for a Chabrier IMF.} 
As defined in sec. 5 of R11, they are determined from SDSS $i$-band absolute magnitudes and $g-r$ colours (both uncorrected for internal dust extinction), 
using stellar mass-to-light ratio estimates from \citet{2003ApJS..149..289B} (reduced by 0.15 dex to account for the difference in the normalization of the IMF). Absolute magnitudes $M_i$ are based on Petrosian apparent magnitudes
and galaxy colours $g-r$ are based on model apparent magnitudes, described in \citet{2002AJ....123..485S} and \citet{2004AJ....128..502A}. 
Absolute magnitudes and colours used to determine $M_*$ were corrected for Galactic extinction using the dust maps of \citet{1998ApJ...500..525S}
and $k$-corrected to $z=0$ using the {\verb kcorrect } product version {\verb v4_1_4 } of \citet{2007AJ....133..734B}.
We propagate errors from $M_i$ and $g-r$ to determine the error in $M_*$; for the TFR sample, the mean statistical uncertainty in $M_*$ (at fixed Kroupa IMF) is 0.041 dex.
R11 found the best-fitting TFR to be
\begin{eqnarray} \nonumber 
\log \vopt &=& (2.142 \pm 0.004) + (0.278 \pm 0.010) \\
&&\times\ (\log M_* - 10.102),
\label{eq:tfr}
\end{eqnarray}
with an intrinsic scatter of $0.036\ \pm\ 0.005$ dex and a total measured scatter of $0.056$ dex in $\log \vopt$. 

\section{Lens sample}
\label{sec:data_lens}

First, we briefly describe SDSS imaging and spectroscopy that we use for both our lens and source galaxy samples (Sec.~\ref{subsec:data_sdss}). Then, we describe the selection of our lens sample (Sec.~\ref{subsec:data_disk}).

\subsection{SDSS data} 
\label{subsec:data_sdss}

The SDSS \citep{2000AJ....120.1579Y} imaged roughly $\pi$ steradians
of the sky, and followed up approximately one million of the detected
objects spectroscopically \citep{2001AJ....122.2267E,
  2002AJ....123.2945R,2002AJ....124.1810S}. The imaging was carried
out by drift-scanning the sky in photometric conditions
\citep{2001AJ....122.2129H, 2004AN....325..583I}, in five bands
($ugriz$) \citep{1996AJ....111.1748F, 2002AJ....123.2121S} using a
specially-designed wide-field camera
\citep{1998AJ....116.3040G}. All of
the data were processed by completely automated pipelines that detect
and measure photometric properties of objects, and astrometrically
calibrate the data \citep{2001ASPC..238..269L,
  2003AJ....125.1559P,2006AN....327..821T}. The SDSS I/II imaging
surveys were completed with a seventh data release
\citep{2009ApJS..182..543A}, though this work will rely as well on an
improved data reduction pipeline ({\sc Photo v5\_6}) and updated
photometric calibration (ubercalibration,
\citealt{2008ApJ...674.1217P}) that was part of the eighth data
release, from SDSS-III
\citep{2011ApJS..193...29A,2011AJ....142...72E}.

Objects are targeted for spectroscopy using the imaging data
\citep{2003AJ....125.2276B}. Main galaxy sample targets are selected
as described by \citet{2002AJ....124.1810S}. The Main galaxy sample
target selection includes a Petrosian (\citeyear{1976ApJ...209L...1P}) apparent
magnitude cut of $r_{\rm P}=17.77$ mag, with slight variation in this
cut across the survey area. Targets are observed with a double
320-fibre spectrograph on the same telescope
\citep{2006AJ....131.2332G}. 
 
Specific subsets of the Main
spectroscopic galaxy sample, to be described in
Sec.~\ref{subsec:data_disk}, will be used as the lens galaxies for the
lensing analysis described in this work.  The new source catalogue
used here, derived from the SDSS imaging data, is described in
Sec.~\ref{sec:data_source_shape}.


\subsection{Disk lens sample}
\label{subsec:data_disk}

We need to select a disk galaxy sample that is adequate in size and at sufficiently high redshift for the lensing signal to be measurable, given the relatively shallow SDSS imaging. R11 defined a parent disk sample adequate for this purpose, selected from the SDSS DR7 NYU-Value Added Galaxy Catalog \citep[VAGC;][]{2005ApJ...629..143B}. In this work, we define a lens sample that is made up of a majority (76 per cent) of the parent disk sample defined in R11, as described below. 


We begin with a brief summary of the selection criteria used to define the parent disk sample (we refer the reader to sec.~3 of R11 for details). First, galaxies were selected to have redshifts $0.02<z<0.10$ and absolute magnitudes $-22.5<M_r<-18.0$ mag (before internal extinction correction). Then, star-forming galaxies were selected by imposing a lower limit on the H$\alpha$ emission-line flux observed through the SDSS spectroscopic fibre, as well as mild cuts in Sersi{\'c} index and emission line ratios (to exclude active galaxies). Applying these cuts yields a sample of 175~920 galaxies.\footnote{
The sample presented in R11 had 169~563 galaxies (or 3.6 per cent fewer). Here, we have updated the sample to include galaxies that had been incorrectly excluded due to a failure in the runs of the emission line fitting code. These failures were not related to any of the galaxy properties, so the addition of these galaxies does not introduce any biases to the sample.}

To select our lens sample from this parent disk sample, we first remove galaxies in areas of the sky where there are no available source galaxies. This removes around 10 per cent of the parent disk sample, leaving a total of 158~735 galaxies. We note that a substantial percentage of galaxies were removed because of the very stringent data quality cuts applied to the source galaxy sample (c.f. Appendix A), and not because of a lack of data or extreme spatial variation in the source number density.

Second, we aim to construct a lens sample that is dominated by central and isolated galaxies so that the observed lensing profiles will be simpler to interpret and analyse (as described in Sec.~\ref{subsec:density_profiles}). To do this, we identify those galaxies that are most likely to be satellites and exclude them from the sample. For each galaxy, we count the number of brighter neighbors $N_{\rm bright}$ within a fixed physical transverse radius of 1.14 Mpc and a redshift width of $\Delta z=0.006$ (following \citealt{2009ApJ...702..249R}). Then, we scale $N_{\rm bright}$ by the number of brighter neighbors of a random galaxy with the same luminosity and redshift.\footnote{For both the real and random galaxy samples, we use the same comparison galaxy sample (namely, the DR7 SDSS NYU-VAGC).} We identify ``satellite galaxies'' as those that have a scaled number count greater than or equal to $7$. This cut excludes 15 per cent of the remaining galaxies and yields a disk lens sample of 133~598 galaxies. 

We note that this average satellite fraction is consistent with \citet{2006MNRAS.372..758M}, who found using halo occupation modelling that typically 10--15 per cent of blue/late-type galaxies were satellites. Moreover, we find, reassuringly, that the satellite fraction we calculate is very weakly dependent on redshift, and dependent on luminosity in the sense that brighter galaxies are more likely to be satellites (as expected since they are more strongly biased than fainter galaxies). 
Finally, we note that we apply the same criteria on the random galaxy catalogue used in the lensing analysis; the satellite cut excludes only 2.4 per cent of the random galaxies.

\input{section4_RM.tex}

\section{Galaxy-galaxy lensing signal} 
\label{sec:lens_sig}

In this section, we present the galaxy-galaxy lensing signal of the lens sample (Sec.~\ref{subsec:gg_results}). Then, we describe and present the results of several tests of systematics on the lensing signal: namely, the random points test, the 45-degree test, and the ratio test (Sec.~\ref{subsec:lens_sys}).

\subsection{Lens sample}
\label{subsec:gg_results}

\begin{figure}
\includegraphics[width=3in]{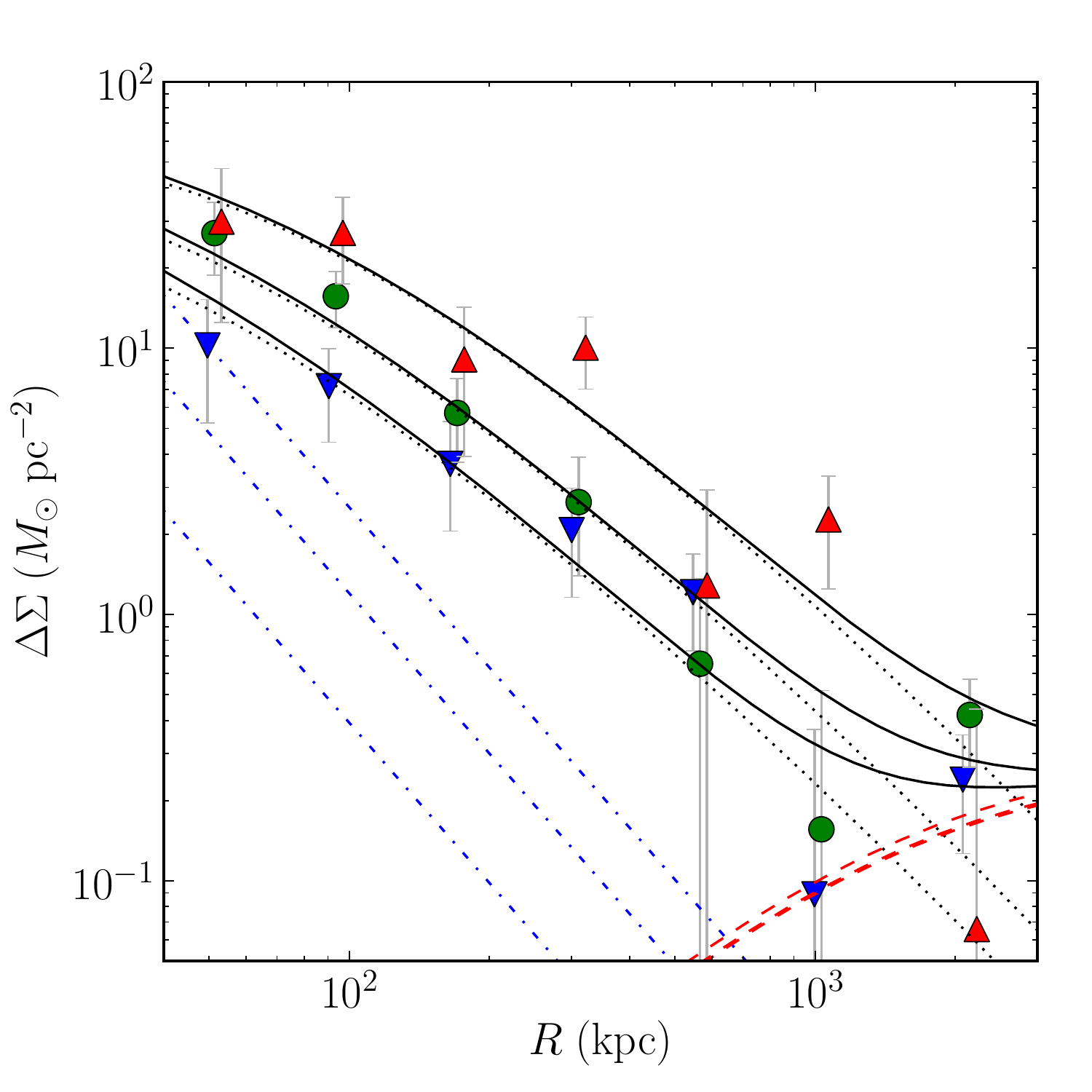}
\caption{Measured lensing signals $\ds(R)$ around stacked disk galaxies in three $\mstr$ bins with weighted mean stellar masses of
0.62, 2.68, and $6.52 \times 10^{10} M_\odot$,
shown as blue inverted triangles, green circles, and red triangles, respectively. Also shown are the best-fitting
one-halo and halo-halo profiles (black dotted and red dashed curves, respectively), the estimated baryonic component (blue dot-dashed curves)
and the sum of these three (black solid curves). 
The range of scales used for the fits is $R=$ 50--2000 kpc. For this range of scales, the baryonic contribution is
negligible and the halo-halo contribution is sub-dominant to the
one-halo term. We model the lensing signal as a sum of the one-halo
and halo-halo profiles (as described in Sec.~\ref{sec:lens_fits}).} 
\label{fig:ds_baryoncont}
\end{figure}

Figure~\ref{fig:ds_baryoncont} shows the measured lensing signals $\ds(R)$ for our lens sample in three stellar mass bins with lensing-weighted mean stellar masses of 0.62, 2.68, and 6.52 $\times 10^{10} M_\odot$ 
(inverted blue triangles, green circles, and red triangles, respectively). The amplitude of the measured lensing signal increases with stellar mass, indicating increasing halo mass, as expected. We perform fits to model density profiles (shown by different curves in Fig.~\ref{fig:ds_baryoncont}) to determine halo masses in Sec.~\ref{sec:lens_fits}.

Table~\ref{tab:mstr_bins} lists the basic properties of each stellar mass bin, including the range in $M_*$, the number of galaxies in the bin $N_{\rm gal}$, the unweighted and lensing-weighted mean stellar mass (cols.~2--5). The lensing-weighted mean rotation velocity (col.~6) is calculated using the TFR, Eq.~\ref{eq:tfr}.

\begin{figure}
\includegraphics[width=3in]{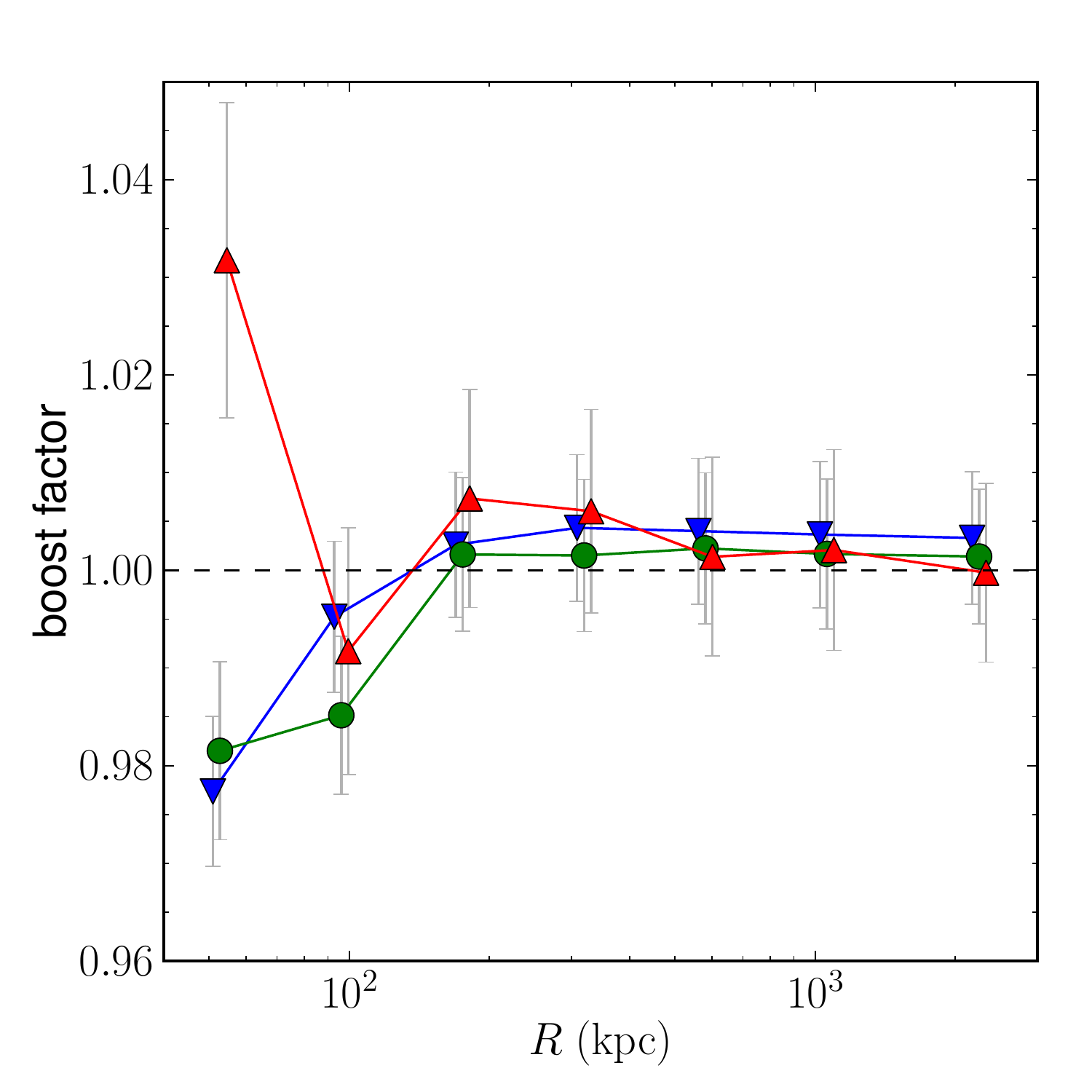}
\caption{Boost factors $B(R)$ applied to the lensing signals shown in Fig.~\ref{fig:ds_baryoncont}, 
for three $\mstr$ bins with weighted mean stellar masses of
0.62, 2.68, and $6.52 \times 10^{10} M_\odot$,
(blue inverted triangles, green circles, and red triangles, respectively). As expected, the boost factors are 
consistent with unity for $R \gtrsim 200$ kpc.}
\label{fig:boost}
\end{figure}

Figure~\ref{fig:boost} shows the boost factors $B(R)$ applied to these lensing signals (symbols refer to the same bins). We find that $B(R)$ is consistent with 1 for $R\gtrsim 200$ kpc in all three cases, as expected. 

\begin{table*}
\begin{tabular}{rrrrrrrr}
\hline
Range in $\log \mstr$ & $N_{\rm gal}$ & $\langle\log \mstr \rangle$& $\log \langle \mstr \rangle_{\rm L}$ & $\log \langle \vopt\rangle_{\rm L}$ & $\langle z \rangle$ & $\langle M_i \rangle$ & $\langle g-r\rangle$ \\
$(M_\odot)$ & & $(M_\odot)$ & $(M_\odot)$ & $(\kms)$ & & (mag) & (mag) \\
\hline
9.00--10.22 & 78419 &     9.808 & 9.792         & 2.06 &  0.0555 & $-19.87$ & 0.4608 \\
10.22--10.70 & 47419  &  10.421 & 10.428    &  2.23 &  0.0744 & $-21.06$ & 0.6124 \\
10.70--11.00  & 7760   & 10.807 & 10.814    &   2.34 &  0.0762 & $-21.86$ & 0.6821 \\
\hline
\end{tabular}
\caption{Basic properties of the stellar mass bins used in the lensing analysis. The subscript ``L'' indicates lensing-weighted means, with weighting factors $w_{\rm ls}$.} 
\label{tab:mstr_bins}
\end{table*}

\input{systests_RM.tex}


\section{Fits to the lensing signal}
\label{sec:lens_fits}

We perform fits to the measured lensing signals to determine halo masses and halo-to-stellar mass ratios as a function of stellar mass.
We describe these fits in this section, and present their results in the following section, Sec.~\ref{sec:results}.
We begin by describing the different components of the model lensing profile (Sec.~\ref{subsec:density_profiles}). Then, we describe the fiducial fits from which we obtain our main results (Sec.~\ref{subsec:fiducial_fits}). Finally, we describe various alternative fits that we use to test the robustness of these results (Sec.~\ref{subsec:alt_fits}). 
Overall, we find that our results are not very sensitive to the particular choices made in these fits.

\subsection{Density profiles} 
\label{subsec:density_profiles}

There are several contributions to the observed lensing signal $\Delta\Sigma(R)$. The dominant contribution comes from the density profile of the dark matter halo in which the galaxy lives (also called the one-halo term). On large scales, beyond 1 Mpc, there is a contribution from statistical correlations between dark matter halos from large scale structure (also called the halo-halo term). On small scales, below several hundred kpc, there is a contribution from the baryons (stars and gas) in the galaxy. 

Figure~\ref{fig:ds_baryoncont} shows these three different contributions to the lensing signal as (black) dotted, (red) dashed, and (blue) dot-dashed curves, respectively. 
The one-halo and halo-halo terms shown are the best-fitting halo profiles from fits to the data (described in Sec.~\ref{subsec:fiducial_fits}). The baryonic contributions shown are for galaxies with stellar mass equal to the lensing-weighted average $M_*$ for each stellar mass bin, and gas masses estimated from the average relation between gas-to-stellar mass ratios and $M_*$ (based on data from R11, c.f. their fig.~11).\footnote{The estimated gas-to-stellar mass ratios are 0.99, 0.41, and 0.24, for the three stellar mass bins, respectively, based on the best-fitting relation from a weighted fit to the 189 galaxies in the TFR sample of R11: $\log (M_{\rm gas}/M_*) = (-0.19 \pm 0.02) + (-0.60\pm 0.04)(\log M_* - 10.102)$.}

We find that over the range of scales used for the lensing analysis, 50--2~000 kpc, the contribution to the lensing signal from the baryonic component
of the galaxy is negligible.\footnote{Because the lensing signal at a given radius includes contributions from the mass at all smaller radii, it is not obvious {\it a priori} that the relative contribution to the lensing signal from the baryonic component of the galaxy (with a typical scale length of several kpc) will be negligible on the scales used in our fits (50--2000 kpc). Since the baryons are predominantly on scales $R< 10$ kpc, their contribution to the lensing signal on the relevant scales is simply $\Delta\Sigma_{\rm bar} = \mbar/(\pi R^2)$.}
On the other hand, the contribution from the halo-halo term is negligible below $\sim 1$ Mpc, and its contribution to the total lensing signal for the outermost radial bin is comparable to the 1$\sigma$ uncertainty in the measurement.  

Recall that we have excluded satellite galaxies from the lens sample (as described in Sec.~\ref{subsec:data_disk}) to simplify the analysis and interpretation of the lensing signal. For a satellite galaxy, which resides in a dark matter halo that is a subhalo in some larger host halo, the lensing signal will also have a contribution on several hundred kpc to 1 Mpc scales from the host halo. Usually, this term is interpreted statistically in terms of a halo model describing the fraction of the lens galaxy sample that are satellites, and using the halo mass function to estimate the typical lensing signal due to the host halos. By excluding satellites from our lens sample, we remove the need to model this term, thus simplifying our analysis.


We model the halo mass distribution as a Navarro-Frenk-White (1996; hereafter NFW) profile of cold dark matter haloes 
\beq \label{eq:nfw}
\rho(r)=\frac{\rho_{\rm s}}{(r/r_{\rm s})(1+r/r_{\rm s})^2},
\eeq
defined by two parameters, a characteristic density $\rho_{\rm s}$ and scale radius $r_{\rm s}$, or alternatively, 
the virial mass $\mvir$ and concentration $\cvir = \rvir/r_{\rm s}$. In this work, we adopt the definition of the virial mass as the mass enclosed within the virial radius $\rvir$ within which the average density is equal to 200 times the critical density of the Universe today $\rho_{\rm crit}$,
\beq \label{eq:mvir}
\mvir = \frac{4\pi}{3}\rvir^3(200\rho_{\rm crit}),
\eeq 
where the subscript denotes that this mass definition uses 200$\rho_{\rm crit}$. Compared to the other commonly adopted mass definition of $\mvirb$ using $200\bar{\rho}=200\Omega_{\rm m}\rho_{\rm crit}$, the virial masses and concentrations using 200$\rho_{\rm crit}$ are lower by roughly 30 and 60 per cent, respectively (for the range of halo masses we study).

The NFW halo concentration is a weakly decreasing function of halo mass with a typical dependence given by
\beq \label{eq:cofM}
\cvir(\mvir) = c_{\rm 200c,0}\left(\frac{\mvir}{10^{12} M_\odot}\right)^{-\betac},
\eeq
with $\betac \sim 0.1$ \citep{2001ApJ...555..240B,2007MNRAS.381.1450N,2008MNRAS.391.1940M}. 
In our fiducial fits, we adopt the results of $N$-body simulations from Maccio et~al. (2008) with cosmological parameters from WMAP5 \citep{2009ApJS..180..330K}. For all (both relaxed and unrelaxed) haloes with $10^{10} \la \mvir/M_\odot \la 10^{15}$, they found $\betac=0.110$. Following their redshift evolution model for the concentration-mass relation, we scale their relation by $H(z)^{-2/3}$ to account for the difference in the effective redshifts of our two samples ($z=0$ vs. $0.07$); this yields $c_{\rm 200c,0}=6.00$.\footnote{Using $H(z) \approx H_0\sqrt{\Omega_{\Lambda} + \Omega_{\rm m}(1+z)^3}$, we find that the evolution in redshift amounts to only a 2 per cent growth from $z=0.07$ to 0.}
They also report a log-normal scatter in the halo concentrations at fixed halo mass of 0.130 dex (this reduces to 0.105 dex when the sample is restricted to only relaxed haloes). 
Our modelling of the lensing signals does not take into account the scatter in the concentration-mass relation at fixed halo mass. However, we do not think that this will significantly affect the results of the fits since we find that they are not very sensitive to the overall normalization of the concentration-mass relation itself, as shown in Sec.~\ref{subsec:alt_fits}.

The halo-halo contribution to the lensing signal is modeled using the galaxy-matter cross-power 
spectrum as in, e.g., \cite{2005MNRAS.362.1451M}. It is proportional
to the bias $b$, the ratio of the galaxy-matter correlation function
to the matter autocorrelation function. We estimate the bias using the 
fitting formulae from \cite{2004MNRAS.355..129S}. At the mean redshift of our galaxy sample $z=0.07$, 
the growth factor $D(z)=0.9665$ and the non-linear mass is $M_{\rm nl}=10^{12.619} M_\odot$. 
We find the typical bias of galaxies in our sample to be approximately 0.7--0.9. 
Note that our removal of satellite galaxies from the lens sample may modify the halo-halo term in such a way that this model is no longer accurate,
but we also note that this effect may not be significant because the satellite cuts remove only $\sim 15$ per cent of the sample. 
Reassuringly, we find that our fits are not very sensitive to the assumptions that go into the modelling of the halo-halo term, as shown in Sec.~\ref{subsec:alt_fits}. 

In principle, intrinsic alignments of galaxy
shapes towards the lens \citep{2004PhRvD..70f3526H} can mimic a
negative lensing signal.  This effect can be important if there is
significant weight in the lensing signal given to ``sources'' that are
actually physically associated to the lens.  In that case, this uncertainty in the
theory would have to be modeled. As shown in Fig.~\ref{fig:boost}, the boost factors, which indicate
what fraction of the sources are actually physically associated
galaxies at the lens redshift, are within $\sim 2$ per cent of unity on
all scales. This finding is unsurprising for a disk galaxy lens population,
which should be dominated by field galaxies.  It is possible that
there are very small numbers of source galaxies that are close enough to be
affected by the same tidal field as the lens (i.e., within a few tens
of Mpc) without causing a boost factor $>1$, because we expect
some galaxies at these distances even for a random distribution.
However, the weighting of the sources by $1/\Sigma_c^2$ means that
such nearby galaxies receive an extremely low weight, in addition to
the fact that they are already quite rare given the relatively small
comoving volume around $\zlens$ compared to that around the
source median redshift of $\sim 0.4$.  Thus, we expect intrinsic
alignments to be completely subdominant compared to all other errors
described in this work, and we neglect them in the modelling.

\subsection{Fiducial fits}
\label{subsec:fiducial_fits}

Here, we describe the fits from which we obtain the results presented in Sec.~\ref{sec:results}. First, we individually fit to the measured lensing signals for the three stellar mass bins, to obtain the best-fitting halo-to-stellar mass ratio $\mvir/M_*$ for each bin (Sec.~\ref{subsubsec:mvir_fits}). Second, we model the HSMR with four free parameters and simultaneously fit the measured lensing signals for the three stellar mass bins to obtain the best-fitting HSMR (Sec.~\ref{subsubsec:hsmr_fits}).

\subsubsection{Fits to $\mvir/M_*$}
\label{subsubsec:mvir_fits}

The model density profiles described in Sec.~\ref{subsec:density_profiles} (and shown in Fig.~\ref{fig:ds_baryoncont}) represent the lensing signal of a single halo. We model an individual halo profile as a sum of the one-halo and halo-halo contributions, assuming a fixed concentration-mass relation given by Eq.~\ref{eq:cofM}. On the other hand, the measured lensing signal for each stellar mass bin has contributions from a large number of haloes, with a distribution of halo masses. To take this into account, we fit the measured lensing signals to the mean profile of an ensemble of individual haloes with a distribution of halo masses based on the actual distribution of stellar masses in each bin. 

For our fiducial fits, we assume a log-normal scatter of $\sigma_{\log \mvir}=0.1$ dex at a fixed $M_*$, and convolve the distribution in $\log M_*$ by a Gaussian distribution with this width before the conversion to a distribution of halo masses. Typically, this scatter is expressed in terms of the log-normal scatter in stellar mass at a fixed $\mvir$, $\sigma_{\log M_*}$. Previous analysis of SDSS galaxies found that the observed stellar mass function (for all galaxy types) can be adequately fit by assuming $\sigma_{\log M_*}=0.15$ dex \citep{2010ApJ...710..903M} and 0.175 dex \citep{2010ApJ...717..379B}. We expect the amount of scatter to be somewhat smaller for our sample including only disk galaxies. Moreover, $\sigma_{\log \mvir}$ is smaller than $\sigma_{\log M_*}$ because of the shallow slope of the HSMR over the range of stellar masses we study. Therefore, our choice of $\sigma_{\log \mvir}=0.1$ dex is a conservative one, even after accounting for the statistical uncertainty in $M_*$ of 0.04 dex. We consider alternative fits with no scatter and with twice the fiducial value in Sec.~\ref{subsec:alt_fits}.


Given the above modelling assumptions, the predicted lensing signal for some stellar mass bin is determined by a single parameter, the halo-to-stellar mass ratio $\mvir/M_*$, after one assumes a certain dependence of $\mvir/M_*$ on $M_*$. For the fits to the lensing signals for each $M_*$ bin, we assume a constant $\mvir/M_*$ (over the stellar mass range of each bin). We use a Levenberg-Marquardt minimization routine \citep{levenberg,marquardt} to fit for $\log \mvir/M_*$ using the measured lensing signal from $\sim$ 50--2000 kpc (27 radial bins in all). For each stellar mass bin, we perform an independent fit to each of the 500 bootstrap resamplings of the 200 subregions. From the mean and width of the distribution of bootstrap parameters, we obtain the best-fitting value of $\mvir/M_*$ and its 1$\sigma$ bootstrap error. We take the 5 per cent uncertainty in the shear calibration into account in the bootstrap errors as follows. For each bootstrap dataset, we multiply the predicted lensing signal by a random number sampled from a Gaussian distribution centered at 1 and with a standard deviation of 0.05 (we note though that the inclusion of this uncertainty has a negligible effect on the error in $\mvir/M_*$).


\subsubsection{Fits to the HSMR}
\label{subsubsec:hsmr_fits}

There is a growing consensus that the halo-to-stellar mass ratio of galaxies shows variation with stellar mass, with a minimum at $\mvir \sim 10^{12} M_\odot$ and $M_* \sim 5\times 10^{10} M_\odot$ and increasing towards lower and higher masses \citep{2006MNRAS.368..715M,2009ApJ...696..620C,2010ApJ...710..903M,2010ApJ...717..379B,2010MNRAS.404.1111G,2012ApJ...744..159L}. Based on the halo occupation distribution (HOD) modeling of \citet{2010ApJ...717..379B}, the relation between the halo-to-stellar mass ratio $\mvir/M_*$ and $M_*$ (or HSMR) can be modelled as 
\beqa \nonumber
\log \left(\frac{\mvir}{M_*} \right) &=& \log \left( \frac{M_1}{M_{*,0}} \right)+ (\beta -1) \log \left(\frac{M_*}{M_{*,0}}\right) \\   \label{eq:hsmratio}
&+& \frac{\left(M_*/M_{*,0}\right)^\delta}{1+\left(M_*/M_{*,0}\right)^{-\gamma}} - \frac{1}{2}.
\eeqa
Here, $M_1$ is the characteristic halo mass, $M_{*,0}$ is the characteristic stellar mass, $\beta$ is the faint end slope, and $\delta$ and $\gamma$ control the massive end behavior.
\citet{2012ApJ...744..159L} adopted the same functional form to constrain the HSMR via a joint analysis of galaxy-galaxy lensing, galaxy spatial clustering, and galaxy spatial densities of galaxies in COSMOS. 

It is unclear what the form of the HSMR is for a sample of only late-type galaxies. Moreover, there are not many disk galaxies that have stellar masses larger than the mass where the HSMR is at its minimum (for reference, our galaxy sample ranges from, $M_* \sim 10^9$ -- $10^{11} M_\odot$). We choose to adopt Eq.~\ref{eq:hsmratio} as a fitting function for the HSMR, with the understanding that we cannot strongly constrain its behavior at high stellar masses. 
Since we are not sufficiently sensitive to $\delta$ to fit for it, we fix $\delta=0.566$, following the results of \citet{2012ApJ...744..159L} for their lowest redshift bin $z=[0.22,0.48]$ and {\sc sig\_mod1} run (c.f. their table~5); they also report a $1\sigma$ uncertainty in $\delta$ of 0.086.\footnote{We note that changes in $\delta$ only affect the HSMR over a small portion of the range of stellar masses we study, $M_* \sim 4$ -- $8 \times 10^{10} M_\odot$, and even a 5$\sigma$ change in $\delta$ leads to differences well below our measurement uncertainties. Therefore, our results are not very sensitive to this modelling assumption.}

We fit for four free parameters--- a normalization, break, a faint end slope, and a bright end slope. The basic set-up is similar to the fits to $\mvir/M_*$ described in Sec.~\ref{subsubsec:mvir_fits}. For fits to the HSMR, we simultaneously fit the measured lensing signals for galaxies in the three stellar mass bins using a Levenberg-Marquardt minimization routine. In all, there are 27 (radial bins) $\times$ 3 (stellar mass bins) = 81 data points. We perform an independent fit to each of the 500 bootstrap resamplings of the 200 subregions to obtain the median HSMR as well as 1$\sigma$ and 2$\sigma$ error envelopes. We take into account of the 5 per cent uncertainty in the shear calibration in the bootstrap errors in the same way as described in Sec.~\ref{subsubsec:mvir_fits}; for each bootstrap dataset, we multiply the same random number to the lensing signal for all three stellar mass bins.

\subsection{Alternative fits}
\label{subsec:alt_fits}

In this subsection, we summarize the results of our tests of the robustness of the results of the fiducial fits (described in Sec.~\ref{subsec:fiducial_fits}) to various modeling assumptions. In each case below, we change a single parameter in the fiducial fit and check the effect of this change on the results. In all cases, we find that the results are robust to reasonable changes in the modelling assumptions, and we quote the magnitude of each effect below.

\begin{itemize}

\item \textit{Fits without the halo-halo contribution} 
\\

In our fiducial fits, the assumed contribution from the halo-halo term depends on a theoretically-uncertain estimate of the galaxy bias, 
as well as an assumed functional form. This may not be an accurate model for our lens sample, from which satellite galaxies have been removed. However, as noted in Sec.~\ref{subsec:density_profiles} (and shown in Figure~\ref{fig:ds_baryoncont}), the halo-halo contribution is negligible below $\sim 1$ Mpc and only affects the lensing signal measurement in the outermost radial bin at $\sim 2$ Mpc. Moreover, the amplitude of the halo-halo term is comparable to the uncertainty in the measurement at that radius, so we do not expect our fits to be very sensitive to changes in this term. 

To test this explicitly, we perform fits to model profiles that do not include the halo-halo contribution at all. As expected, these fits yield slightly higher best-fitting $\mvir/M_*$, by 0.02 -- 0.03 dex, which is well below the 1-$\sigma$ errors, 0.09 -- 0.13 dex.
\\

\item \textit{Fits with an alternate concentration-mass relation} 
\\

To test the sensitivity of the fits on the adopted concentration-mass relation, Eq.~\ref{eq:cofM}, we perform fits with an alternate relation. 
\citet{2008JCAP...08..006M} used lensing measurements of SDSS galaxies brighter than $L_*$ to fit for the relation over halo masses $10^{12} \la \mvirb/M_\odot \la 10^{15}$ and found $c_{\rm 200b}(\mvirb)=(4.6\pm 0.7) \times (\mvirb/(1.56 \pm 0.12) \times 10^{14} h^{-1} M_\odot)^{-(0.13\pm 0.07)}$ at $z=0.22$. Converting to our virial mass definition $\mvir$ and scaling by $(1+z)^{-1}$ to our mean redshift of 0.07, we find the equivalent parameters for Eq.~\ref{eq:cofM} to be $c_{\rm 200c,0}=5.84$ and $\betac=0.14$. Halo concentrations derived from this alternate relation are around 10 per cent lower than the fiducial one. One would expect the lower concentrations to yield systematically higher $\mvir/M_*$; we indeed find higher values but the difference is tiny, $\la 0.01$ dex for the three stellar mass bins. 

We note that \citet{2011arXiv1104.5130P} studied halo concentrations using a set of state-of-the-art $N$-body simulations. They report halo concentrations that are $\sim 10$ per cent lower than those of Maccio et~al. (2008) for Milky Way-size haloes; this difference is of the same order as for the alternate relation we tested, so we do not expect it to have a significant effect on our fits either.\footnote{We note that while the two works agree for Milky Way-size haloes, for cluster-size haloes, Prada et~al. (2011) report substantially larger halo concentrations (by up to 50 per cent) than Maccio et~al. (2008).}
We conclude that our results are not sensitive to our particular choice of concentration-mass relation.
\\


\item \textit{Fits with alternate values of $\sigma_{\log \mvir}$} 
\\

To convert the distribution in stellar masses in each bin into a distribution in halo masses, we assume a log-normal scatter of  $\sigma_{\log \mvir}=0.1$, at a fixed $M_*$. To test the sensitivity of the fits on this choice, we perform fits with alternate values $\sigma_{\log \mvir}=0.0$ and 0.2 dex. Assuming zero scatter yields best-fitting $\mvir/M_*$ that are higher by 0.01 dex (for all three stellar mass bins), while assuming twice the scatter yields values lower by 0.02 dex; in both cases, the difference is much smaller than the 1-$\sigma$ uncertainty in the measurement itself.
\\



\item \textit{Fits to a lens sample including satellites} 
\\

To test the sensitivity of the results to the removal of satellite galaxies from the lens sample (described in Sec.~\ref{subsec:data_disk}), we perform fits to the disk galaxy sample before the satellite cut. We do not attempt to model the contribution of satellites to the lensing signal, but simply restrict the range of the fits to exclude the outermost radii, where the satellite contribution dominates. Using the lensing signal from $\sim$ 50 -- 300 kpc (11 radial bins in all), we find best-fitting $\mvir/M_*$ that are consistent to within $1\sigma$ of those from the fiducial fits to the sample with the satellite cut applied.\footnote{We note that the effect of the inclusion of the satellites varies with stellar mass: the sign is zero, positive, and negative for the lowest, intermediate, and highest stellar mass bins, respectively. Not surprisingly, the sign and amount of the difference also depends on the choice of range of radii used for the fit.}  
\\




\item \textit{Fits to a lens sample with an axis ratio cut} 
\\

Recall that the selection of the TFR sample from which $\vopt$ is derived includes an axis ratio cut $b/a<0.6$. The lens sample from which $\vvir$ is derived has not been subjected to an axis ratio cut. In this section, we show that this difference in selection does not compromise our results for $\vopt/\vvir$.

We calculate halo-to-stellar mass ratios for the subsample with $b/a<0.6$ and found $\log(\mvir/M_*)=1.75\pm 0.14$, $1.43\pm 0.13$, and $1.37\pm 0.20$ for each stellar mass bin, respectively. For a comparison with the full sample that takes into account the correlation between the two samples, we calculate the ratio of halo masses for each bootstrap resampling. We find that the bootstrap distributions in $\mvir(b/a<0.6\ {\rm subsample})/\mvir({\rm full\ sample})$ are approximately Gaussian, with mean values consistent with unity: $1.37\pm 0.36$, $1.21\pm 0.25$, and $0.96\pm 0.33$,
for each stellar mass bin, respectively, indicating that the application of the axis ratio cut does not significantly bias the results presented here. We also note that there is considerable overlap between the two samples being compared; consequently, the derived halo masses are correlated at the 60--70 per cent level, but the bootstrap errors on the ratios above automatically account for these correlations.

\end{itemize}

\section{Results}
\label{sec:results}

In this section, we present our derived constraints on the HSMR and OVVR (Secs.~\ref{subsec:results_hsmratio} \& \ref{subsec:results_vovratio}). Then, we discuss comparisons of our results with previous work (Sec.~\ref{subsec:prev_work}) and with predictions for $\Lambda$CDM haloes (Sec.~\ref{subsec:comp_lcdm}). 

\subsection{Halo-to-stellar mass relation}
\label{subsec:results_hsmratio}

\begin{figure}
\includegraphics[width=3in]{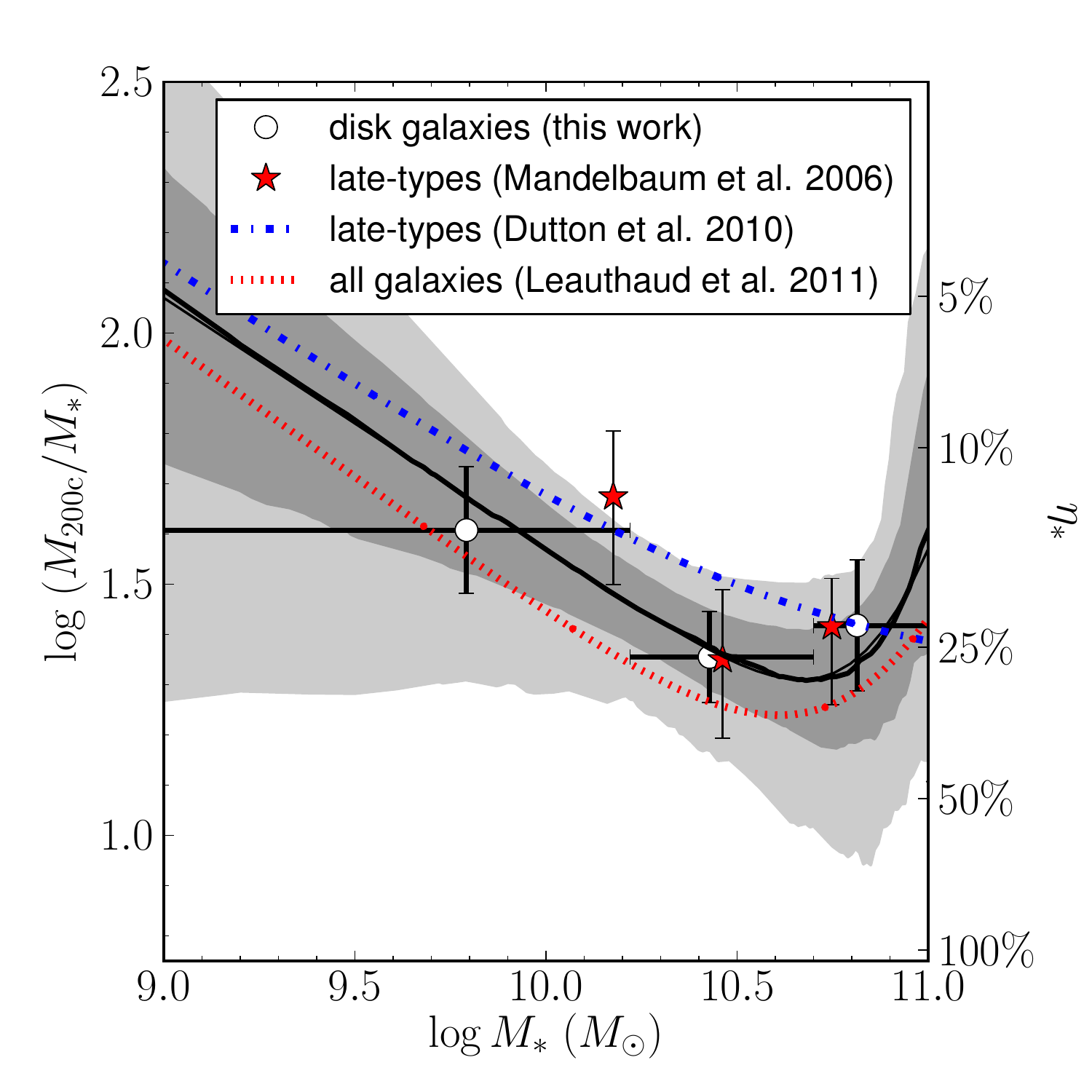}
\caption{Constraints on the HSMR from simultaneous fits to the lensing signals for three stellar mass bins. The thick solid curve and dark and light grey shaded regions show the median relation and its $1\sigma$ and $2\sigma$ error envelopes. Best-fitting $\mvir/M_*$ for the three stellar mass bins are shown by circles with $1\sigma$ error bars; horizontal error bars indicate bin widths. The thin solid curve (which largely overlaps with the \textit{thick} solid curve) shows the analytical fit to the median relation, given by Eq.~\ref{eq:hsmratio} with best-fitting parameters listed in the main text. Results from Mandelbaum et~al. (2006) are shown by red stars with $1\sigma$ error bars. Published HSMRs from Dutton et~al. (2010) and Leauthaud et~al. (2011) are shown by the blue dot-dashed and red dotted curves, respectively, after applying conversions for differences in the choices of stellar IMF and virial mass definition. Numbers on the right vertical axis indicate stellar conversion efficiencies $\eta_* \equiv (M_*/\mvir)f_{\rm b}^{-1}$.}
\label{fig:hsmratio}
\end{figure}

Figure~\ref{fig:hsmratio} shows our constraints on the relation between $\mvir/M_*$ and $M_*$. In qualitative agreement with the literature, our data show a clear variation in $\mvir/M_*$ with $M_*$ and suggest a minimum at $M_* \sim 5 \times 10^{10} M_\odot$. Best-fitting $\mvir/M_*$ for three stellar mass bins are shown as open cirles (and listed in Table~\ref{tab:nfwfit_ratios}); the median relation and $1\sigma$ and $2\sigma$ error envelopes are shown by the thick solid curve and dark and light grey shaded regions, respectively. Recall that the bootstrap error envelopes include the 5 per cent uncertainty in the shear calibration. We note that we use results from 471 out of the 500 bootstrap datasets (of 200 subregions) that resulted in converged fits.

We provide the derived constraints in tabular form in Table~\ref{tab:pdf_hsmratio} for a grid in stellar mass from $M_* = 10^9$ -- $10^{11} M_\odot$ in steps of 0.1 dex. Since each bootstrap dataset is fitted by a HSMR with a different set of parameters, the median curve and error envelopes do not exactly follow the functional form in Eq.~\ref{eq:hsmratio}. To obtain an analytical fitting formula, we fit the derived median relation to Eq.~\ref{eq:hsmratio}, for which we find best-fitting parameters: 
\beqa \nonumber
\log(M_1/M_{*,0}) &=& 1.57\pm 0.04 \\ \nonumber
\log M_{*,0} &=& 11.00 \pm 0.03 \\ \nonumber
\beta-1 &=& -0.50 \pm 0.03 \\ \nonumber
\gamma &=& 2.60 \pm 0.77,
\eeqa
with $\delta=0.566$.\footnote{Allowing $\delta$ to freely vary does not lead to a closer approximation to the median HSMR, so we present the fits with $\delta$ fixed to its fiducial value.} This relation is shown as the thin solid curve in Fig.~\ref{fig:hsmratio}; as shown, it very closely approximates the median curve (shown by the {\it thick} solid curve) over the range of stellar masses we study. For this model and the data from the bootstrap mean of the lensing signals of the three stellar mass bins, we calculate a $\chi^2$ value of 46 for $68-4=65$ degrees of freedom, indicating that the data are consistent with this model. 

We calculate Pearson correlation coefficients $r(P_i,P_j)$ and covariances using the results from the bootstrap datasets, where $P_0=\log(M_1/M_{*,0})$, $P_1=\log M_{*,0}$, $P_2=\beta-1$, and $P_3=\gamma$. We find a significant correlation between two pairs of parameters: $r(P_0,P_2)=0.68$ and $r(P_1,P_2)=0.48$, both with $p$-values $\ll 1$. 
We also find that $P_1$ and $P_3$ correlate in a peculiar manner; their joint distribution does not follow an elliptical contour. We note that these covariances are properly taken into account in our fits through our use of bootstrap analysis.

The right vertical axis of Fig.~\ref{fig:hsmratio} shows stellar conversion efficiencies $\eta_* \equiv (M_*/\mvir)f_{\rm b}^{-1}$, the percentage of the cosmologically available baryons that end up as stars in the galaxy, with $f_{\rm b} \equiv \Omega_{\rm m}/\Omega_{\rm b}=0.169$ \citep[WMAP7,][]{2011ApJS..192...18K}. 
Including the contribution of cold gas, we also estimate ``baryon retention fractions'' $\eta_{\rm b} \equiv [(M_*+M_{\rm gas})/\mvir] f_{\rm b}^{-1}$. For the three bins with lensing-weighted average stellar masses of 0.62, 2.68, and 6.52 $\times 10^{10} M_\odot$, we find $\eta_*=0.15^{+0.05}_{-0.04}$, $  0.26^{+0.06}_{-0.05}$, $  0.23^{+0.08}_{-0.06}$ and $\eta_{\rm b}=0.30^{+0.10}_{-0.08}$, $0.37^{+0.08}_{-0.07}$, $0.29^{+0.10}_{-0.07}$, respectively, with 1$\sigma$ error bars based on the uncertainties in $\mvir/M_*$. Here, we have used gas-to-stellar mass ratios of 0.99, 0.41, and 0.24, respectively, based on the empirical relation described in Sec.~\ref{subsec:density_profiles} (and quoted in a footnote there). 

\begin{table}
\caption{Constraints on $\mvir/M_*$ from individual fits to the lensing signals of galaxies in three stellar mass bins (col.~1), stellar conversion efficiencies $\eta_* \equiv (M_*/\mvir)f_{\rm b}^{-1} $ (col.~2), and $\vopt/\vvir$ from combination with the calibrated TFR (col.~3). Our main results are from the full lens sample (top). We also list here results for the subsample with axis ratio cut $q<0.6$ (bottom); we find no systematic bias between the two, as described in the final subsection of Sec.~\ref{subsec:alt_fits}.}
\begin{tabular}{cccc}
\hline
$\log \langle \frac{\mstr}{M_\odot} \rangle_{\rm L}$ & $\log(\mvir/M_*)$ & $\eta_*$ &$\vopt/\vvir$ \\
\hline
\multicolumn{4}{c}{Full lens sample} \\
\hline
\input{Tables/nfwfit_ratios.v15.tex}
\hline
\multicolumn{4}{c}{$q<0.6$ subsample} \\
\hline
\input{Tables/nfwfit_ratios.v18.tex}
\hline
\end{tabular}
\label{tab:nfwfit_ratios}
\end{table}

\begin{table}
\caption{Median HSMR and $\pm 1$ and $2\sigma$ error envelopes.}
\begin{tabular}{rrrrrr}
\hline
$\log\frac{M_*}{M_\odot}$ & $-2\sigma$ & $-1\sigma$ & $\log\frac{\mvir}{M_*}$ & $+1\sigma$ & $+2\sigma$ \\
\hline
\input{Tables/pdf_hsmratio.tex}
\hline
\end{tabular}
\label{tab:pdf_hsmratio}
\end{table}


\subsection{Optical-to-virial velocity relation}
\label{subsec:results_vovratio}

We derive constraints on the relation between $\vopt/\vvir$ and $M_*$ (or OVVR) as outlined in Sec.~\ref{subsec:overview_comb}.
For each bootstrap dataset, (i) we generate a $\vopt$ vs. $M_*$ relation based on a TFR with the zero-point and slope randomly sampled (independently) from a Gaussian distribution centered on the best-fitting values and of Gaussian widths equal to the $1\sigma$ fit uncertainties in those values. (We note that this scatter in the TFR is negligible compared to the uncertainty in the HSMR, of $\sim13$ per cent.) Next, (ii) we convert the $\mvir/M_*$ vs. $M_*$ relation into a $\mvir$ vs. $M_*$ relation, and then into a $\vvir$ vs. $M_*$ relation via $\vvir =(G \mvir / \rvir)^{1/2}$  Dividing relations (i) and (ii) yields a OVVR for each bootstrap dataset. Finally, we obtain the median relation and its 1$\sigma$ and 2$\sigma$ error envelopes from the bootstrap distributions. 

We also obtain $\vopt/\vvir$ for each stellar mass bin. First, we multiply the best-fitting $\mvir/M_*$ by the lensing-weighted mean $M_*$ to obtain the corresponding $\mvir$, convert that into $\vvir$, and finally, take the ratio of that and the lensing-weighted mean $\vopt$. The results are listed in Table~\ref{tab:nfwfit_ratios}. 

\begin{figure}
\includegraphics[width=3in]{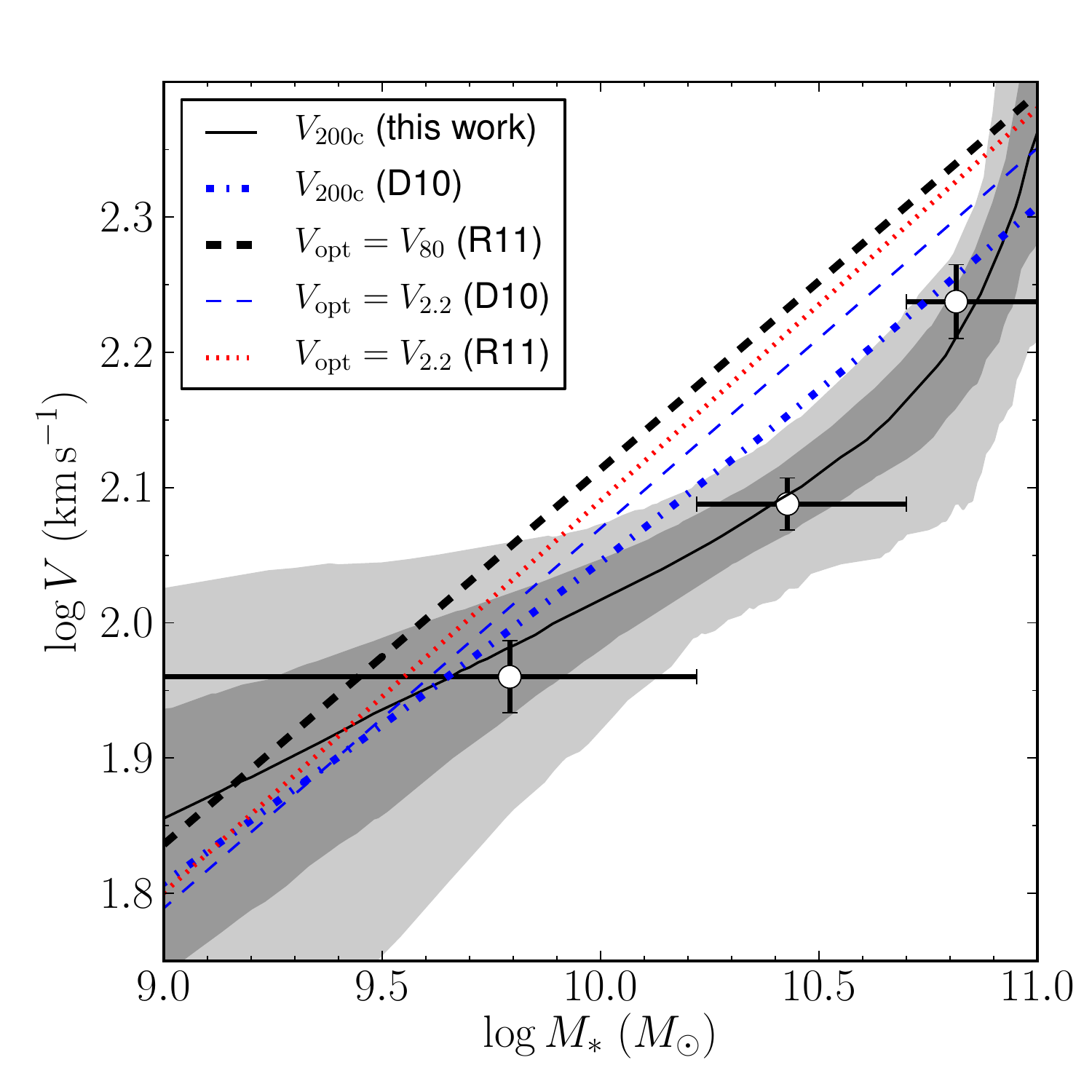}
\caption{Constraints on the $\vvir$ vs. $M_*$ relation are shown by the solid curve and dark and light grey shaded regions (median, 1$\sigma$ and 2$\sigma$ error envelopes, respectively). Virial velocities $\vvir$ for the three $M_*$ bins are shown by circles with 1$\sigma$ error bars; horizontal error bars indicate bin widths. The $\vopt$ vs. $M_*$ relation, Eq.~\ref{eq:tfr} (derived in R11) is shown as the thick dashed line. For comparison, we also plot the $\vvir$ vs. $M_*$ relation from the OVVR in Dutton et~al. (2010) (blue dot-dashed curve) and their $V_{2.2}$ vs. $M_*$ TFR (thin blue dashed line).}
\label{fig:vvir_vopt}
\end{figure}

\begin{figure}
\includegraphics[width=3in]{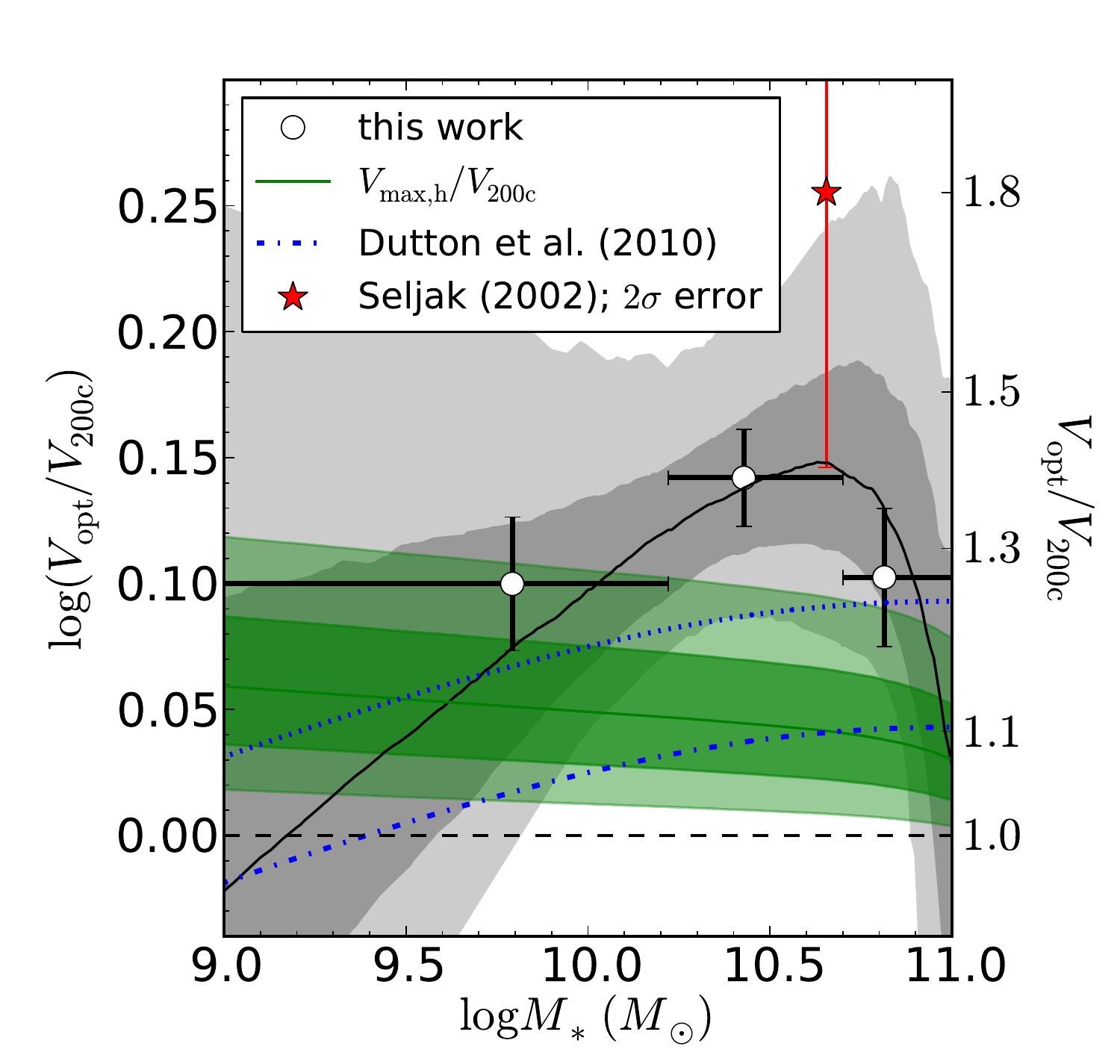}
\caption{Constraints on the OVVR are shown by the solid curve and dark and light grey shaded regions (median, 1$\sigma$ and 2$\sigma$ error envelopes, respectively). $\vopt/\vvir$ for the three $M_*$ bins are shown by circles with $1\sigma$ error bars; horizontal error bars indicate bin widths. The red star symbol is plotted with a 2$\sigma$ error bar and show $\vopt/\vvir$ derived for $L_*$ galaxies by Seljak (2002). The blue dot-dashed and dotted curves show the $V_{2.2}/\vvir$ and $V_{80}/\vvir$ vs. $M_*$ relations derived by Dutton et~al. (2010), before and after ``correcting'' for the differences in the TFRs used in that work vs. this one (note that the two differ by $\approx$ 0.05 dex). The dark and light green shaded regions show the variation in $V_{\rm max,h}/\vvir$ for unmodified pure dark matter NFW haloes, corresponding to $1\sigma$ and $2\sigma$ scatter (0.13 and 0.26 dex) in halo concentrations $\cvir(\mvir)$ (given by Eq.~\ref{eq:cofM}) at a fixed halo mass.} 
\label{fig:vovratio}
\end{figure}

Figure~\ref{fig:vvir_vopt} shows the $\vvir$ vs. $M_*$ relation and its 1$\sigma$ and 2$\sigma$ error envelopes (solid curve, dark and light grey shaded regions, respectively), together with the $\vopt$ vs. $M_*$ relation, Eq.~\ref{eq:tfr} (thick dashed line). Figure~\ref{fig:vovratio} shows the median OVVR and its 1-$\sigma$ and 2-$\sigma$ error envelopes (solid curve, dark and light grey shaded regions, respectively). We provide these constraints in tabular form (Table~\ref{tab:pdf_vovratio}) for a $M_*$ grid from $\log M_*/M_\odot = 9 - 11$ in steps of 0.1 dex. Numbers on the right vertical axis of Fig.~\ref{fig:vovratio} indicate $\vopt/\vvir$. 

We constrain the OVVR to around 6 per cent ($1\sigma$) and find $\vopt/\vvir \approx 1.3$ for stellar masses $10^9-10^{11}$. Recall that since $\vvir \propto \mvir^{1/3}$, the uncertainty on $\log(\vopt/\vvir)$ is a third of that in $\log \mvir$. We note that the shape of the OVVR is partly dictated by the assumed functional form for the HSMR, Eq.~\ref{eq:hsmratio}; although the best-fitting function suggests a turn-over at $\sim 3\times 10^{10} M_\odot$, the data are consistent with a flat $\vopt/\vvir$ with $M_*$. 

\begin{table}
\caption{Median OVVR and $\pm 1$ and $2\sigma$ error envelopes.}
\begin{tabular}{rrrrrr}
\hline
$\log\frac{M_*}{M_\odot}$ & $-2\sigma$ & $-1\sigma$ & $\log\frac{\vopt}{\vvir}$ & $+1\sigma$ & $+2\sigma$ \\
\hline
\input{Tables/pdf_ovvratio.tex} 
\hline
\end{tabular}
\label{tab:pdf_vovratio}
\end{table}

\subsection{Comparison with previous work}
\label{subsec:prev_work}

Figure~\ref{fig:vovratio} shows our constraints on the OVVR together with those from previous work that used a similar methodology as ours, namely, the combination of halo mass measurements (from weak lensing and/or other techniques) with TFR measurements from galaxy rotation curves. 

For late-type $L^*$ galaxies from an early SDSS dataset, \citet{2002MNRAS.334..797S} combined weak lensing measurements from \citet{2002MNRAS.335..311G} with an $I$-band TFR from \citet{1997ApJ...477L...1G} and found $\vopt/\vvir = 1.8$ with a $2\sigma$ lower limit of 1.4 (shown by the red star and error bar in Fig.~\ref{fig:vovratio}); there is no upper limit because the halo masses were consistent with zero. This measurement is only marginally consistent with ours, but our result benefits from significant improvements in the measurements of both weak lensing and the TFR, due to larger available datasets (both lens and source galaxy samples) and improved analysis methods. 

Dutton et~al. (2010, hereafter D10) combined their visual fit to the HSMR for late-type galaxies, modelled as a double power law, based on a compilation of measurements from weak lensing \citep{2006MNRAS.372..758M} and satellite kinematics \citep[][and Klypin, Prada \& Montero-Dorta, in preparation]{2007ApJ...654..153C,2011MNRAS.410..210M} with their fit to the $V_{2.2}$ vs. $M_*$ TFR based on rotation curve data from Pizagno et~al. (2007).
Their central HSMR is shown by the blue dot-dashed curve in Figure~\ref{fig:hsmratio}. For comparison, we also show measurements from \citet{2006MNRAS.372..758M} based on late-type galaxies in SDSS selected via the {\verb frac_deV } parameter (red stars with 1$\sigma$ error bars), as well as the best-fitting HSMR from Leauthaud et~al. (2011), modelled as Eq.~\ref{eq:hsmratio}, and derived from combined early- and late-type galaxies in COSMOS (red dotted curve). The COSMOS result is for galaxies at redshifts $0.2-0.5$, and the rest are for galaxies at a mean redshift of $z\sim 0.1$.

Figure~\ref{fig:vvir_vopt} compares the $\vvir$ vs. $M_*$ relation derived from the HSMR in D10 with ours (blue dot-dashed and black solid curves, respectively). As shown, the two relations are consistent within $\sim 2\sigma$ over the range of stellar masses we study. 
The figure also compares the $V_{2.2}$ vs. $M_*$ TFR used in D10 with the $V_{2.2}$ vs. $M_*$ and $V_{80}$ vs. $M_*$ TFRs from R11. As shown, the $V_{2.2}$ vs. $M_*$ TFR used in D10 is lower than that in R11 (blue dashed vs. red dotted lines). The difference between the TFRs is significant, even for the same definition of $\vopt$; this may be attributed to differences in the galaxy samples and analysis methods in the two analyses.\footnote{There is a large overlap between the TFR sample in R11 and the Pizagno et~al. (2007) sample used in D10. Out of 189 galaxies in the R11 sample, 99 galaxies are from the Pizagno et~al. (2007) sample (those out of the galaxies that passed our selection cuts). However, the analysis methods for deriving both photometric and kinematic quantities and their uncertainties, as well as fits to the TFR, are completely independent.} 
Compared to the $V_{80}$ vs. $M_*$ TFR in R11, the relation used in our derivation of the OVVR, the $V_{2.2}$ vs. $M_*$ TFR in D10 turns out to have a similar slope and a lower normalization by $\approx$ 0.05 dex.

Figure~\ref{fig:vovratio} shows the $V_{2.2}/\vvir$ vs. $M_*$ relation from D10, after converting their $V_{2.2}/\vvir$ vs. $V_{2.2}$ relation into a $V_{2.2}/\vvir$ vs. $M_*$ relation, using {\it their} $V_{2.2}$ vs. $M_*$ TFR (blue dot-dashed curve). D10 estimates that $V_{2.2}/\vvir \simeq 1$ for stellar masses $M_* = 5 \times 10^{9}$ -- $2 \times 10^{11} M_\odot$. For a fairer comparison with our results, we calculate the $\vopt/\vvir = V_{80}/\vvir$ relation that one \textit{would} get from D10 if the $V_{80}$ vs. $M_*$ TFR from R11 is used instead (blue dotted curve); this is simply offset by $+0.05$ dex from the former curve.\footnote{If one used our $V_{2.2}$ vs. $M_*$ TFR to derive the $V_{2.2}/\vvir$ vs. $M_*$ relation, the curve will lie between the two blue curves.} This relation is consistent within $2\sigma$ of our derived OVVR (black solid curve) for the range of stellar masses we study. 

\subsection{Comparison with $\Lambda$CDM haloes}
\label{subsec:comp_lcdm}

Dark matter haloes in a $\Lambda$CDM cosmology form a remarkably tight relation 
between halo mass and the maximum circular velocity of the halo $\vmaxh$ \citep{1997ApJ...490..493N}.\footnote{The tightness of the relation holds for other cosmologies as well, but especially for $\Lambda$CDM.}
The question of whether the tightness of this relation 
translates into the tightness of the TFR is a key to understanding disk galaxy formation, and therefore, the relationship between $\vmaxh$ and $\vopt$ is of crucial interest.

For a NFW halo profile, the ratio $\vmaxh/\vvir$ depends only on the halo concentration \citep{1997ApJ...490..493N}:
\beq
\frac{\vmaxh}{\vvir} \simeq 0.465 \sqrt{\frac{\cvir}{A(\cvir)}}
\eeq
where $A(x) = \ln(1+x) - x/(1+x)$. 
The peak of the halo velocity curve occurs at a radius $\rmaxh \simeq 2.163 \ r_{\rm s} = 2.163\ \rvir/\cvir$. 
For the range of halo masses we consider, $\cvir$ ranges from $\simeq$ 5 -- 8, so $\rmaxh \simeq 0.3$ -- $0.4\ \rvir$; the velocity curves of these haloes rise gradually out to an appreciable fraction of their virial radii. 

The dark and light green shaded regions in Figure~\ref{fig:vovratio} show the variation in $\vmaxh/\vvir$ for unmodified pure dark matter NFW haloes, corresponding to $1\sigma$ and $2\sigma$ scatter (0.13 and 0.26 dex) in halo concentrations $\cvir(\mvir)$ (given by Eq.~\ref{eq:cofM}) at a fixed halo mass. Note that we have ignored the scatter in the horizontal direction (in other words, we have used the central HSMR to directly translate a grid of halo masses to the corresponding stellar masses).

When comparing $\vmaxh$ and $\vopt$ (i.e., green vs. grey shaded regions in Fig.~\ref{fig:vovratio}, showing $\vmaxh/\vvir$ and $\vopt/\vvir$, respectively), note that only dark matter contributes to $\vmaxh$, while both baryons and dark matter contribute to $\vopt$. Also note that $\rmaxh$ is larger than the optical radius $R_{80}$, by a factor of $\sim 15$; between these two radii, the halo circular velocity decreases by around $20$ to 30 per cent.\footnote{For an NFW halo, the halo velocity curve is given by $V_{\rm h}^2(r)=\vvir^2 [c/A(c)][A(x)/x]$, with $x=r/r_{\rm s}$. Thus, $V_{\rm opt,h}/\vmaxh$ depends only on $\cvir$ through the scale length $r_{\rm s}$ appearing in the argument $x$.}  
Despite these differences, the comparison is instructive: $\vopt > \vmaxh$ indicates that the baryons have modified the potential well in the optical region of the galaxy, either by their own gravity and/or by modifying the structure of the dark matter halo (e.g., through adiabatic contraction). 

We find that $\vopt \ga \vmaxh$ over the range of stellar masses covered by our sample, $M_* = 10^9$ -- $10^{11} M_\odot$. 
For the three stellar mass bins with lensing-weighted stellar masses of 0.62, 2.68, and $6.52 \times 10^{10} M_\odot$, we find $\vopt/\vmaxh = 1.11\pm 0.06$, $1.25\pm 0.05$, and $1.16\pm 0.07$,
respectively (with $1\sigma$ uncertainties propagated from the uncertainties in $\vopt/\vvir$, i.e., not including the uncertainty corresponding to the scatter in halo concentrations at a given halo mass, among others).
Assuming an NFW profile for the halo unmodified by the baryons yields $\vmaxh/V_{\rm opt,h} = 1.27, 1.30, 1.43$.
Multiplying this by $\vopt/\vmaxh$ gives $\vopt/V_{\rm opt,h} = 1.41\pm 0.08$, $1.61\pm 0.05$, and $1.64\pm 0.11$,
with 1$\sigma$ uncertainties from the uncertainties in $\vopt/\vmaxh$. In terms of (3-dimensional) mass, this corresponds to a dark matter contribution of 56, 47, and 45 per cent, suggesting that dark matter and baryonic contributions are comparable at the optical radius. If the dark matter halo undergoes adiabatic contraction, this contribution will be even higher.

It is interesting to compare the expected contribution from the different components--- dark matter halo, stellar and gas disks--- against the observed total rotation velocity at the optical radius $V_{\rm opt}$. In particular, the sum of the different contributions should not exceed the observed velocity; if they do, then one or more assumptions in the modeling must be wrong. For a quick and crude comparison, we estimate the contribution of the baryons to the rotation velocity using our stellar mass estimates (based on Bell~et al. 2003 $M_*/L$ ratios and assuming a fixed Kroupa IMF) and gas mass estimates (based on the Kannappan et~al 2004 relation between gas-to-stellar mass ratio and from $u-r$ colour). We assume that the stellar disk scale length is given by the mean relation with stellar mass (derived in R11, eq.~35), and that the gas disk has the same scale length.\footnote{This assumption is not valid for our low $M_*$ galaxies for which the gas is typically more extended than the stars \citep[e.g.,][]{1994AJ....107.1003C,1994A&AS..107..129B,1996A&AS..115..407R,1997A&A...324..877B,2002A&A...390..829S,2008MNRAS.386.1667B}. The stellar and gas disks will be modelled separately in our future analysis.} Assuming, as above, an unmodified NFW halo, we find $\left(V_{\rm opt,h}^2 + V_{\rm opt,b}^2\right)^{1/2}/\vopt \approx 1.01$, 1.00, 0.98 for the mean stellar masses of our three bins, $M_* = 0.62$, 2.68, and $6.52 \times 10^{10} M_\odot$, respectively. This suggests that the observations are consistent with a model in which the radial profile of the dark matter halo is close to an unmodified NFW profile, i.e., without the effect of adiabatic contraction. This result is of course sensitive to many other assumptions in the model, including halo concentrations, the inner profile of the halo (e.g., NFW vs. Einasto), the stellar IMF, stellar and gas mass estimates, among others. In future work, we will perform a careful comparison with a proper accounting of the full distributions and uncertainties (Reyes et al. {\it in prep}). 

\section{Summary and future work}
\label{sec:summ}

In this work, we use measurements of the average halo masses $\mvir$ of disk galaxies from galaxy-galaxy lensing to direct constrain the relation between halo-to-stellar mass ratios $\mvir/M_*$ and stellar mass $M_*$ for a large sample of disk galaxies from the SDSS with $\langle z\rangle \sim 0.07$ and $10^9 < M_*/M_\odot < 10^{11}$. Moreover, we combine these measurements with the Tully-Fisher relation (TFR), which relates disk rotation velocities at the optical radius $\vopt$ and stellar mass $M_*$, to constrain the relation between optical-to-virial velocity ratios $\vopt/\vvir$ and stellar mass $M_*$. Unlike previous measurements of $\vopt/\vvir$, we use similarly-selected galaxy samples and consistent definitions in both the lensing and TFR measurements to enable a fair combination of the two. In particular, we use the minimal-scatter Tully-Fisher relation from Reyes et~al. (2011) based on a galaxy sample that is, by construction, a fair subsample of the lens sample we use here. 

We model the relation between $\mvir/M_*$ and $M_*$ as a functional form based on halo occupation modelling, Eq.~\ref{eq:hsmratio}, and find that the ratio $\mvir/M_*$ varies over the range of stellar masses we study, with a minimum of $\approx 20$ at $\sim 5\times 10^{10} M_\odot$. For our three $M_*$ bins with lensing-weighted stellar masses of 0.62, 2.68, and $6.52 \times 10^{10} M_\odot$, we find $\mvir/M_* = 41$, 23, and 26, respectively (with $1\sigma$ uncertainties of around 0.1 dex). These correspond to stellar conversion efficiencies $\eta_* = (M_*/\mvir) f_{\rm b}^{-1} = 15^{+5}_{-4}$, $26^{+6}_{-5}$, and $23^{+8}_{-6}$ per cent, respectively (assuming a cosmic baryon fraction of $f_{\rm b}=0.169$). Adding information from the Tully-Fisher relation, we find $\vopt/\vvir = 1.27\pm 0.08$, $1.39\pm 0.05$, and $1.27\pm 0.08$, respectively.

We find that the maximum halo circular velocity $\vmaxh \la \vopt$ over the range of stellar masses we study. 
For the three stellar mass bins we use, we find  $\vopt/\vmaxh = 1.11\pm 0.06$, $1.25\pm 0.05$, and $1.16\pm 0.07$,
respectively (with quoted $1\sigma$ uncertainties accounting solely for the uncertainty in $\vopt/\vvir$). 
Assuming an unmodified pure NFW halo profile, we find that the halo contribution to the rotation velocity at the optical radius is given by
$\vopt/V_{\rm opt,h} = 1.41\pm 0.08$, $1.61\pm 0.05$, and $1.64\pm 0.11$ 
(again, with quoted $1\sigma$ uncertainties accounting solely for the uncertainty in $\vopt/\vvir$).
This corresponds to a halo contribution in mass of roughly half, suggesting that dark matter and baryonic contributions are comparable at the optical radius. A crude accounting of the contribution of the baryons and the dark matter halo to the rotation velocity at the optical radius suggests that, given the many modelling assumptions made, the data are consistent with a radial halo profile that is close to an unmodified NFW halo (i.e., with no adiabatic contraction). 
This result will be refined after a more detailed analysis, in which the mass distribution of the dark matter halo, stars, and gas, as well as the adiabatic contraction of the halo, will be modelled separately, and the distributions and uncertainties in the different parameters will be properly taken into account. 

The observational constraints derived in this work will serve as input to our models of disk galaxy formation. The ultimate goal is to construct models that simultaneously satisfy all the available observational constraints (including those presented here and in R11). We will also investigate the degeneracies between model parameters, and identify other observations that may help to eventually break them.

\section*{Acknowledgements}
We thank Michael Strauss, David Weinberg, and David Spergel for their comments on this work.
C.H. is supported by the U.S. Department of Energy under contract DE-FG03-02-ER40701 and the David \& Lucile Packard Foundation.

Funding for the SDSS and SDSS-II has been provided by the Alfred P. Sloan Foundation, the Participating Institutions, the National Science Foundation, the U.S. Department of Energy, the National Aeronautics and Space Administration, the Japanese Monbukagakusho, the Max Planck Society, and the Higher Education Funding Council for England. The SDSS is managed by the Astrophysical Research Consortium for the Participating Institutions. The Participating Institutions are the American Museum of Natural History, Astrophysical Institute Potsdam, University of Basel, Cambridge University, Case Western Reserve University, University of Chicago, Drexel University, Fermilab, the Institute for Advanced Study, the Japan Participation Group, Johns Hopkins University, the Joint Institute for Nuclear Astrophysics, the Kavli Institute for Particle Astrophysics and Cosmology, the Korean Scientist Group, the Chinese Academy of Sciences (LAMOST), Los Alamos National Laboratory, the Max-Planck-Institute for Astronomy (MPIA), the Max-Planck-Institute for Astrophysics (MPA), New Mexico State University, Ohio State University, University of Pittsburgh, University of Portsmouth, Princeton University, the United States Naval Observatory, and the University of Washington. 

\bibliography{ms_RR} 

\appendix

\input{appA_RM.tex}

\input{appB_RM.tex}

\end{document}

%% file: macros_RM.tex
\newcommand{\ssh}{\ensuremath{\cal R}}
\newcommand{\erms}{\ensuremath{e_\mathrm{rms}}}
\newcommand{\rmd}{\ensuremath{\mathrm{d}}}
\newcommand{\photoz}{photo-$z$}
\newcommand{\zphot}{\ensuremath{z_\mathrm{phot}}}
\newcommand{\zspec}{\ensuremath{z_\mathrm{spec}}}
\newcommand{\zlens}{\ensuremath{z_\mathrm{lens}}}
\newcommand{\putcite}{\textbf{(CITE)}}
\newcommand{\fillinrm}{\textbf{??RM??}}
\newcommand{\fillinrr}{\textbf{??RR??}}
\newcommand{\fillinRR}{\textbf{??RR??}}

%% file: section4_RM.tex
{\section{Source and shape catalogue}
\label{sec:data_source_shape}


This work introduces a new source galaxy catalogue that is meant to be
an improvement, both in area coverage and quality, over the one
introduced in M05 
and used for subsequent science papers.  This catalogue, like that from M05,
utilises a method of PSF-correction known as re-Gaussianization
\citep{2003MNRAS.343..459H}.  Re-Gaussianization is a method based on
the use of the moments of the image and of the PSF to correct for the
effects of the PSF on the galaxy shapes. However, unlike many other
moments-based corrections, it includes corrections for the
non-Gaussianity of the galaxy profile
\citep{2002AJ....123..583B,2003MNRAS.343..459H} and of the PSF (to
first order in the PSF non-Gaussianity).

Details about how the catalogue was generated, and an explicit
contrast with the catalogue from M05 are in
Appendices~\ref{S:generation} and~\ref{S:differences}, respectively.  A description of its
properties, and systematics tests, will be presented in the following
subsections (Secs.~\ref{subsec:new_cat}--\ref{subsec:scaledep_sys}).

\subsection{New catalogue properties}
\label{subsec:new_cat}

The new shape catalogue covers an area of 9~243 deg$^2$, with 39~267~029 
unique galaxy detections in the SDSS DR8 area passing all cuts on photometry, shape
measurements, and \photoz\ described in
Appendix~\ref{S:generation} (or an average source number density of
$1.2$ arcmin$^{-2}$).  For this work, we use a subset of that area (7~131 
 deg$^2$) corresponding to the DR7 lens catalogue used for this work.

The relevant areas are shown in
Fig.~\ref{fig:show_area}; the majority of the ``ratty'' areas result
from imposition of the cut on $r$-band extinction $A_r<0.2$ mag, since
there are regions that are close to that limiting value and that get shredded
by this cut.  For science work that might be dominated by those areas,
a reprocessed version of the catalogue might be necessary to ensure better coverage.  However, for the lens catalogue used for this work,
those areas are not necessary.
\begin{figure}
\begin{center}
\includegraphics[width=\columnwidth,angle=0]{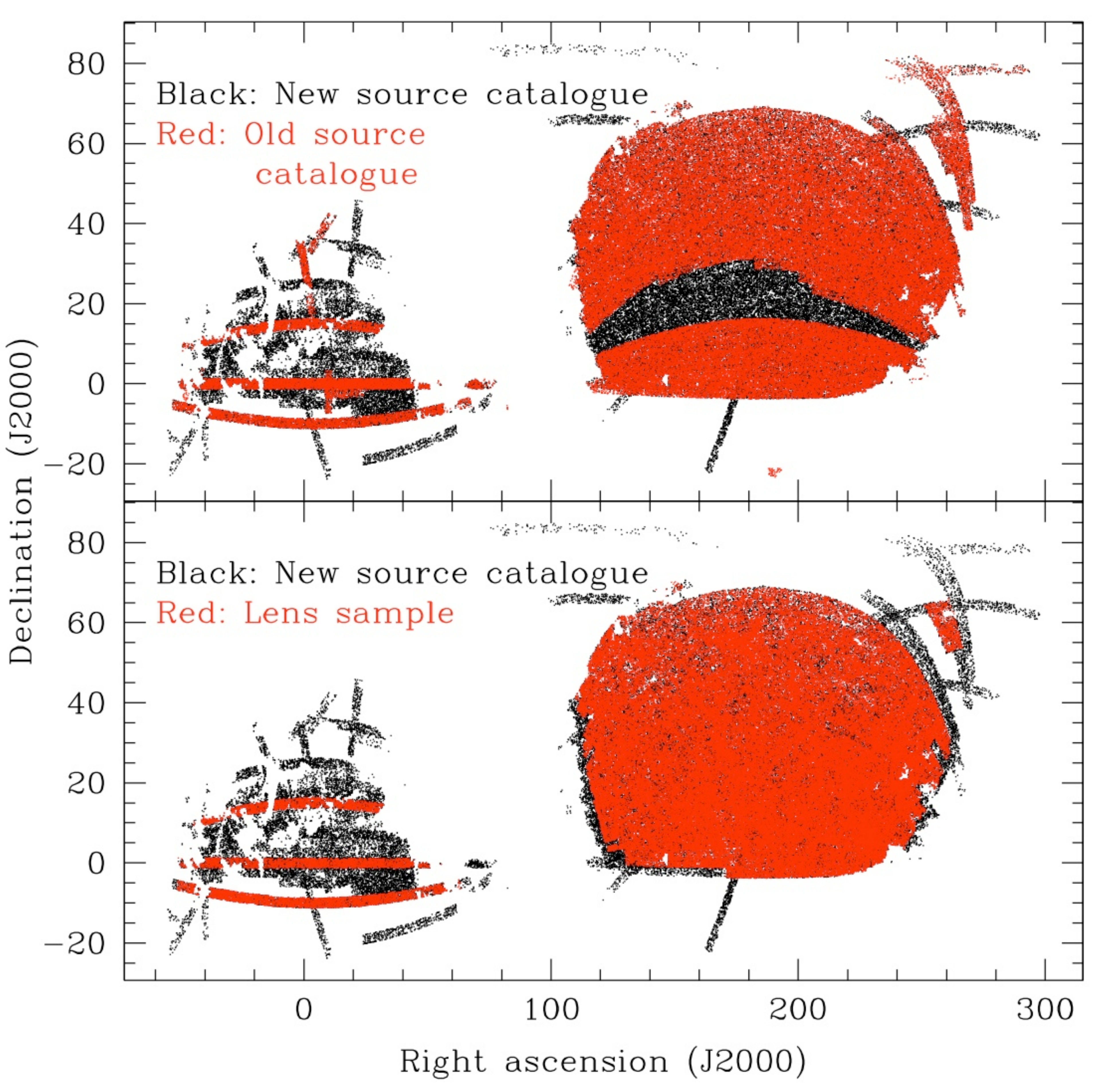}
\caption{\label{fig:show_area} Area coverage of the new shape
  catalogue, compared to the old catalogue from M05 (top) and the lens
sample used for this work (bottom).}
\end{center}
\end{figure}

\begin{figure*}
\begin{center}
$\begin{array}{c@{\hspace{0.5in}}c}
\includegraphics[width=0.8\columnwidth,angle=0]{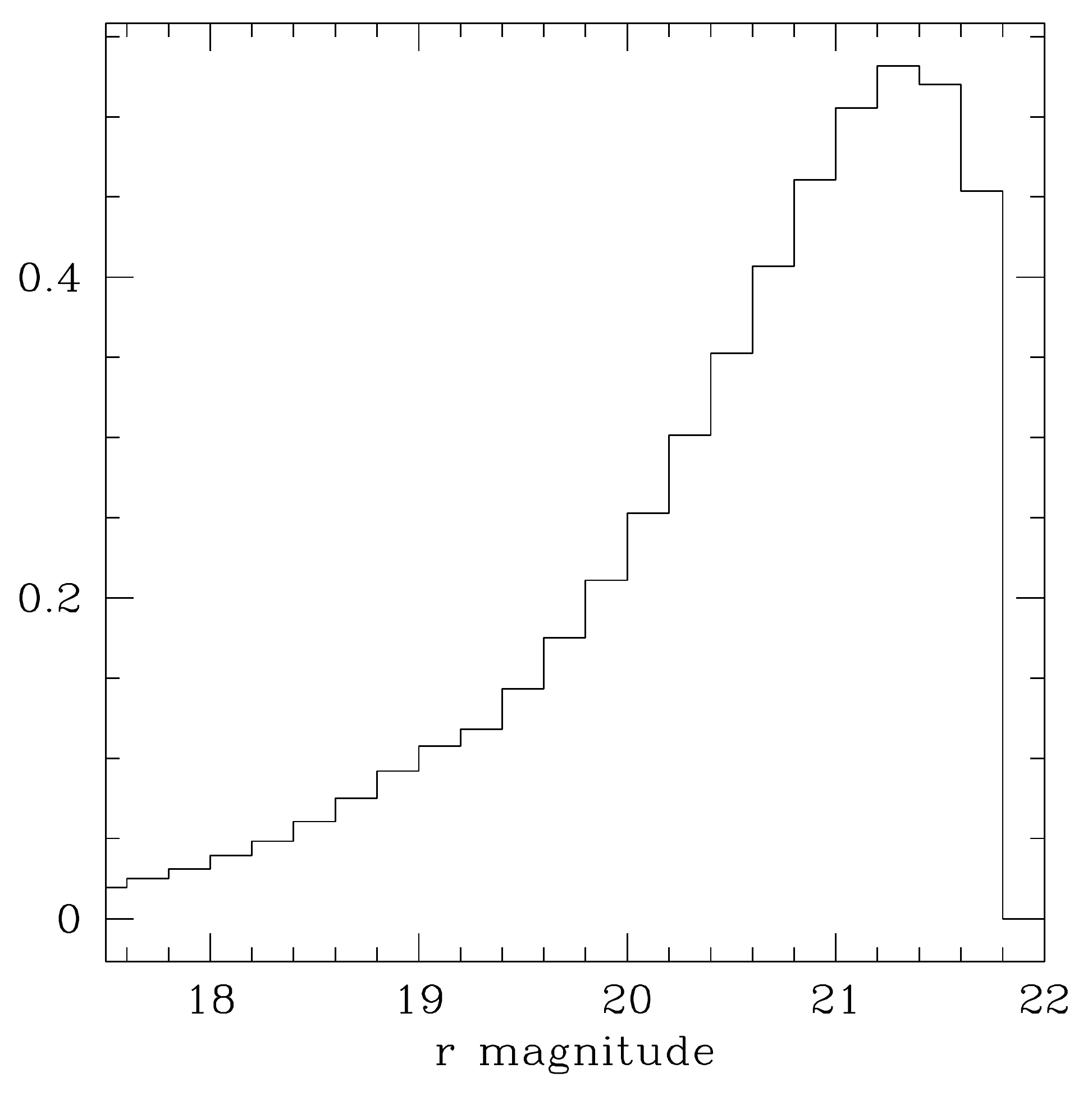} &
\includegraphics[width=0.8\columnwidth,angle=0]{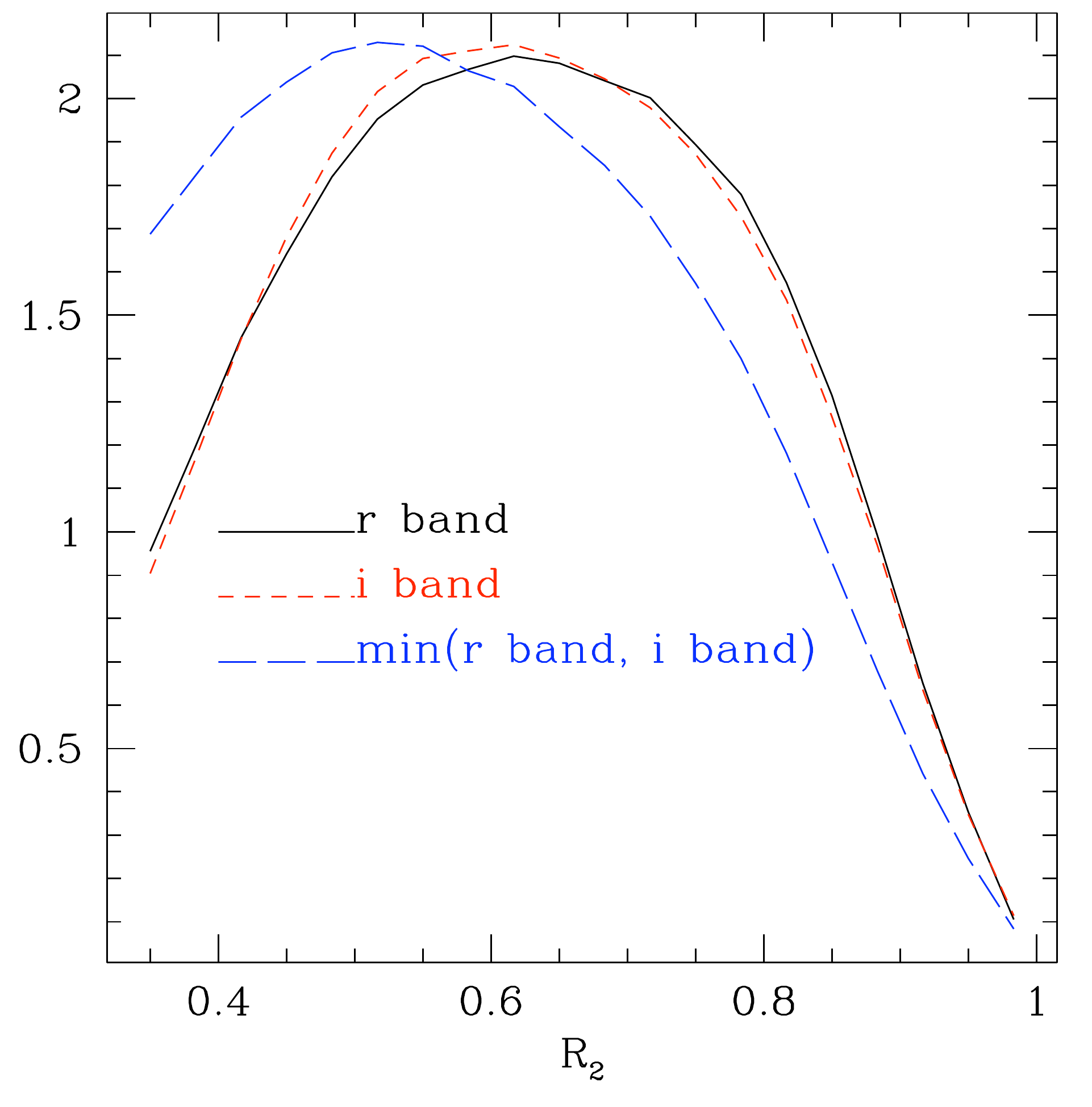} \\
\includegraphics[width=0.8\columnwidth,angle=0]{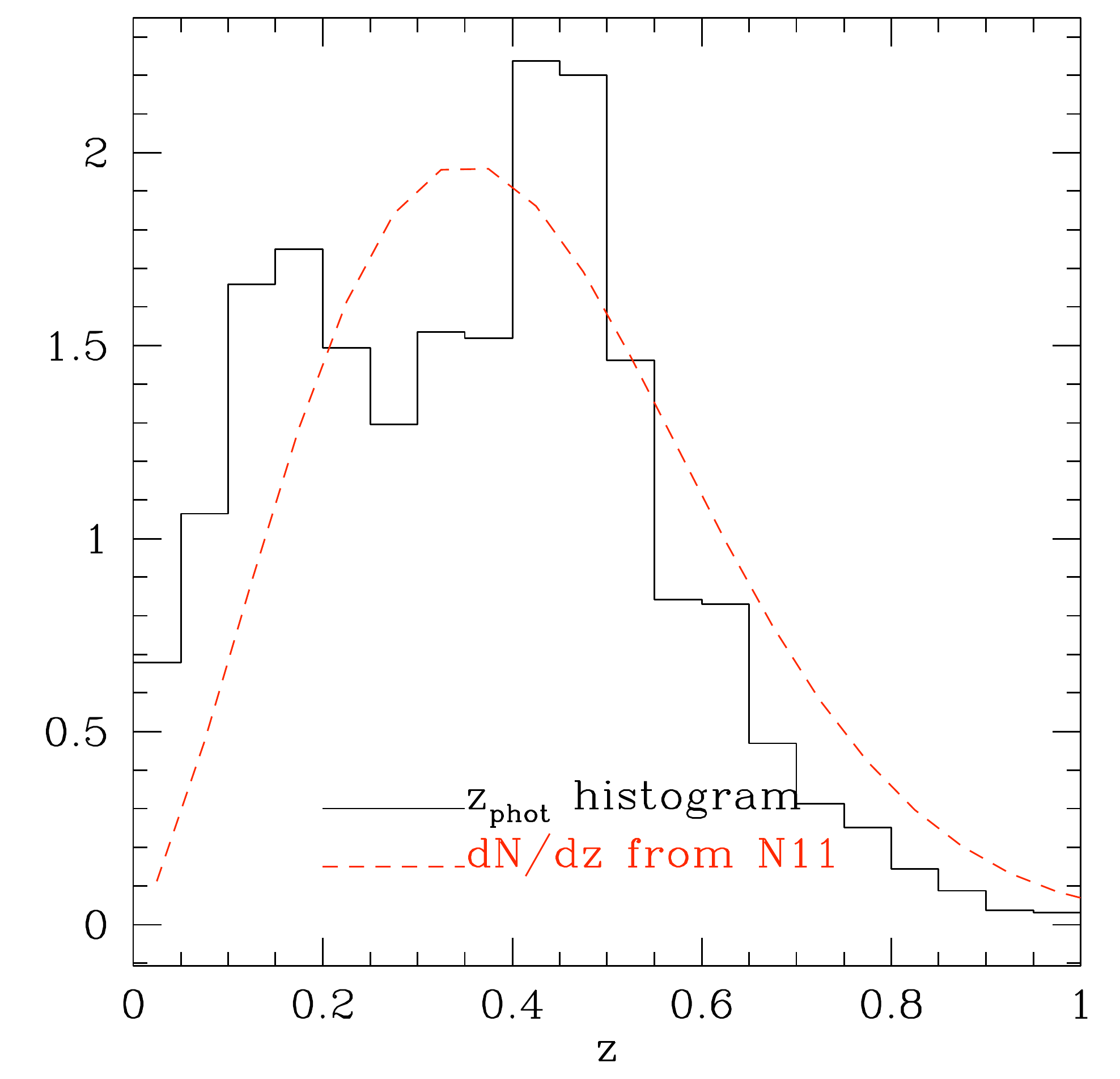} &
\includegraphics[width=0.8\columnwidth,angle=0]{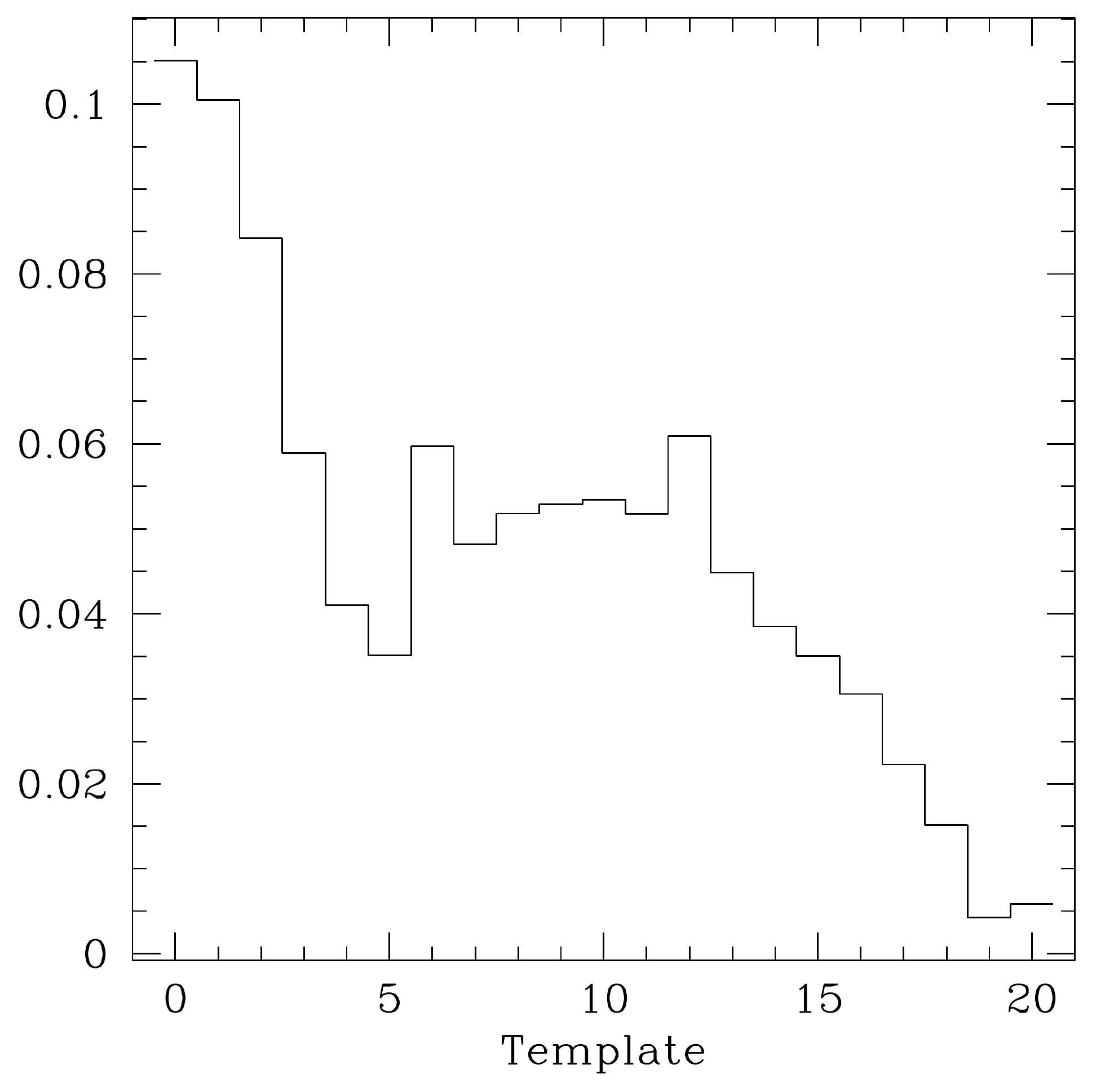} \\
\end{array}$
\includegraphics[width=0.8\columnwidth,angle=0]{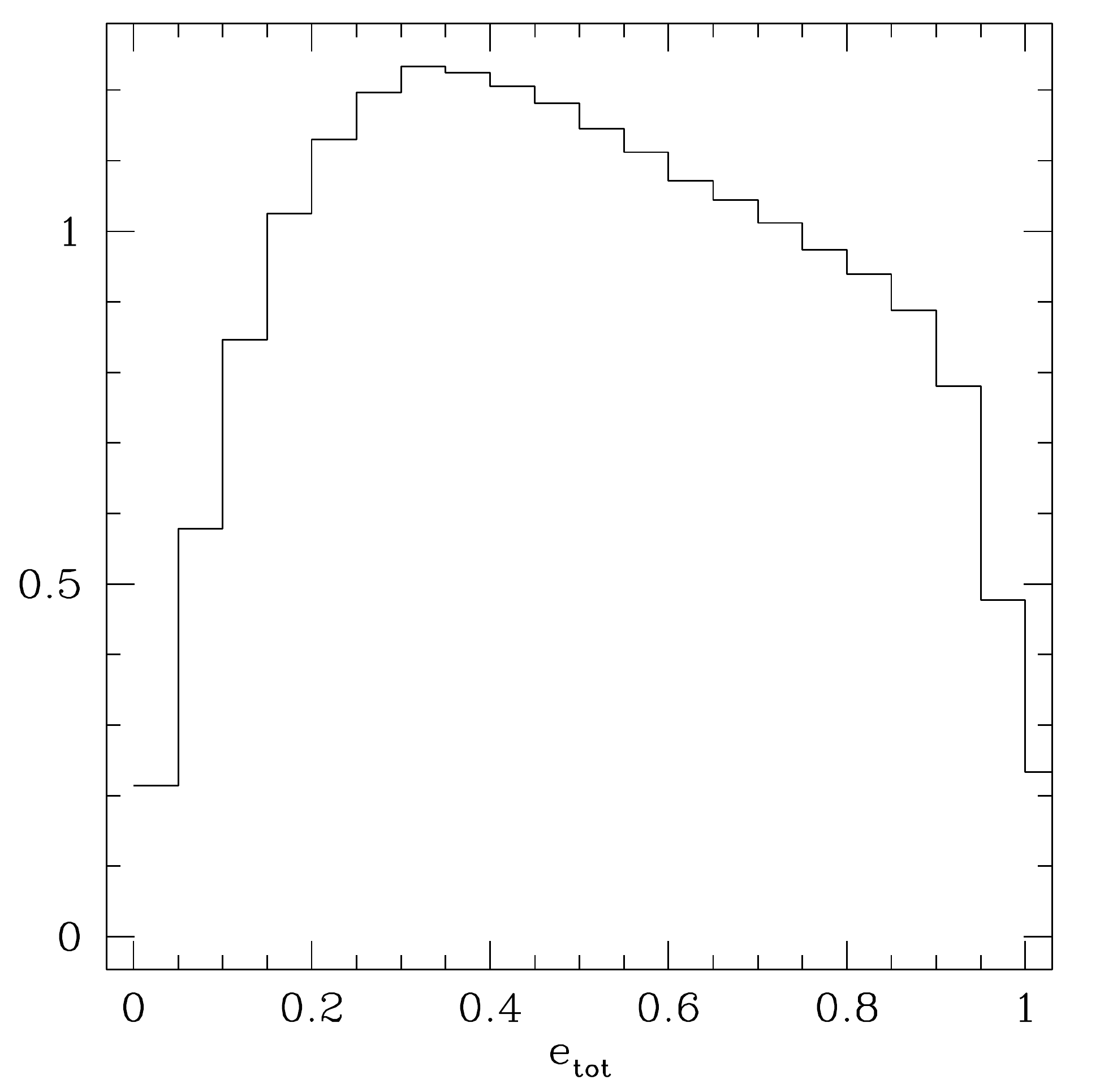}
\caption{\label{fig:basic_hist} Each panel shows the histogram of
  source galaxy properties, derived from a random subsample of 5 per
  cent of the catalogue after imposing all shape and \photoz\ cuts
  ($\sim 2\times 10^{6}$ galaxies).  {\em Top left:} Histogram of $r$-band
  extinction corrected model magnitude. {\em Top right:} Same, for
  resolution factor $R_2$.  Lines are shown for the $r$ and $i$ band
  resolution factors separately, and for $\mathrm{min}(R_{2,r},
  R_{2,i})$ which will be important when we consider selection
  biases. {\em Middle left:} \photoz\ histogram, and the inferred true
  $\rmd N/\rmd z$ from N11.  {\em Middle right:} Histogram of template
  values used for inferring \photoz, where lower values correspond to
  redder/earlier-type galaxies.  {\em Bottom:} Histogram of total ellipticity value
  ($\sqrt{e_1^2+e_2^2}$), which includes a significant contribution
  from noise.}
\end{center}
\end{figure*}

Before describing the catalogue properties, we begin by introducing the
quantities used to describe each galaxy.  These include the following: 
\begin{enumerate}
\item The extinction-corrected $r$-band model magnitude, which is a measure of the total galaxy flux.
\item The \photoz, which is calculated using the Zurich Extragalactic
Bayesian Redshift Analyzer (ZEBRA, \citealt{2006MNRAS.372..565F}). Its
usage for SDSS lensing studies was explored thoroughly by 
\citet[][hereafter N11]{2012MNRAS.420.3240N}.
\item The
galaxy spectral energy distribution (SED) template corresponding to that \photoz\ (more information about the
templates that were used are in Appendix~\ref{S:generation}; the
templates for galaxies that are used for science range from 0 to 20,
with zero corresponding to early types and 20 to late types). 
\item The galaxy resolution factor
$R_2$, which expresses how resolved it is compared to the PSF.  A
given galaxy's resolution factor thus depends on the conditions under
which it was observed.  The detailed definition of resolution factor
is given in Appendix~\ref{S:generation}, Eq.~\ref{eq:r2}; for the purpose of this
section, it suffices to know that $R_2$ approaches zero for completely
unresolved galaxies, one for perfectly resolved galaxies, and we
require $R_2>1/3$ in both $r$ and $i$ bands to avoid excessive
systematic errors in the galaxy shapes.
\item The galaxy shape $(e_1, e_2)$  and
  the estimated shape measurement error $\sigma_e$ per component.  The
  shapes are rotated to a coordinate system in
  which positive $e_1$ corresponds to East-West elongation and
  positive $e_2$ corresponds to Northeast-Southwest elongation.  Our
  shape definition corresponds to 
\beq\label{eq:shapedef}
|e| = \frac{1-q^2}{1+q^2}
\eeq
for minor-to-major axis ratio $q$.  
\end{enumerate}
Details of the
derivation of these quantities are in Appendix~\ref{S:generation}.

Table~\ref{tab:catalogue_info} gives a summary of the basic catalogue properties
with respect to these quantities.  In addition,
Fig.~\ref{fig:basic_hist} shows the histogram of apparent magnitude,
resolution factor, \photoz, template, and total ellipticity.  As
shown, the number counts do not rise as steeply as for a flux-limited
sample, because of the loss of galaxies at the faint end due to both
the difficulty in measuring shapes at low $S/N$ (the flux limit
$r=21.8$ corresponds to $S/N \gtrsim 9.5$) and the difficulty in
resolving such faint galaxies given the typical SDSS seeing.

\input{Tables/catalogue_info.tex} 

While the \photoz\ histogram in Fig.~\ref{fig:basic_hist} does not
match the true $\rmd N/\rmd z$ as well as one might like, the impact
of the significant \photoz\ errors, $\sigma_z/(1+z)=0.113$, on galaxy-galaxy lensing
measurements has been quantified by N11 
using a training sample consisting of 9~631~galaxies, and the
effects on the lensing signal calibration are well understood.

Fig.~\ref{fig:shape_frac} shows the fraction of the galaxies that satisfy all the cuts 
we impose on the source catalogue (i.e., the magnitude, flag and \photoz\ cuts listed in Appendix~A)
that have usable shape measurements;
the dependence on apparent magnitude
is fairly strong for galaxies fainter than $r\approx 21$ mag.

Fig.~\ref{fig:basic_contour} shows density contour plots
(logarithmically spaced, with factors of 2.5 in the density) relating
the resolution factor $R_2$, shape measurement uncertainty $\sigma_e$,
and the \photoz\ to the $r$-band apparent magnitude.  First, we see that the
$r$-band magnitude and the resolution factor $R_2$ are weakly
correlated.  Naturally, the $r$-band magnitude and the shape
measurement errors are significantly (positively) correlated.
Finally, the $r$-band magnitude correlates with the \photoz\ as well
because fainter objects are more likely to be at high redshift.

\begin{figure}
\begin{center}
\includegraphics[width=\columnwidth,angle=0]{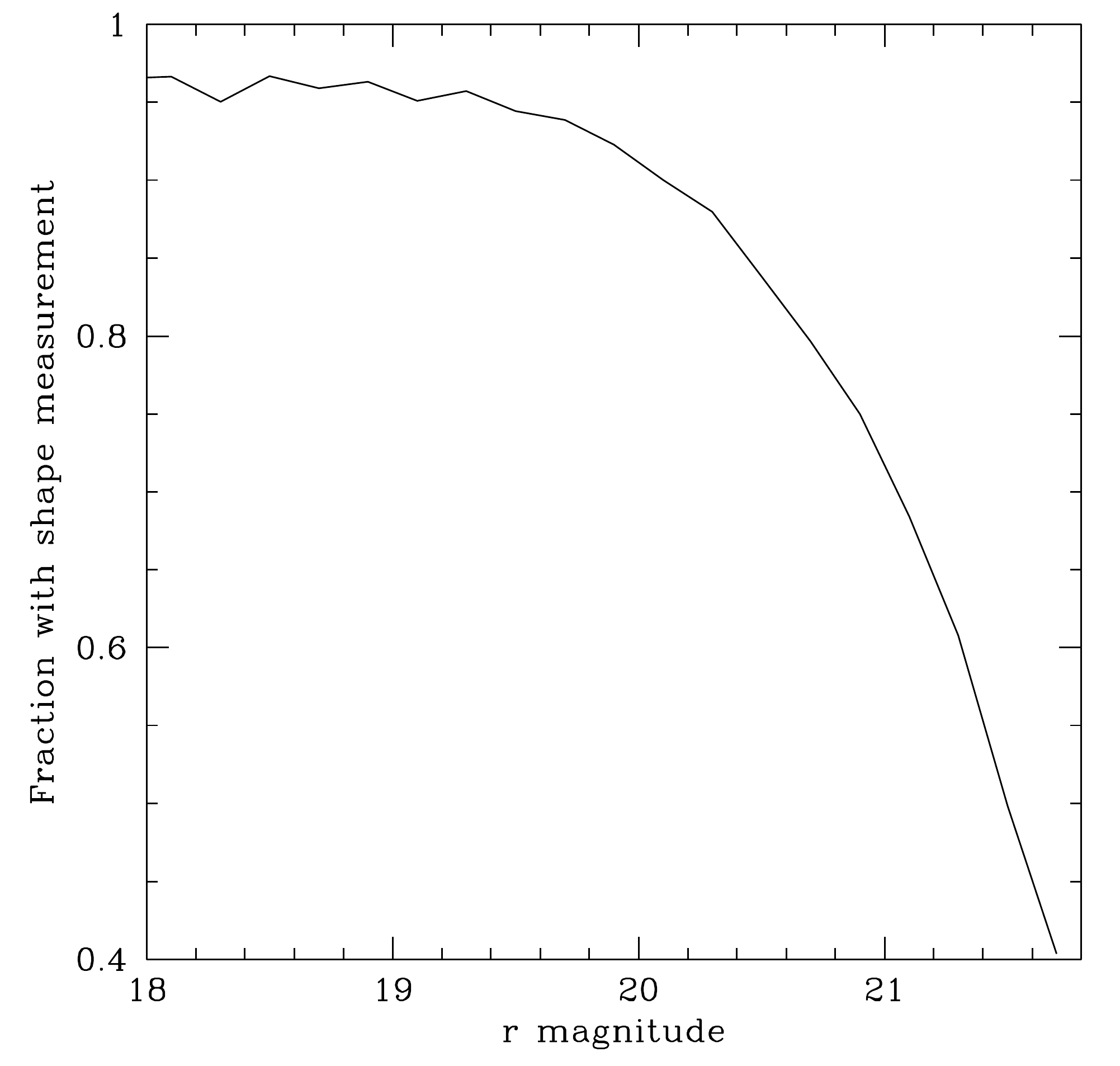}
\caption{\label{fig:shape_frac} 
Fraction of the galaxies that satisfy all the cuts imposed on the source catalogue (i.e., in magnitude, flags, and \photoz) that have usable shape measurements.}
\end{center}
\end{figure}

\begin{figure}
\begin{center}
\includegraphics[width=0.8\columnwidth,angle=0]{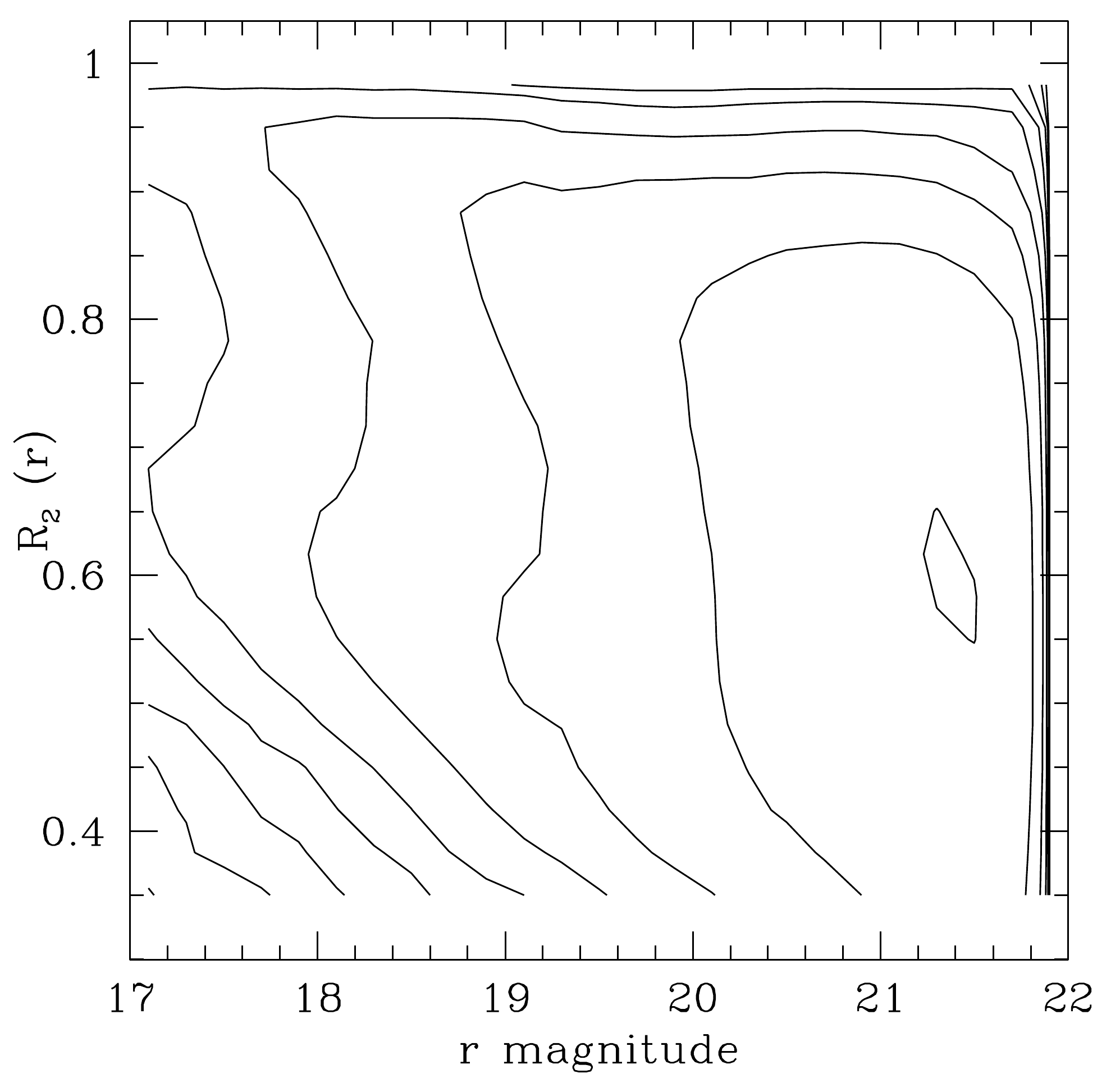} 
\includegraphics[width=0.8\columnwidth,angle=0]{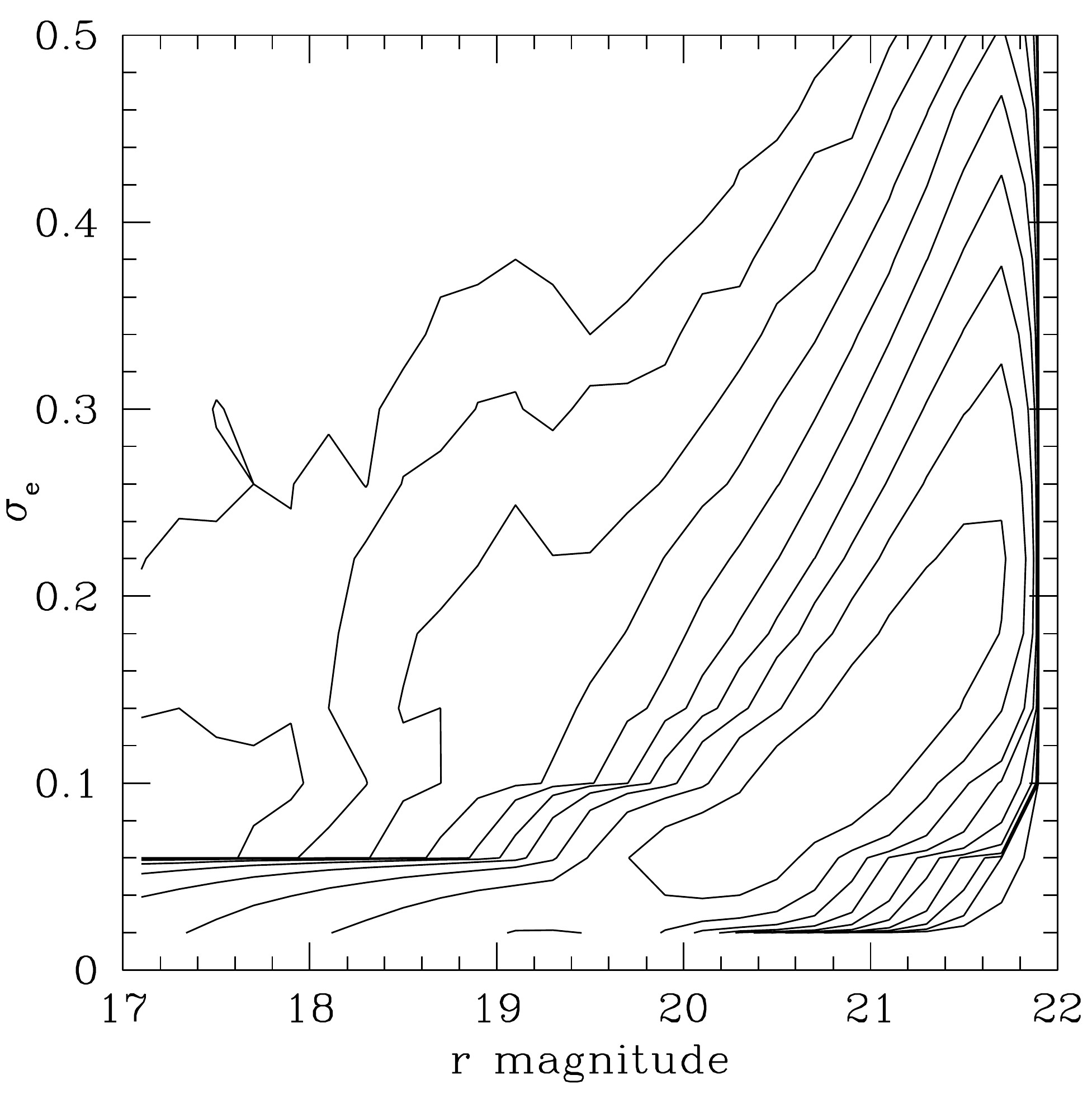}
\includegraphics[width=0.8\columnwidth,angle=0]{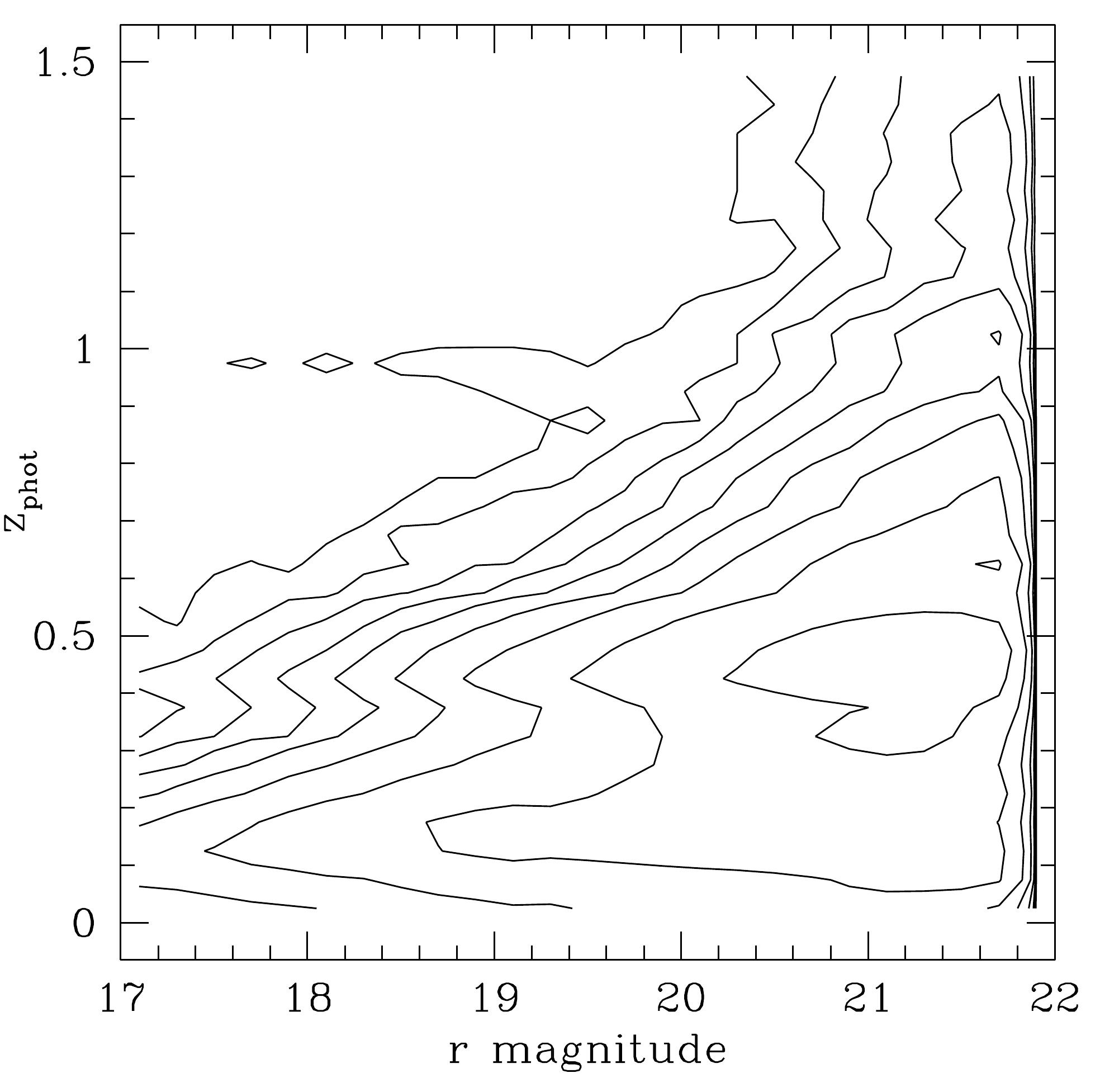} 
\caption{\label{fig:basic_contour} Each panel shows contour plots of
  the density of source galaxies in a different 2d parameter space.
  In all cases, contour levels are logarithmically spaced by a factor
  of 2.5.  {\em Top:} Density contour plot for $r$-band magnitude and
  resolution factor $R_{2,r}$.  {\em Middle:} Same, for $r$-band
  magnitude and the total (band-averaged) shape measurement error
  $\sigma_e$, estimated from the sky variance and simple assumptions
  in Eq.~\ref{eq:sigmaedef}.  {\em Bottom:} Same, for $r$-band magnitude and \photoz.}
\end{center}
\end{figure}

\subsection{Dependence on imaging conditions}\label{subsec:imagecond}

Given that the catalogue generation procedure entails placing a cut on
the resolution of the PSF-convolved galaxy image with respect to the
PSF, the source number density is clearly dependent on the imaging
conditions.  

In order to explore this effect, we carry out several tests on stripe
82 ($-50<$RA$<+60$, $-1.25<$Dec$<1.25$ degrees). 
Conveniently, this area has enough observations at any given point that 
allows us to make multiple independent versions of the source catalogue
(using the procedure from Appendix~\ref{S:generation}) 
with different observing conditions.

\subsubsection{Realistic range of conditions}

Here we consider the realistic range of observing conditions covered
by the full source catalogue, including the seeing FWHM and the relevant combination
of sky noise and extinction that determines the $S/N$, or
$10^{0.4 A_r}\sigma_\mathrm{sky} $ in nanomaggies (nmgy)\footnote{This unit
  of flux is defined such that the apparent magnitude $m=22.5 - 2.5
  \log_{10}{[\textrm{flux (nmgy)}]}$.}.
Fig.~\ref{fig:obsconddist} shows histograms of these properties for
uniformly distributed random points within the physical boundaries of
the source catalogue, and for the source galaxies, which are biased in
the direction of better seeing (where the seeing in the two bands used
for the shape measurement is highly correlated).  It is clear from
this plot that the sky noise level has a minimal effect on the
probability of a source galaxy being included in the
catalogue. 

\begin{figure}
\begin{center}
\includegraphics[width=0.8\columnwidth,angle=0]{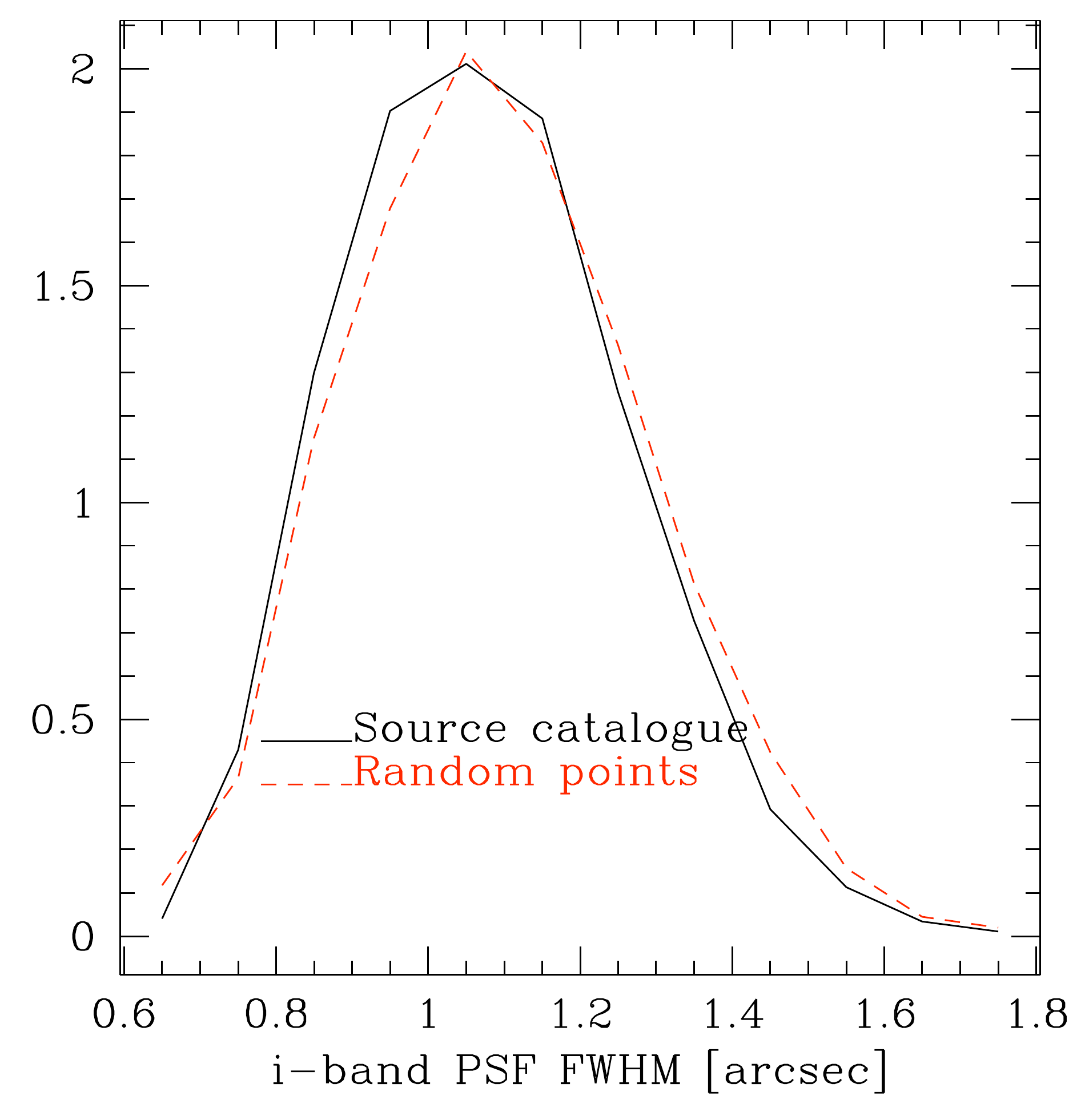} 
\includegraphics[width=0.8\columnwidth,angle=0]{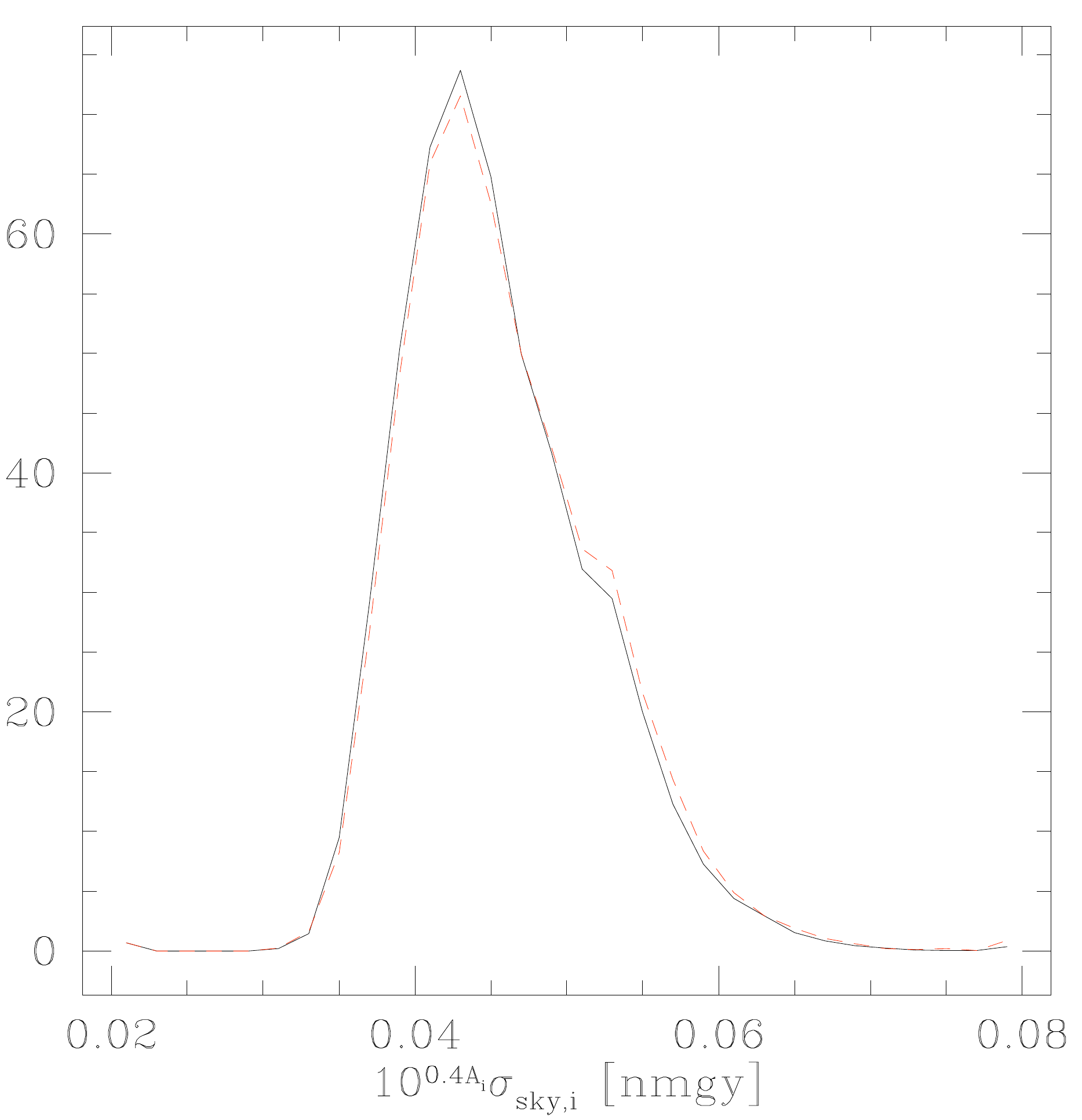}
\includegraphics[width=0.8\columnwidth,angle=0]{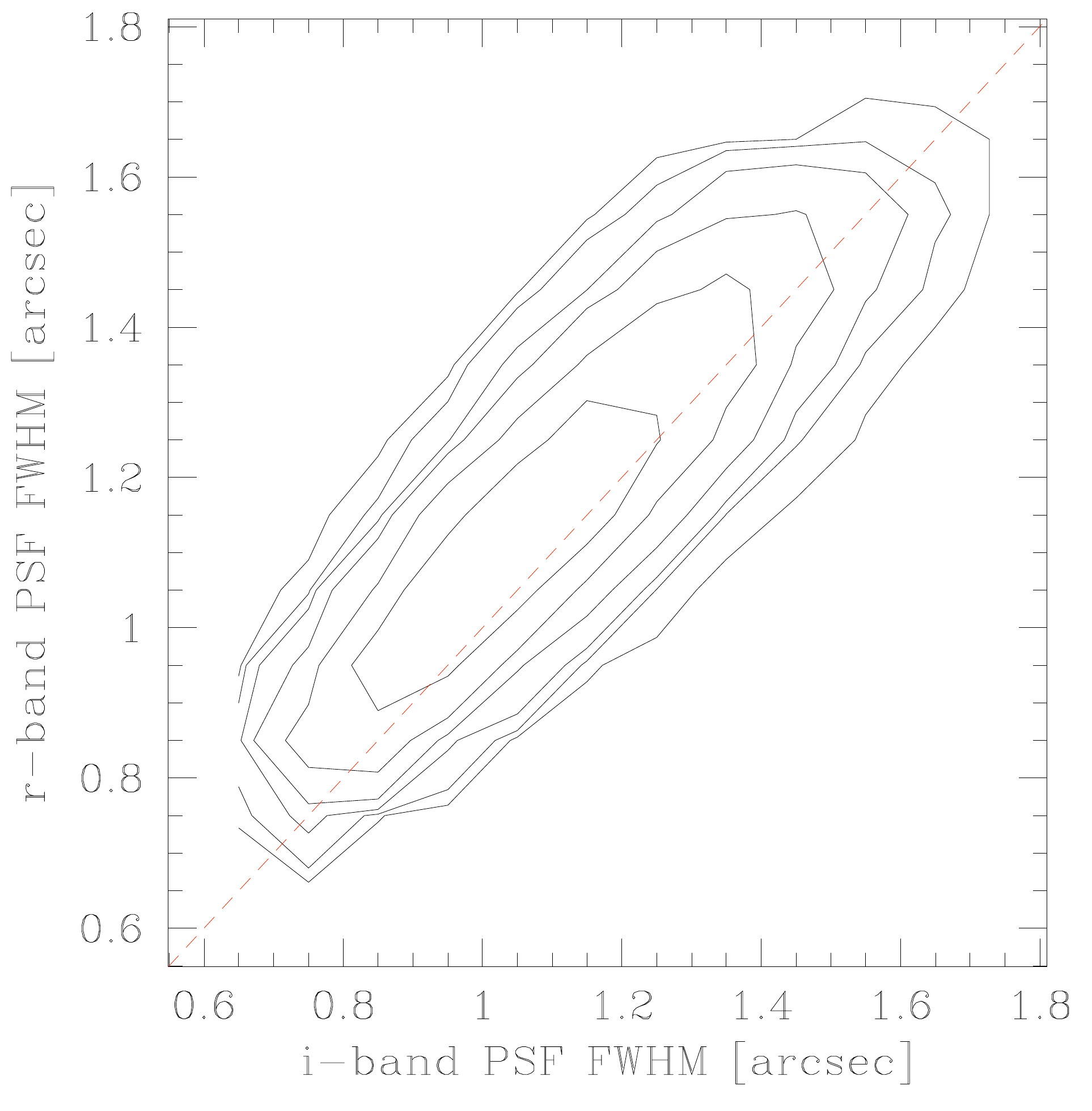} 
\caption{\label{fig:obsconddist} Observing conditions in the entire
  source catalogue, as compared to that for randomly distributed
  points within the same area coverage.  {\em Top:} Histogram of
  $i$-band PSF FWHM for source galaxies (solid) and random points
  (dashed lines).    {\em Middle:} Same, for the sky noise (standard
  deviation).   {\em Bottom:} Density contour plots of $r$ versus $i$
  band PSF FWHM, with contours that are logarithmically spaced by
  factors of 2.5.}
\end{center}
\end{figure}

\subsubsection{Seeing}
\label{subsubsec:seeing}

Given our finding that the catalogue covers a broad range of seeing
conditions, which do have a noticeable impact on the source number
density, we generated independent versions of the source catalogue on
stripe 82 using 3 sets of imaging runs with good (run numbers 2700,
2728, 4263), typical (4128, 4203), and poor seeing (250, 3434, 4253,
4288).  In all cases, the sky levels are fairly typical for stripe 82,
so the difference in PSF is the predominant difference between the
catalogues.  These catalogues cover an area of 125 deg$^2$ ($-10
<$RA$<40$ for the full width of the stripe) and have identical
coverage, so any differences in the observed distribution of galaxy
properties is due to a difference in what we can detect given the
observing conditions, and {\em not} some underlying difference in the
galaxy populations.

The histograms of PSF FWHM in the three versions of the catalogue are
shown in Fig.~\ref{fig:seeing}; the median $r$ ($i$) band PSF FWHM values are
0.90\arcsec (0.87\arcsec), 1.15\arcsec
(1.09\arcsec), and 1.32\arcsec (1.26\arcsec), going
from best to worst seeing.  For the three versions of the catalogue,
the observed source number density is 1.25, 1.08, and
1.04 arcmin$^{-2}$. 

The distribution of observed galaxy properties is also shown in
Fig.~\ref{fig:seeing}, including histograms of apparent magnitude,
resolution factor, and \photoz.  As shown, one impact of poor seeing
is the inability to resolve shapes preferentially for the fainter
galaxies.  As a consequence, in good seeing, the typical source
apparent magnitude is fainter, indicating that we are sampling a
different intrinsic galaxy population.  This effect was also observed
in N11.  Moreover, in good seeing the typical
resolution factor $R_2$ of the sources is higher. While this should
obviously be the case for an identical sample of galaxies, it was not
necessarily obvious that this would happen since the galaxy
populations are intrinsically fainter and smaller in the catalogue
that has good seeing.  Finally, the \zphot\ distributions only change
marginally as we go from typical to poor seeing.  

We can therefore conclude that the seeing is a major factor that
determines the underlying galaxy population
that gets included in the catalogue.  However,
Fig.~\ref{fig:seeing} does not answer the question of whether the
observed galaxy properties depend on the seeing, or whether the errors
in the galaxy properties in different observations are
correlated.\footnote{This could conceivably occur even with two
  observations that have the same seeing; one could imagine that noise
  fluctuations might correlate the magnitudes, resolutions, shapes,
  \photoz, and other properties.}  To answer those questions, we turn
to a direct comparison of the observed galaxy properties for those
that appear both in the ``best seeing'' and the ``worst seeing''
catalogues.  Due to noise and the different seeing values, only a
subset of these catalogues are matched with each other (66 per cent of
the poor seeing catalogue).  Still, we compare the observed
magnitudes, \photoz, resolution factors, and shapes, for
the matched sample in Fig.~\ref{fig:seeingdiffs}. 

As shown in Fig~\ref{fig:seeingdiffs}, the resolution factor $R_2$
is the only parameter that clearly varies substantially when
comparing a matched sample with observations in good versus poor
seeing, as one might expect given that it is determined by a
combination of the galaxy size and the seeing.  The shape, apparent
magnitude, and \photoz\ contour plots are quite symmetric about the
1:1 line, which is reassuring.  However, given the large matched sample
($>2\times 10^5$ galaxies), it should be possible to check explicitly
for correlations between the different observations of these galaxy
properties via an estimate of the Pearson cross-correlation
coefficient $r_{xy}$. 

To do so, we calculate the {\em differences} between these observed
quantities ($\Delta r$, etc.) and the difference between the seeing
values in the two observations, and correlate the differences with
each other.  The correlation coefficients between the various
properties are shown in Table~\ref{tab:rxy}.  As shown, aside from the
expected significant negative correlation ($r_{xy}=-0.329\pm 0.002$)
between the seeing and the resolution factor (larger FWHM results in a
lower resolution factor), there is another correlation of similar
magnitude, between the resolution factor and the apparent magnitude.
The sign of this correlation implies that in the observation in which
the galaxy appears to be larger (better-resolved) according to the
second moments, it also appears to
be brighter when measuring flux using the model magnitudes.  This
correlation is expected, and in the Gaussian case, should have
magnitude $+0.707$ \citep{2004MNRAS.353..529H}. We do not expect exactly this value
because the magnitudes are obtained via model fits, and there are
selection criteria imposed on these quantities which will change the
expected correlation.  Thus we accept the observed, significantly
positive correlation as expected despite its being weaker than the
naive Gaussian calculation would predict.

There are also several parameter correlations in Table~\ref{tab:rxy}
that are small, at the few per cent level, yet statistically
significant given the large sample size.  For example, when the seeing
is worse, the total ellipticity tends to be larger.  It is unclear
whether this is a true bias based on the different observation
conditions (there are several types of calibration bias that could
result from the decreased resolution or increased noise, as explained
in Sec.~\ref{subsec:shear_calib}), or whether it is merely due to the
fact that a galaxy observed in worse seeing is spread out over more
pixels and therefore has a noisier shape.  

There is also a small positive correlation
between noise fluctuations in apparent magnitude and \photoz: those
galaxies that appear fainter due to noise fluctuations also appear to
be at higher redshift.  Given the lack of luminosity prior on the
\photoz, it is not completely obvious that this correlation should
exist.  However, it may be tentatively attributed to the fact that if the galaxy
SED that the \photoz\ code is trying to use has a break in the $r$
band (as for the $z>0.4$ galaxy population), and the galaxy seems too
faint in $r$, 
then the \photoz\ code will move the break even further into
the red (i.e., higher redshift) to compensate.  

There are a few conclusions we can draw from the calculations in this
section.  First, the observed galaxy properties that do not depend
explicitly on the seeing (such as the ellipticity, apparent magnitude,
and \photoz), but that could have acquired some dependence on seeing
due to issues with data processing, do not have a strong dependence on
seeing, as one would hope.  This statement is true even for a fairly
significant difference in seeing values, from $0.9$\arcsec\ to
$1.4$\arcsec, which as shown in Fig.~\ref{fig:obsconddist} and
Table~\ref{tab:catalogue_info} are around the 16 and 84 percentiles of
the observing conditions of the source catalogue.  The second point is
that there is no significant correlation between noise
fluctuations (or seeing-dependent systematic effects) in the galaxy
ellipticity and the \photoz.  This point is very important, since we
would like to treat \photoz\ systematics and shape systematics on the
lensing signal independently (Sec.~\ref{subsec:shear_calib}), and the results in this section confirm
the validity of that approach.

\input{Tables/rxy.tex}

\begin{figure*}
\begin{center}
$\begin{array}{c@{\hspace{0.5in}}c}
\includegraphics[width=0.95\columnwidth,angle=0]{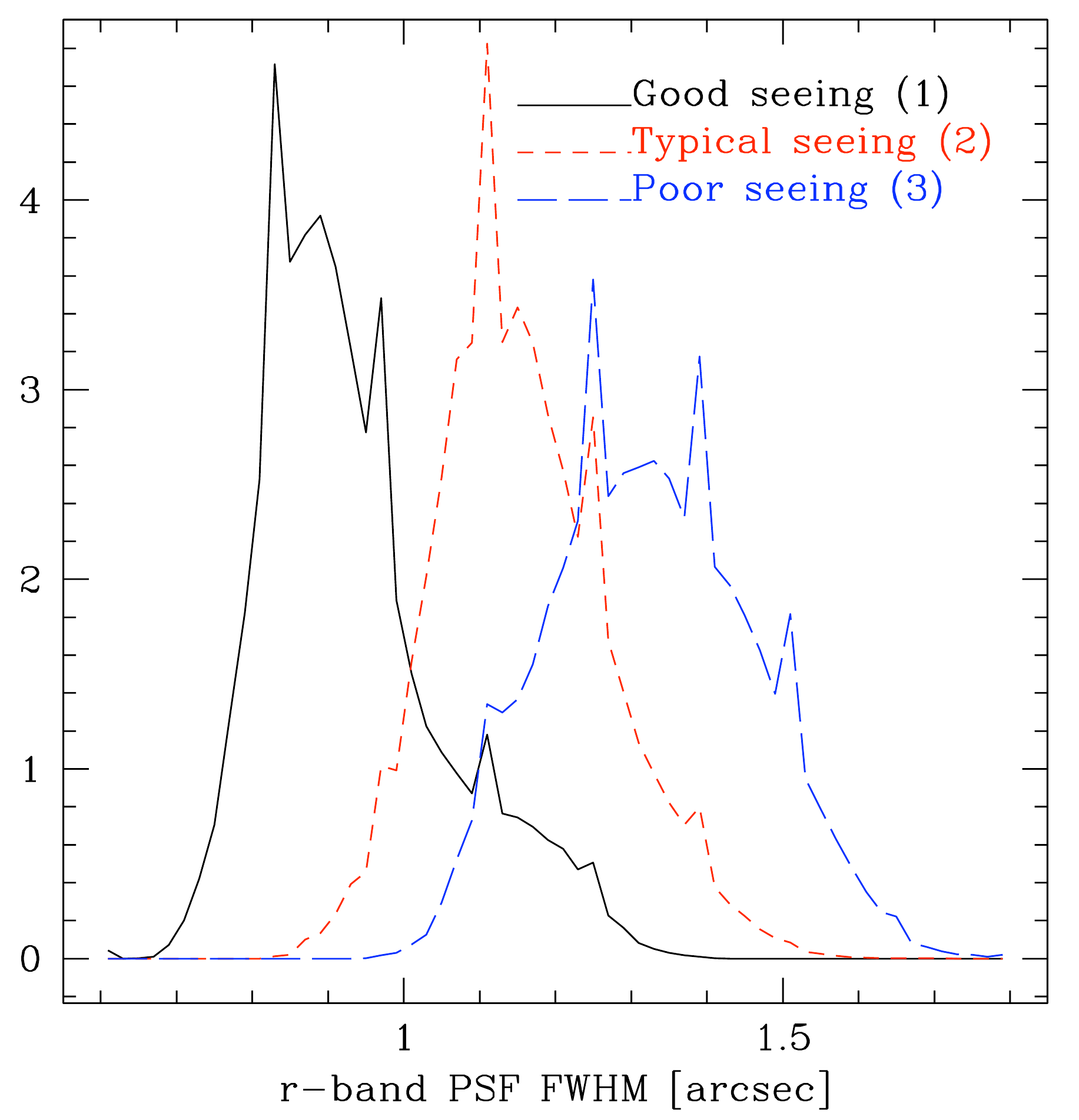} &
\includegraphics[width=0.95\columnwidth,angle=0]{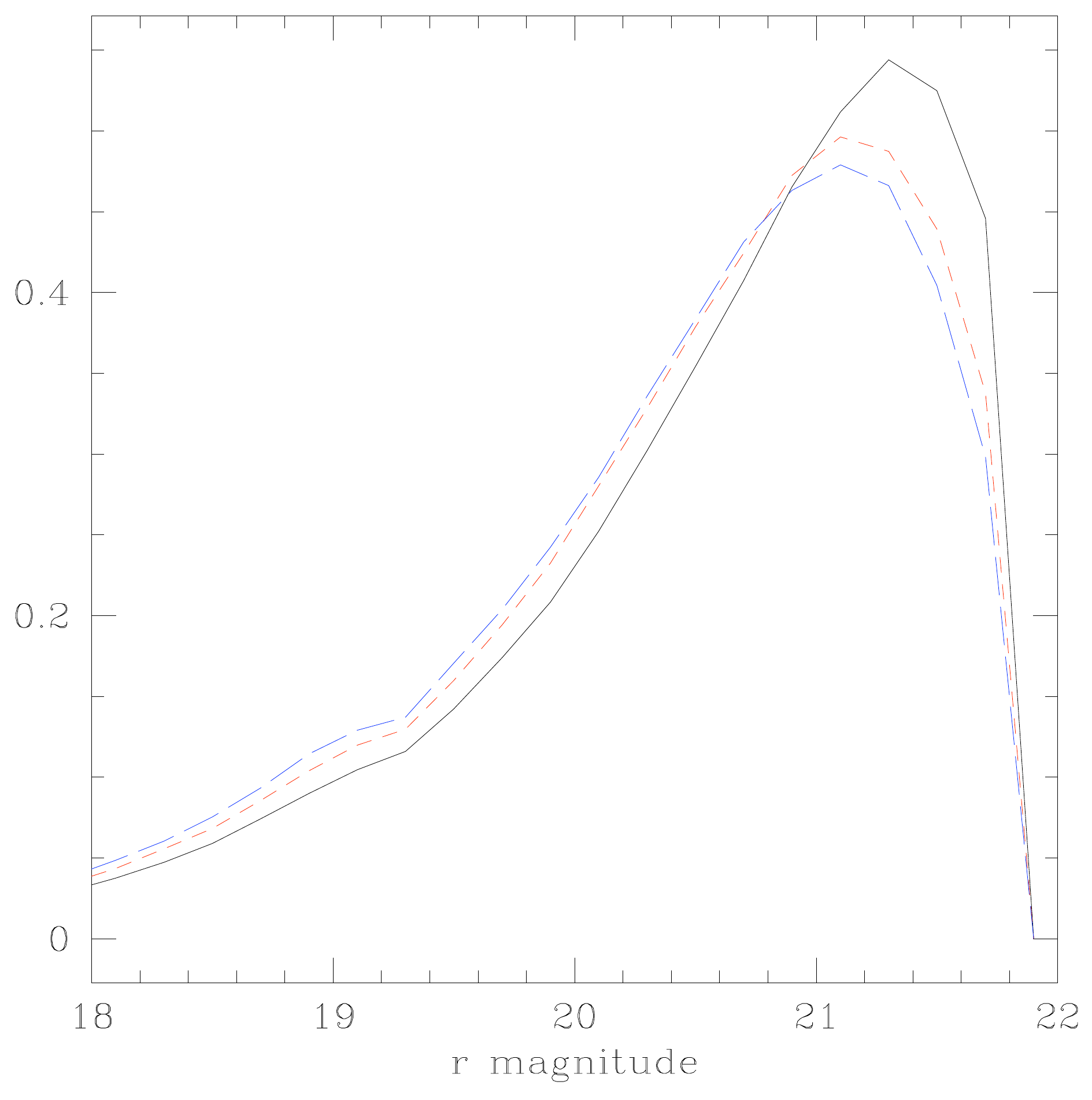} \\
\includegraphics[width=0.95\columnwidth,angle=0]{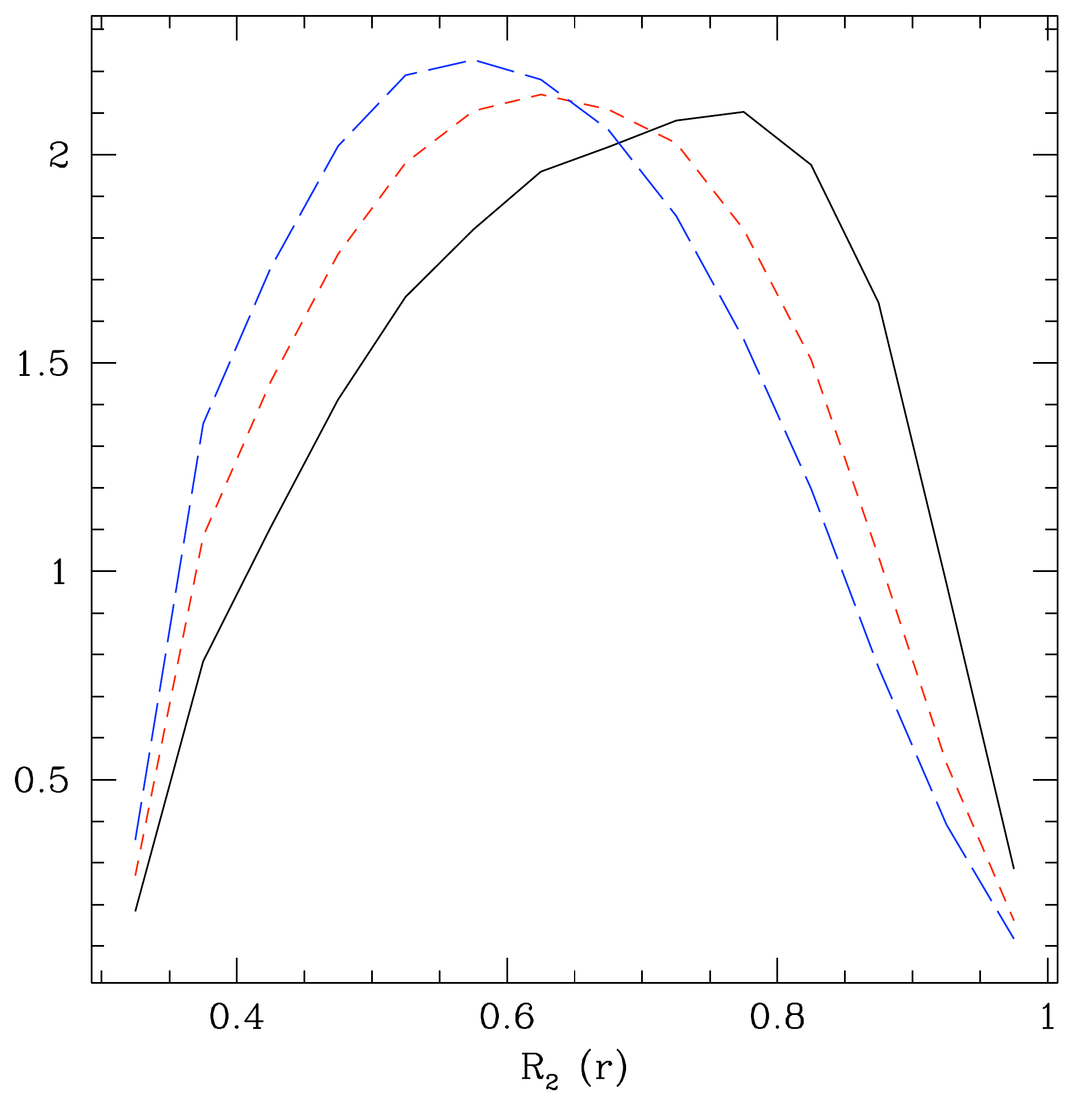} &
\includegraphics[width=0.95\columnwidth,angle=0]{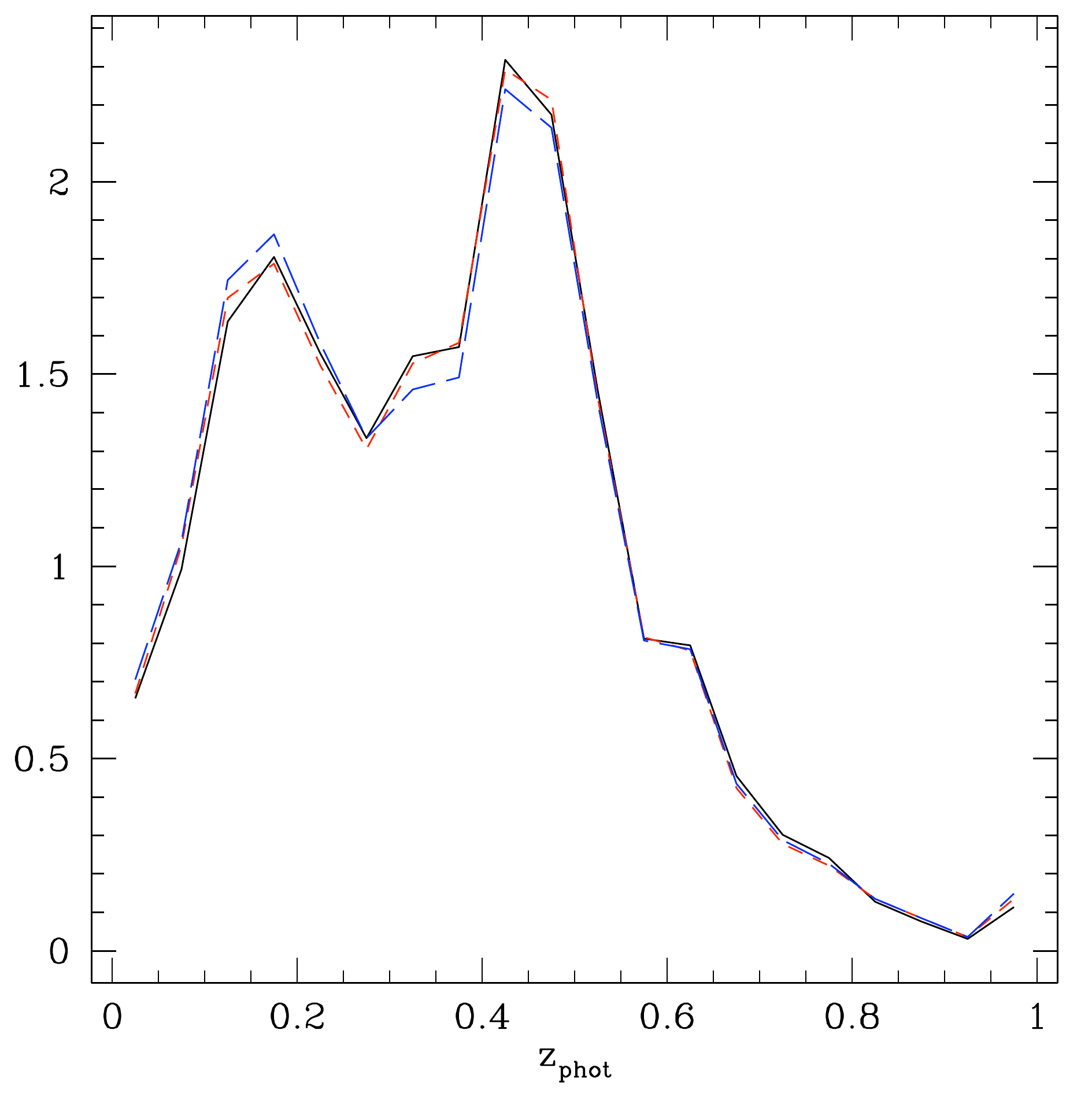} \\
\end{array}$
\caption{\label{fig:seeing} Each panel compares three independent
  versions of the source catalogue covering 125 deg$^2$ of stripe 82,
  with good, typical, and poor seeing.   {\em Top left:} Histogram of $r$-band
  PSF FWHM. {\em Top right:} Same, for
  $r$-band apparent magnitude.  {\em Bottom left:} Same, for the
  $r$-band resolution factor $R_2$.  {\em Bottom right:} Same, for \zphot.}
\end{center}
\end{figure*}

\begin{figure*}
\begin{center}
$\begin{array}{c@{\hspace{0.5in}}c}
\includegraphics[width=0.95\columnwidth,angle=0]{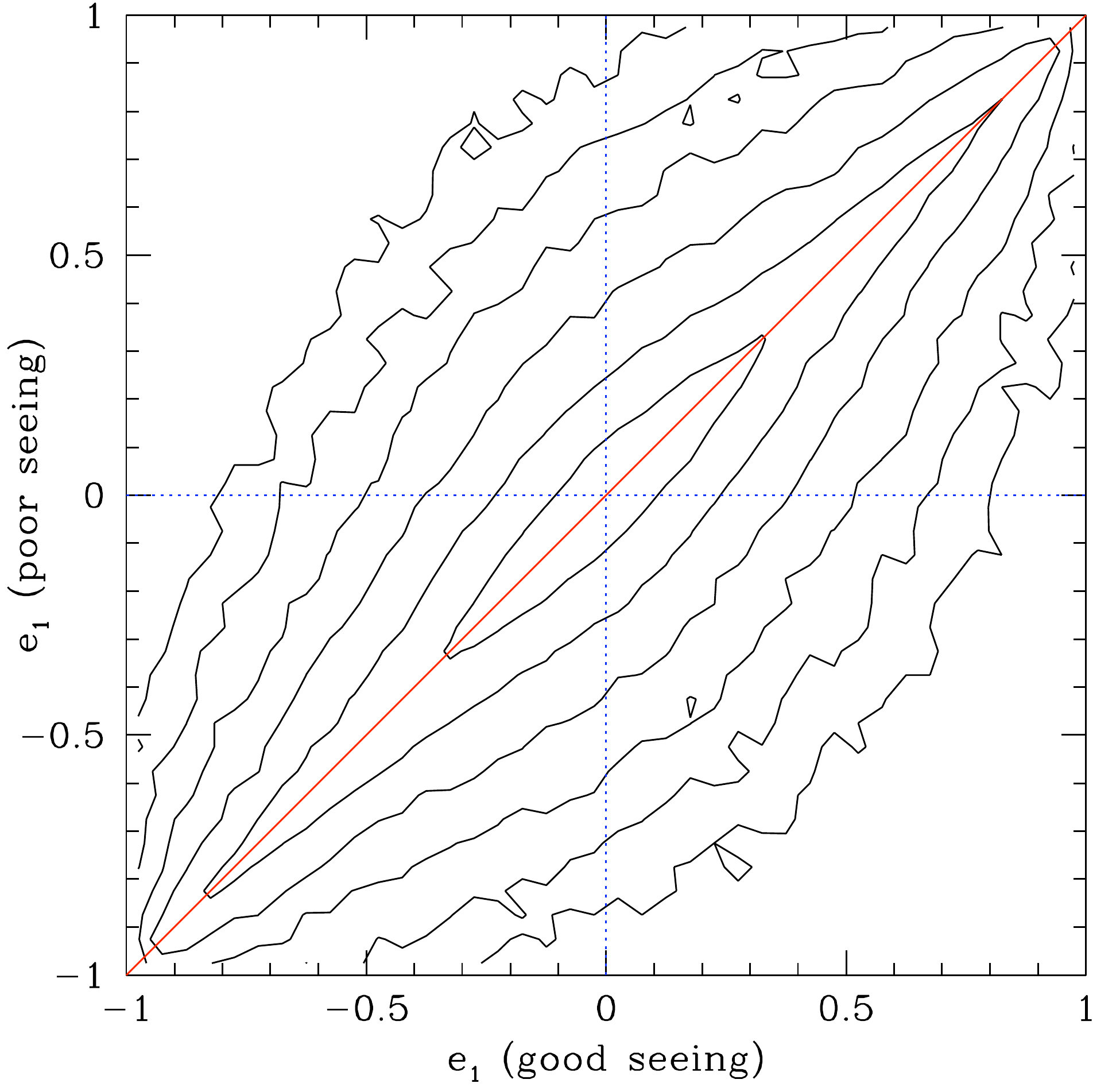} &
\includegraphics[width=0.95\columnwidth,angle=0]{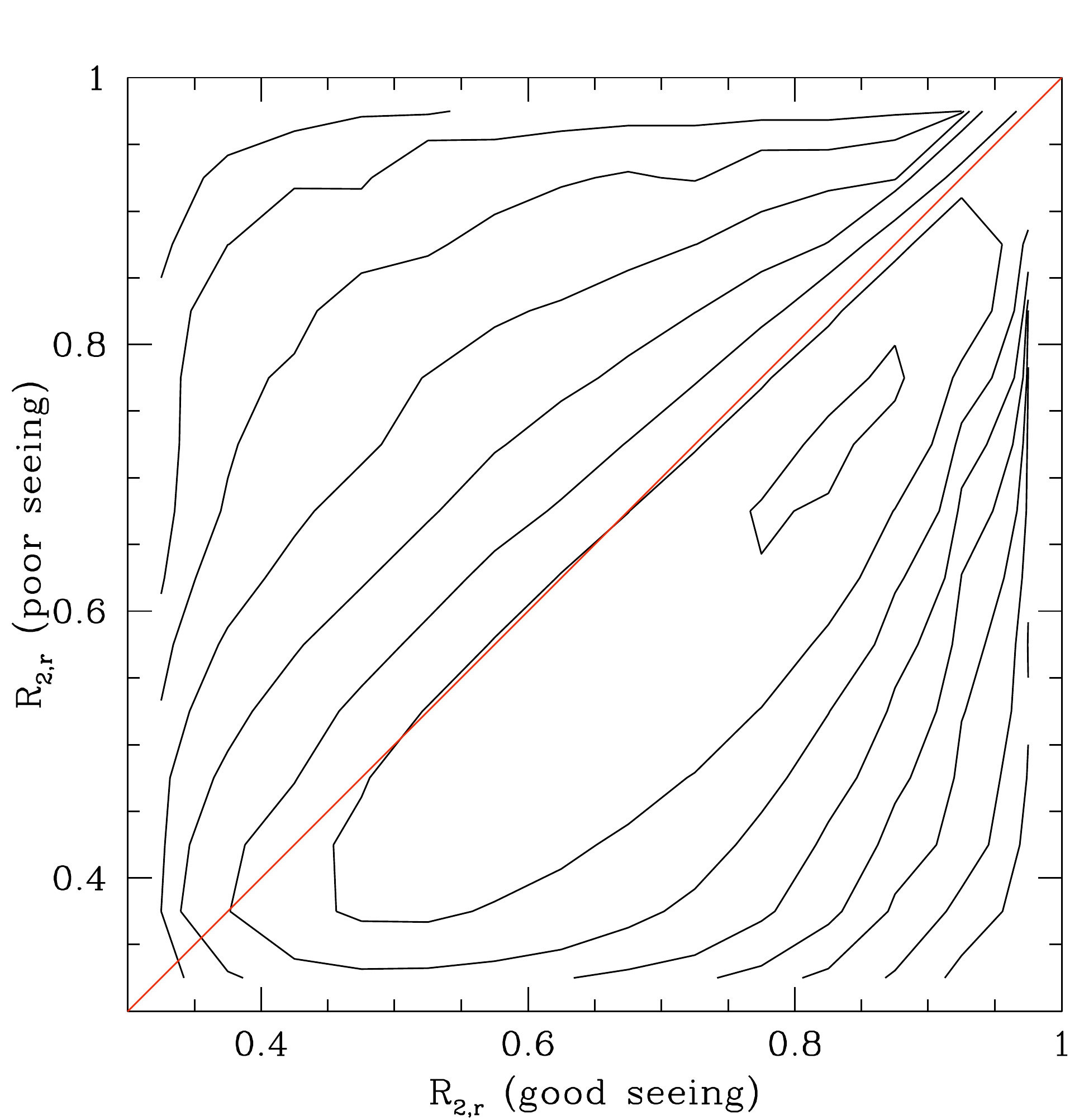} \\
\includegraphics[width=0.95\columnwidth,angle=0]{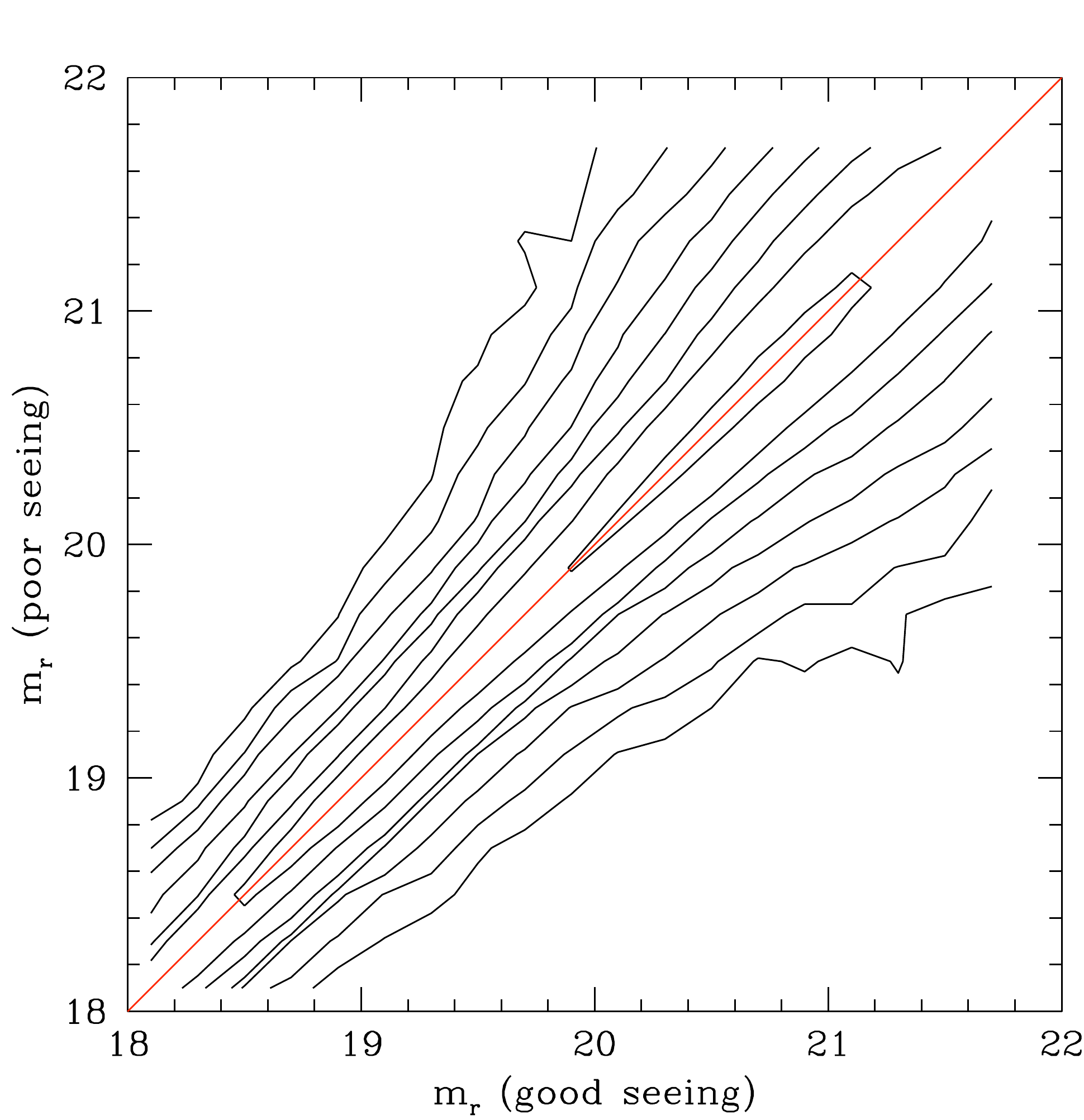} &
\includegraphics[width=0.95\columnwidth,angle=0]{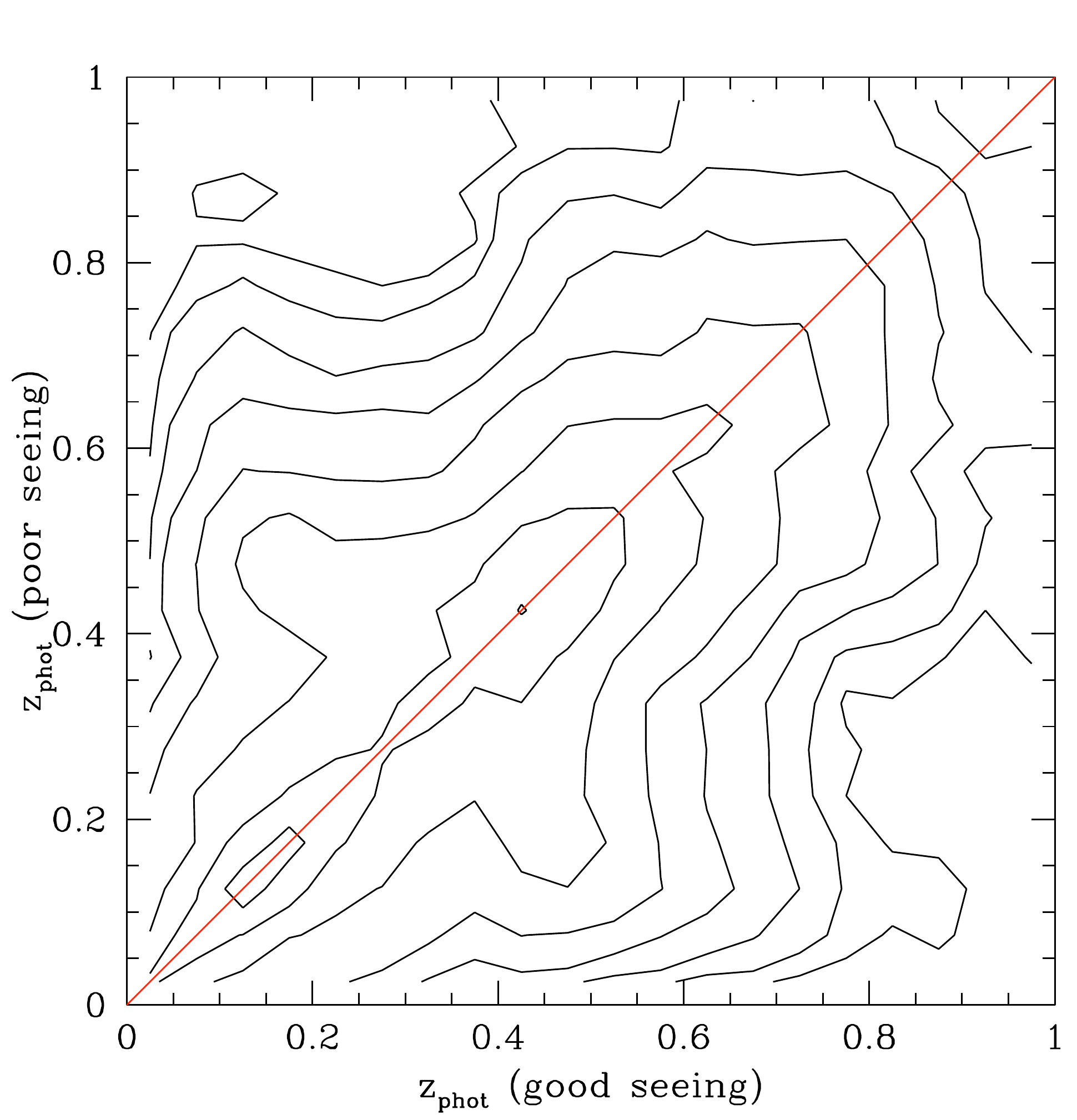} \\
\end{array}$
\caption{\label{fig:seeingdiffs} A comparison of observed galaxy
  properties for those galaxies appearing in the ``good'' and the
  ``poor'' seeing versions of the catalogues covering 125 deg$^2$ on
  stripe 82.  {\em Top left:} Density contours for the comparison
  between the $e_1$ shape components in the good- and poor-seeing
  catalogues, with logarithmically-spaced contours representing
  factors of 2.5 in density.  The plot for $e_2$ (not shown) looks
  nearly identical. {\em Top right:} Same, for
  the $r$-band resolution factor, $R_2$.  {\em Bottom left:} Same, for the
  apparent model magnitude $r$.  {\em Bottom right:} Same, for \zphot.}
\end{center}
\end{figure*}

\subsection{Systematics tests: shape catalogue}
\label{subsec:syst_cat}

\subsubsection{Tests of individual galaxy shapes}

The first systematics test that we present is a comparison between
this and the M05 shape catalogue, for galaxies that appear in both.

For this purpose, we carried out a matching process between the two
catalogues, using a tolerance of 1\arcsec.  Due to the effects of noise
on the observed resolutions and magnitudes, and the different versions
of software used to process the two catalogues, the match rate is
typically 73 per cent.  The results of matching are shown for a $\sim
100$ deg$^2$ patch of sky in
Fig.~\ref{fig:compoldnew}: we compare the $r$ band magnitudes, the
$r$-band resolution factors, and the band-averaged $e_1$ (comparison
of $e_2$ is similar). 

As shown in the top panel of Fig.~\ref{fig:compoldnew}, in this
particular patch of the sky, there is a small zero-point offset of
$0.02$ magnitudes in the $r$ band, in addition to some scatter
resulting from differences in processing.  The typical zero-point
offset between the old and new catalogue varies from location to
location, because the old catalogue used older, less reliable, and
run-dependent photometric calibrations than the new catalogue.  For
nearly all of the area covered by both catalogues, the zero-point
offset is typically several hundredths of a magnitude in either
direction.

The middle panel of Fig.~\ref{fig:compoldnew} compares the $r$-band
resolution factors in the M05 and new catalogue.  There is a clear
tendency for galaxies to be estimated as being better resolved in the
new catalogue than in the old one, which is more significant for
poorly resolved galaxies.  This results from one of the bugs in
determining the PSF for the shape measurement that affected the old
catalogue (as described in Appendix~\ref{S:differences}).  The result
of this problem is that the galaxy selection cut of $R_2>1/3$ was
intrinsically more conservative in the old catalogue than in the new
one. 

Finally, the bottom panel of Fig.~\ref{fig:compoldnew} compares the
ellipticities in the new and M05 catalogue, and reveals a slight
additive offset and a multiplicative calibration difference.
Both of these issues are due to the bug (Appendix~\ref{S:differences}) that led to
the PSF size being slightly overestimated in the old catalogue. 
The multiplicative offset is such that the galaxy ellipticities in the old
catalogue are $\sim 1.5$ per cent larger, because 
the bug had resulted in overestimated sizes for the PSFs, so the
dilution corrections were too large.  Fortunately, this 1.5 per cent effect
is not very large compared to the typical systematic calibration
uncertainties (8 per cent) that is considered as part of the error budget
in the science analyses with
the old catalogue.  The slight additive offset between
$\langle e_1\rangle$ in the old and new catalogues likewise arises due
to the PSF size misestimate in the old catalogue, which meant that the
correction for the PSF ellipticity (which is systematically nonzero along
the $e_1$ direction) was also incorrect in the old catalogue.

\begin{figure}
\begin{center}
\includegraphics[width=0.8\columnwidth,angle=0]{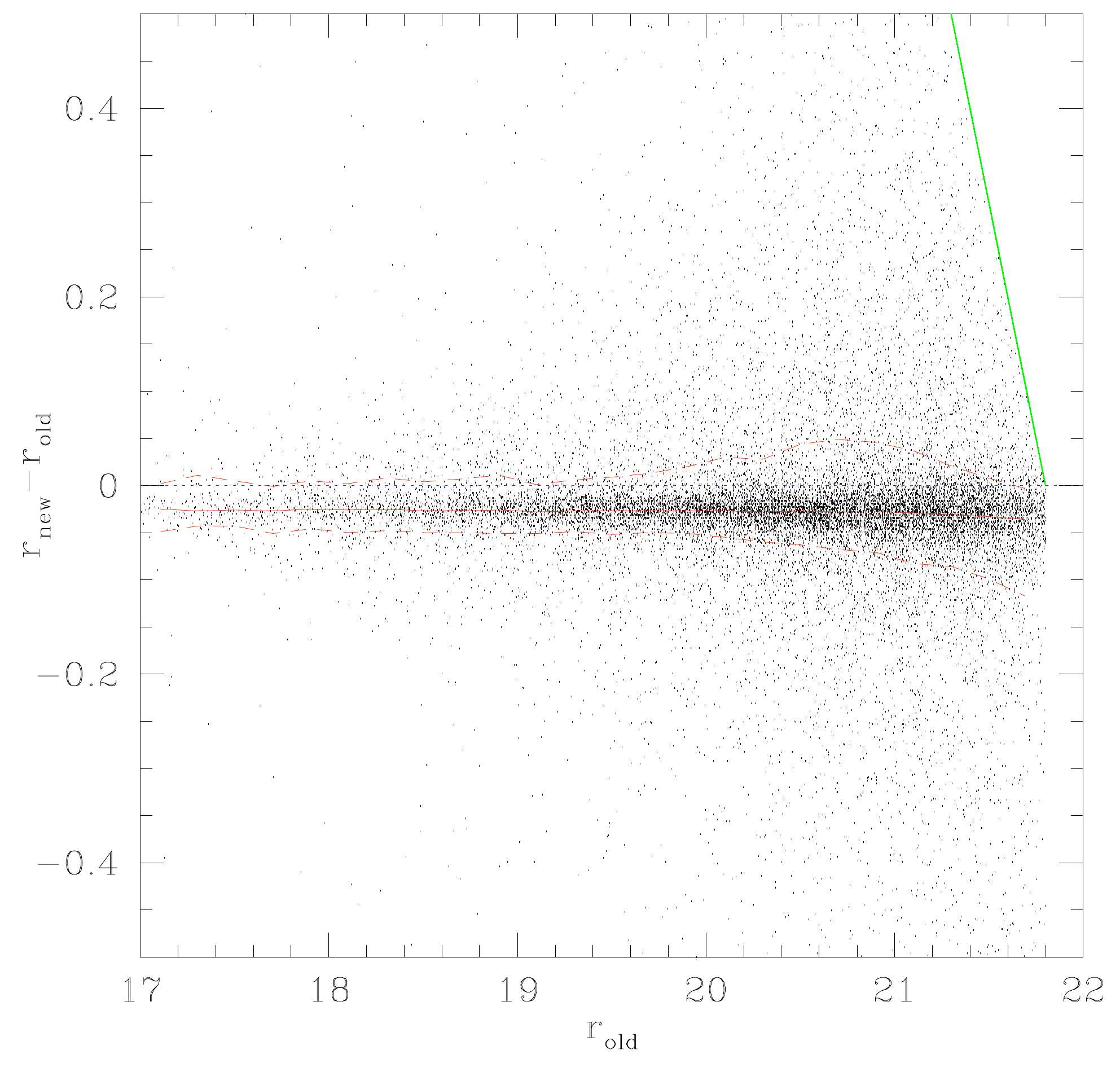} 
\includegraphics[width=0.8\columnwidth,angle=0]{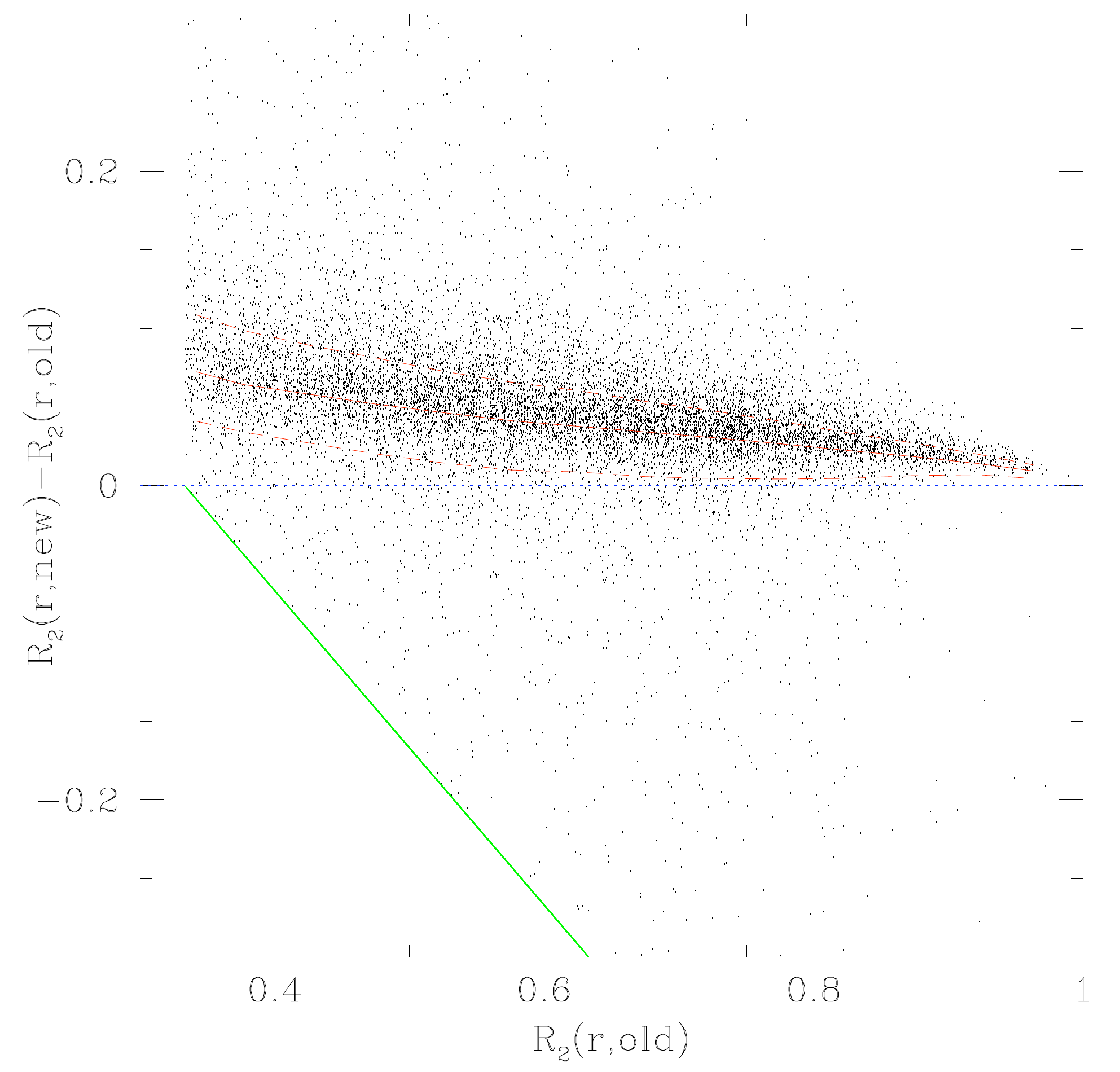}
\includegraphics[width=0.8\columnwidth,angle=0]{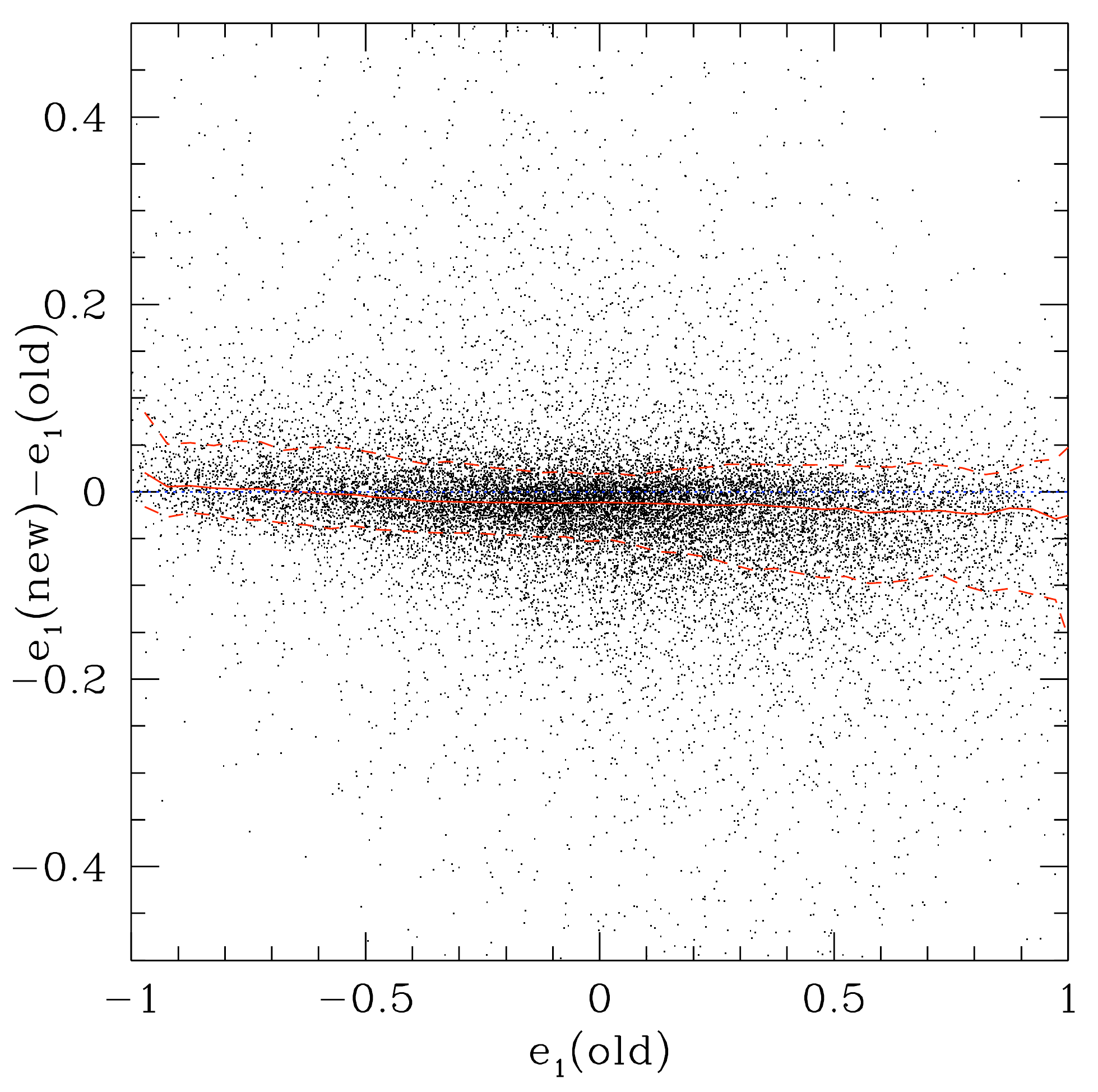} 
\caption{\label{fig:compoldnew} Each panel shows a comparison between
  observed galaxy properties in the M05 and the new source catalogue
  presented here, for a representative subsample of the area.   {\em
    Top:} Difference in $r$ band apparent model magnitudes, where the
  points show values for randomly subsampled galaxies, the solid line
  shows the median trend line as a function of magnitude in the old
  catalogue, and the dashed lines show the 68 per cent CL.  The heavy
  diagonal line shows a limit imposed by the fact that we required
  $r<21.8$ in both catalogues.   The horizontal dotted line shows the
  ideal value of $0$. {\em Middle:} Same, for $r$-band
  resolution factor $R_2$.  {\em Bottom:} Same, for the band-averaged
  first ellipticity component $e_1$.}
\end{center}
\end{figure}

We also consider, for galaxies in the new source catalogue, the
comparison between $r$ and $i$ band shapes.  Since the resolution
factor and $S/N$ is not necessarily the same in the two bands, we
might expect there to be some systematic offset even if the light
profile is the same in the two bands. Moreover, for bulge$+$disk
galaxies, we expect the bulge to be more prominent in $i$ band than in
$r$, which might lead to generally rounder shapes.
Fig.~\ref{fig:rvsi} shows a comparison of the $r$ vs. $i$ band shapes
for a randomly selected subsample (5 per cent) of galaxies in the
source catalogue.  As shown, there is no indication of any systematic
offset between the shape measurements in the two bands.

\begin{figure}
\begin{center}
\includegraphics[width=\columnwidth,angle=0]{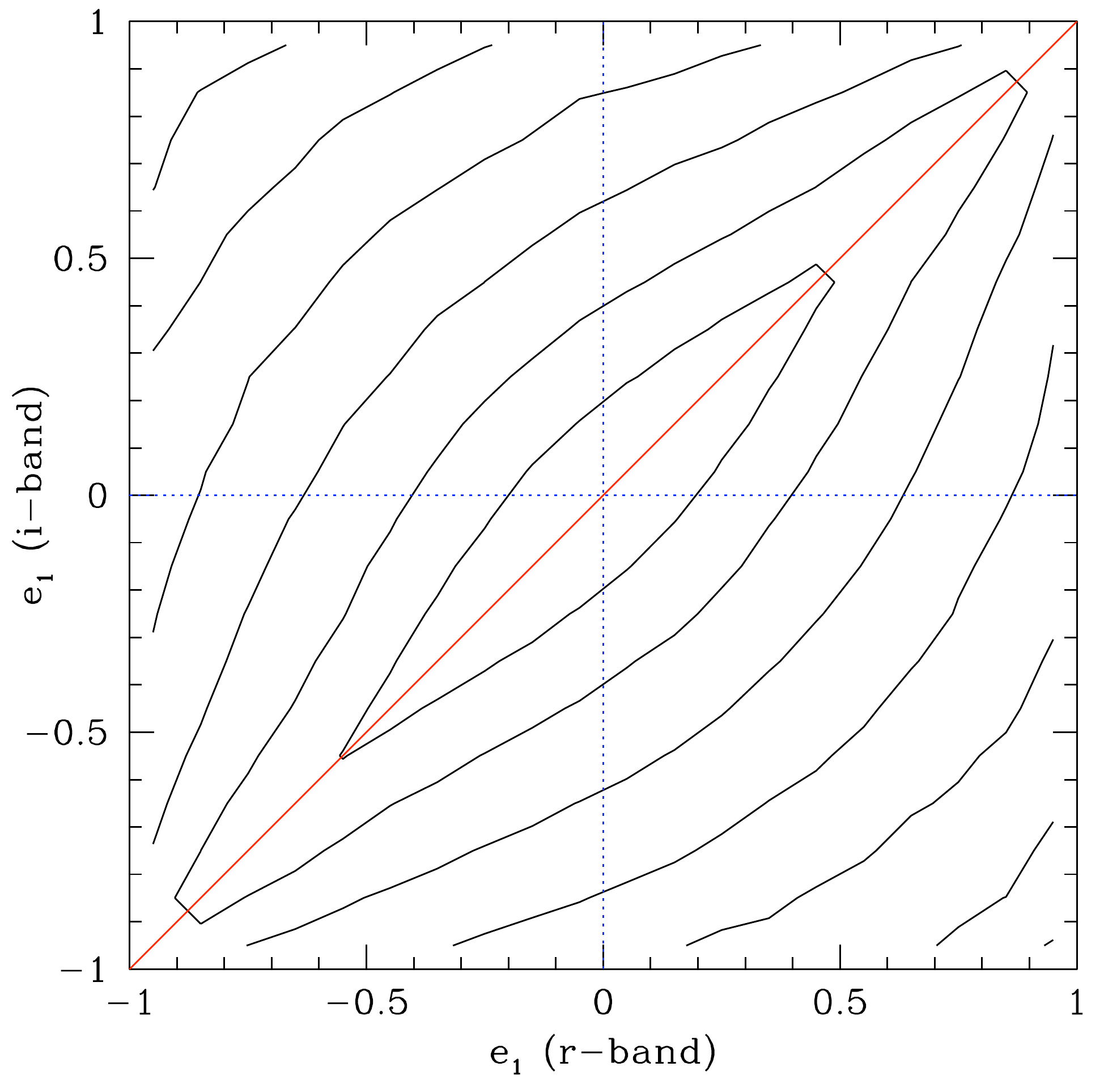} 
\caption{\label{fig:rvsi} A comparison of the $r$-band $e_1$ and the
  $i$-band $e_1$ on a per-galaxy basis.  The contours shown are
  logarithmically spaced by factors of 2.5 in density; the line shows
  the ideal 1:1 line.  A corresponding plot for the $e_2$ component
  (not shown) looks identical.}
\end{center}
\end{figure}

\subsubsection{RMS ellipticity and shape measurement errors}\label{subsubsec:erms}

One quantity which is important for estimating errors, for
optimally weighting the galaxies for shear measurement, and for
deriving shears from the ensemble (via the shear responsivity
$\ssh$$\approx 1-\erms^2$) is the RMS
ellipticity, \erms, which is defined per component, i.e.
\beq
\erms^2 \equiv \frac{1}{2N}\sum_{i=1}^N (e_{1,i}^2 + e_{2,i}^2).
\eeq
However, the complication in estimating \erms\ from the data itself is
that measurement error in the shapes can artificially inflate the
estimated \erms.  Thus, in practice it is common to estimate
\beq
\hat{e}_\mathrm{rms}^2 = \frac{1}{2N} \sum_{i=1}^N (e_{1,i}^2 +
e_{2,i}^2 - 2\sigma_{e,i}^2),
\eeq
subtracting off the estimated measurement error. This equation relies
crucially on accurate estimation of shape measurement error, and was
used in M05 to estimate that \erms\ for SDSS galaxies is a function of
magnitude, ranging from $0.35$ for bright galaxies to $0.42$ for faint
ones. 

However, recent work with simulated SDSS data \citep{2012MNRAS.420.1518M} based on
realistic galaxy models from the Cosmic Evolution Survey \citep[COSMOS;][]{2007ApJS..172....1S} strongly suggest that the
$\sigma_e$ values used in this shape catalogue are underestimated, such
that when we correct for the underestimation, the \erms\ is no longer
a function of magnitude (or at most, is a weak function of
magnitude).  Here, we present additional evidence for this issue based
on the data itself.

The key part of this analysis is that we have shape measurements in
two bands ($r$ and $i$) that we expect to be the same (modulo very
tiny differences due to colour gradients).  Thus, we can consider the
$r$ and $i$ band shape measurements to be measures of the same
intrinsic quantity, with two different noise realisations.  This means
that if we estimate 
\beq
\Delta e_{\alpha, j} = \frac{e_{r,\alpha,j} - e_{i,\alpha,j}}{\sqrt{\sigma_{e,r,j}^2+\sigma_{e,i,j}^2}}
\eeq
for $\alpha = (1,2)$ (components), $r$ and $i$ representing the bands,
and $j$ representing the galaxy, then we expect a histogram of the
$\Delta e_{\alpha,j}$ values to be approximately a Gaussian with a
mean of zero and a standard deviation of $1$.  

As shown in Fig.~\ref{fig:sige} for a random subsample (5 per cent) of the
source catalogue, the histogram differs from the expected one in two
ways: first, the best-fitting Gaussian has a standard deviation of
$1.43$ rather than one, and second, there are tails to $|\Delta e|
\ge 5$ that exceed our expectations for a Gaussian distribution.  This
plot represents empirical evidence that our shape errors are underestimated
and non-Gaussian.
\begin{figure}
\begin{center}
\includegraphics[width=\columnwidth,angle=0]{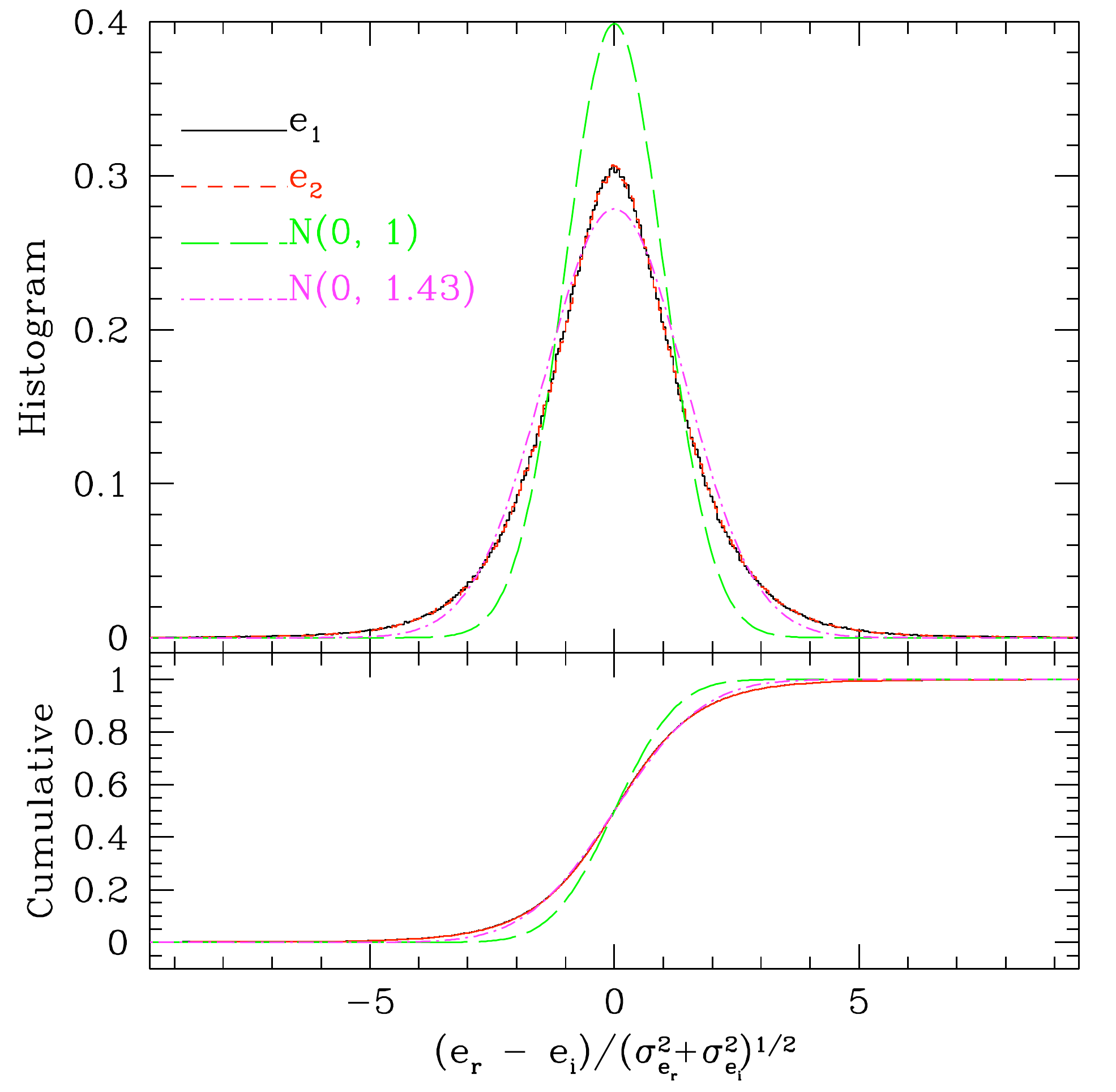}
\caption{\label{fig:sige}{\em Top:} Histograms of difference between
  the shape measurements in the two bands, normalised by the expected
  shape measurement errors.  As shown, the histograms are essentially
  identical for the two shape components, and differ drastically from
  the expected Gaussian with standard deviation of $1$ (long-dashed
  line).  The best-fit Gaussian instead has a standard deviation of
  $1.43$, but the distributions clearly deviate from a Gaussian in
  that they have too much weight in the central peak and in the tails,
  and not enough at intermediate values (large kurtosis). {\em
    Bottom:} Cumulative probability distributions corresponding to the
  histograms in the top panel.}
\end{center}
\end{figure}

We determine an approximate correction for this underestimation of the
shape measurement errors (still in the Gaussian approximation)
assuming that it is a function of apparent magnitude, resolution, and
ellipticity itself.  The correction takes the form of independent third-order
polynomials in each of those three quantities, with the correction
being more important for bright, poorly-resolved, and/or round galaxies.  Once
we apply the correction, we can estimate the \erms\ as a function of
apparent magnitude and other properties, for which the results are
shown in Fig.~\ref{fig:ermsmag}.  
\begin{figure}
\begin{center}
\includegraphics[width=\columnwidth,angle=0]{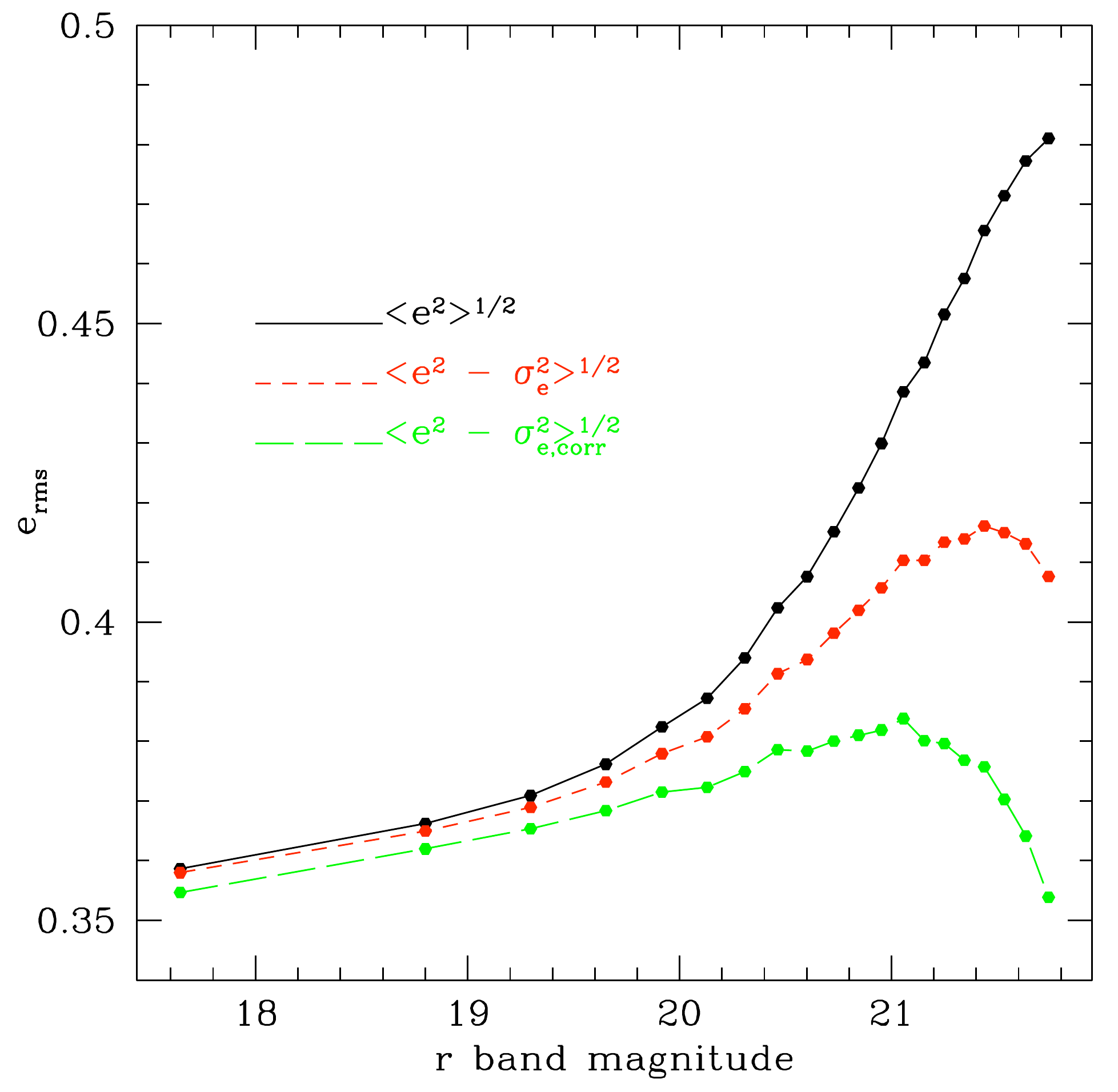}
\includegraphics[width=\columnwidth,angle=0]{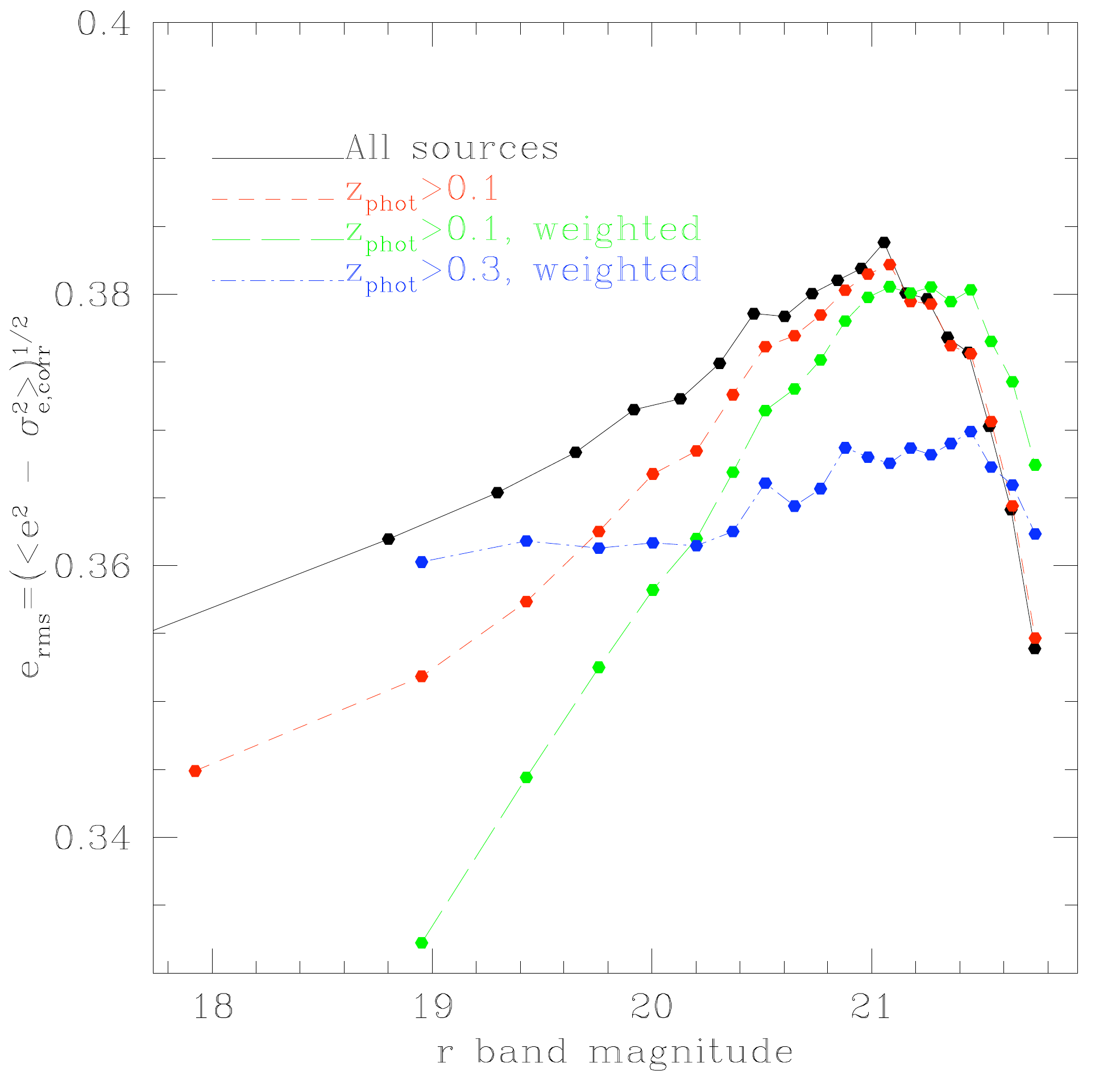}
\caption{\label{fig:ermsmag}{\em Top:} RMS ellipticity estimated
  estimated as a function of magnitude in three ways: without
  subtracting the shape measurement errors (solid line), after
  subtracting the ones estimated using Eq.~\ref{eq:sigmaedef} (dashed line),
  and after subtracting the ones that are corrected for their
  underestimation as in Sec.~\ref{subsubsec:erms} (long-dashed line).
   {\em Bottom:} RMS ellipticity estimated using the corrected shape
   measurement errors, for all sources (same as top plot); sources
   passing a $\zphot>\zlens$ cut for $\zlens=0.1$; sources
   with the same \photoz\ cut and including the source weighting from
   Eq.~(\ref{eq:wls}); and same as the previous but with $\zlens=0.3$. }
\end{center}
\end{figure}

As shown in the top panel of Fig.~\ref{fig:ermsmag}, which uses all
sources in a random 5 per cent of the source catalogue without
imposition of any additional cuts or weight factors, the underestimated shape
measurement errors led to what seemed like a significant evolution in
the value of \erms\ with magnitude, from 0.36--0.42.  However, after
correcting for the underestimation of the shape measurement errors, we
see that the \erms\ curve is much closer to flat, ranging from
0.36--0.38.  

However, what really matters when computing the signal is the \erms\
for the source galaxy population behind a given lens, including all
weight factors from Eq.~\ref{eq:wls}.  In the bottom panel of
Fig.~\ref{fig:ermsmag}, we show what happens to the \erms\ (including
the corrected shape measurement errors) when we impose cuts on the
\photoz, requiring $\zphot>0.1$--- a cut that corresponds to the
maximum lens redshift for the sample used in this paper--- and also
when we include the lensing weight factors when calculating \erms.  As
shown, when we include the $\zphot>0.1$ cut, the estimated \erms\
decreases somewhat, indicating that the source galaxy population
becomes somewhat rounder.  The inclusion of lensing weights
exaggerates this effect even further.  

This discussion ignores the non-Gaussianity in the shape measurement
errors, demonstrated by the kurtosis leading to tails to large values
in Fig.~\ref{fig:sige}.  Using the noise probability distribution from
that plot, rather than a Gaussian one, we estimate that the
non-Gaussian nature of the noise has an equal impact on \erms\ as if
the shape measurement errors were Gaussian but with a standard
deviation that is a factor of $1.18$ larger.  When we include this in
the calculations that go into Fig.~\ref{fig:ermsmag}, the long-dashed
line in the top panel becomes even closer to flat with magnitude than it already is.

Ultimately, we will need to use these results about the \erms\ when we
determine the shear calibration bias using the new catalogue in
Sec.~\ref{subsec:shear_calib}.  We conclude based on the corrected
shape measurement error estimates and our assessment of the noise
probability distribution that an appropriate value of \erms\
(including the source \photoz\ cut implied by the requirement that
$\zphot>\zlens$, and the source weighting), is in the range
$0.35$--$0.37$.  This implies that the true shear responsivity should
be \ssh $= 1-\erms^2=0.863$--$0.878$.  

As discussed in \cite{2012MNRAS.420.1518M}, a flat \erms\ with
magnitude is also consistent with space-based data from COSMOS
\citep{2007ApJS..172..219L}.  While the actual values of \erms\ differ
($0.27$ there, versus $0.36$ here), this is a consequence of the
different shape estimators used for the COSMOS versus for the SDSS
data, rather than a true disagreement in the intrinsic galaxy
shapes \citep{2012MNRAS.420.1518M}.  The circularly-weighted shape
estimators used for the COSMOS galaxies will tend to less elliptical
(rounder) measurements than the adaptive moments used for SDSS.



\subsection{Lensing signal calibration}
\label{subsec:shear_calib}

To estimate the calibration of the lensing signal 
measured as weighted sums over $\hat{\gamma}_t \hat{\Sigma}_c$
(Sec.~\ref{subsec:lens_calc}), we consider several types of calibration biases,
from M05 and subsequent work (e.g. \citealt{2008MNRAS.386..781M}).  We
then combine the estimated biases by multiplying them all together,
assuming that they are independent.  We combine the $1\sigma$
calibration uncertainties in the following way: those that are roughly
independent are added in quadrature, whereas those that cannot be
considered independently from each other (since they arise due to
e.g. related issues in the data reduction or galaxy selection) are
added linearly.  We therefore consider calibration biases and
uncertainties due to all of the following:

\begin{enumerate}
\item Calibration uncertainty due to misestimation of $\Sigma_c$ due
  to use of \photoz\ for sources.
\item Stellar contamination due to incorrect inclusion of stars in the
  shape catalogue.
\item PSF model uncertainty.
\item Shear responsivity error due to mistaken \erms\ estimate
  resulting from incorrect shape measurement error estimates (as in
  Sec.~\ref{subsubsec:erms}).
\item Three different shear calibration biases that we {\em cannot}
  consider independently, and that have uncertainties that add
  linearly:
\begin{enumerate}
\item PSF dilution
\item Noise rectification bias
\item Selection biases
\end{enumerate}
\end{enumerate}

We describe each of these separately and conclude this subsection with
a final tally of the systematic error budget.

\subsubsection{Photometric redshift errors}

N11 addressed the question of how \photoz\ errors for these ZEBRA
\photoz\ impact the calibration of the galaxy-galaxy lensing signal, as a
function of lens redshift.  Here, we simply use the results from N11
directly, which results in a calibration bias estimate of $2.0 \pm
0.5$ ($1\sigma$) per cent (the same for each stellar mass bin due to
their similar redshift distributions).

\subsubsection{Stellar contamination}

Using the calibration sample from N11 and space-based data from
COSMOS, we can constrain the stellar contamination in the galaxy-galaxy lensing
signal.  Out of 4~290 source galaxies overlapping the COSMOS region,
32 are stars, or $0.75$ per cent.  However, to constrain their impact
on the shear calibration, we cannot simply use the fractional
contamination; we have to include the lensing weights to calculate the
fractional weight given to stars, and account for the fact that we
require source $\zphot > \zlens$.  When we take this into
account, we estimate stellar contamination of $0.64$ per cent, with
$1\sigma$ Poisson uncertainties of $[-0.15, +0.12]$ per cent.

However, there is an additional uncertainty having to do with the fact
that (a) the stellar density is not always the same as in the COSMOS
field, and (b) the observing conditions in the COSMOS field, including
the atypically high sky noise (see N11 for details), may affect the
influence of incorrectly including stars in the catalogue.  
The first issue is likely not very significant, since the stellar number density
depends on galactic latitude according to $1/|\sin{b}|$. 
On average, this quantity is equal to $1.43$ over the COSMOS field versus 1.40 over the whole catalogue
(the difference is not severe because we exclude regions near $|\sin{b}|\sim 0$ with the $A_r<0.2$
cut).   Because of systematic uncertainty associated with the second
point, we double the $1\sigma$ statistical error, which ends up giving
us a $50$ per cent uncertainty in the stellar contamination: $0.64\pm
0.3$ per cent. Note that stellar contamination biases the shear to be lower, 
so the sign of the bias is negative, i.e., $-0.64$ per cent.

\subsubsection{PSF model uncertainty}

We follow the method of \cite{2004MNRAS.353..529H} for handling PSF
model uncertainty.  The basic idea is if the PSF model is not correct,
then the PSF correction can introduce systematics into the shape
measurement.  If these errors in the PSF model are purely statistical,
then we can expect that they will lead to shear calibration biases
that will ultimately cancel out using a large enough
area.  However, systematic issues with the PSF model size will lead to
coherent tendencies to over- or under-correct for the dilution of the
galaxy shapes due to the PSF, so we would like to constrain such
systematics.\footnote{Systematic issues with the PSF ellipticity
  typically cause coherent additive shears rather than shear
  calibration biases; see Sec.~\ref{subsubsec:sysshear}.}

We first estimate systematic errors in PSF ellipticity ($\delta e_1$,
$\delta e_2$) and trace, $\delta T^{(P)}/T^{(P)}$, using real stars
drawn from randomly selected fields in the source catalogue.  In each
randomly selected field, we use all stars passing basic flag cuts and
with $18<r<20$.  The motivation behind using this range of magnitudes
is that (a) they are faint enough that this is a non-trivial test of
the PSF model because PSF stars\footnote{R. Lupton, priv. comm.} only go as faint as $r\sim$19, and
(b) they are bright enough that the moments are not too noisy and
contamination of the star sample by galaxies is not overly large
\citep{2001ASPC..238..269L}.  For each of the $5\times 10^5$ stars, we find, in both $r$
and $i$ bands, the following:
\beqa \nonumber
\Delta e_{1} &= e_{1,\mathrm{star}} - e_{1,\mathrm{PSF}} \\\nonumber
\Delta e_{2} &= e_{2,\mathrm{star}} - e_{2,\mathrm{PSF}} \\ \label{eq:deltaq} 
\Delta \ln{T} &= (T_\mathrm{star} - T_\mathrm{PSF})/T_\mathrm{PSF},
\eeqa
where $T$ is the trace of the moment matrix (Eq.~\ref{eq:xxmom}) and the
``PSF'' quantities are those of the full KL PSF model extrapolated to the
position of that star.

The results of this test are shown in Fig.~\ref{fig:psfmodel}.  The
upper left panel shows the histograms of $\Delta e_1$, $\Delta e_2$,
and $\Delta \ln{T}$ in the $r$ band ($i$ band is qualitatively
similar).  Before generating this histogram, we removed the small
fraction of outliers that had $>5\sigma$ discrepancies between the
star and PSF model, assuming that these are due to contamination of
the star sample rather than true modeling failures.  As shown, the PSF
model ellipticities appear to be quite accurately estimated.  However,
the PSF trace is
skewed towards positive values.  Note that this is exactly the
signature we might expect from contamination of our test sample by
poorly resolved galaxies (since we have not imposed precisely the same
set of cuts that are used to select stars for PSF reconstruction).
  As
a result, we cannot definitively state that these characteristics of
the histogram are due to true PSF model failure.  Moreover, even if it
is due to PSF model failure, the actual size of the failure is (in the
mean) quite small, $\lesssim 1$ per cent.  But a difficulty with this explanation is that when we split the star
sample used for these tests into $r<19$, which may have been used for
PSF model estimation, and $r>19$, which should not have been, we find
that the deviations between the star sizes and the PSF sizes are
actually more significant for the brighter sample than for the fainter
one.  

\begin{figure*}
\begin{center}
$\begin{array}{c@{\hspace{0.5in}}c}
\includegraphics[width=0.95\columnwidth,angle=0]{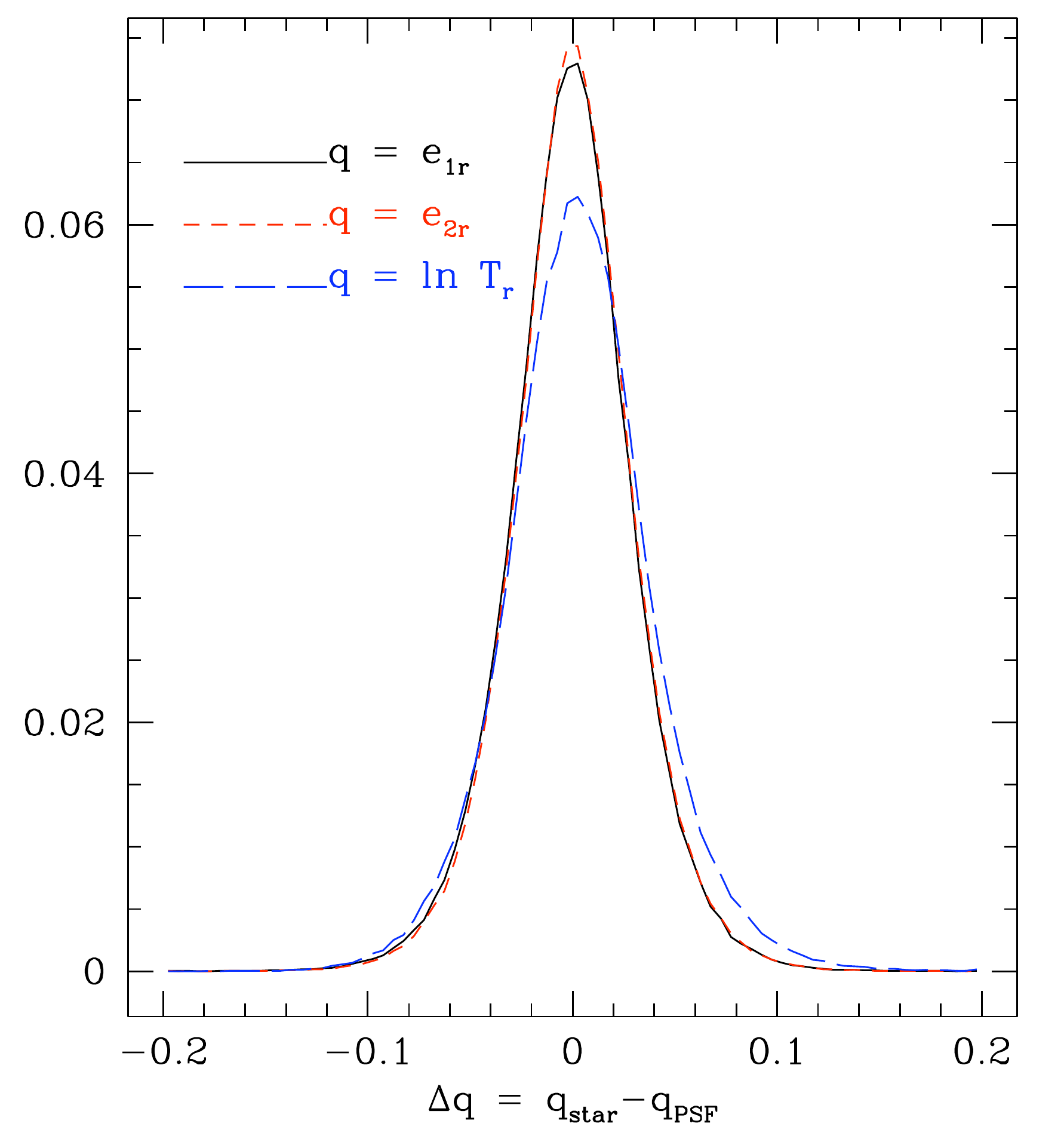} &
\includegraphics[width=0.95\columnwidth,angle=0]{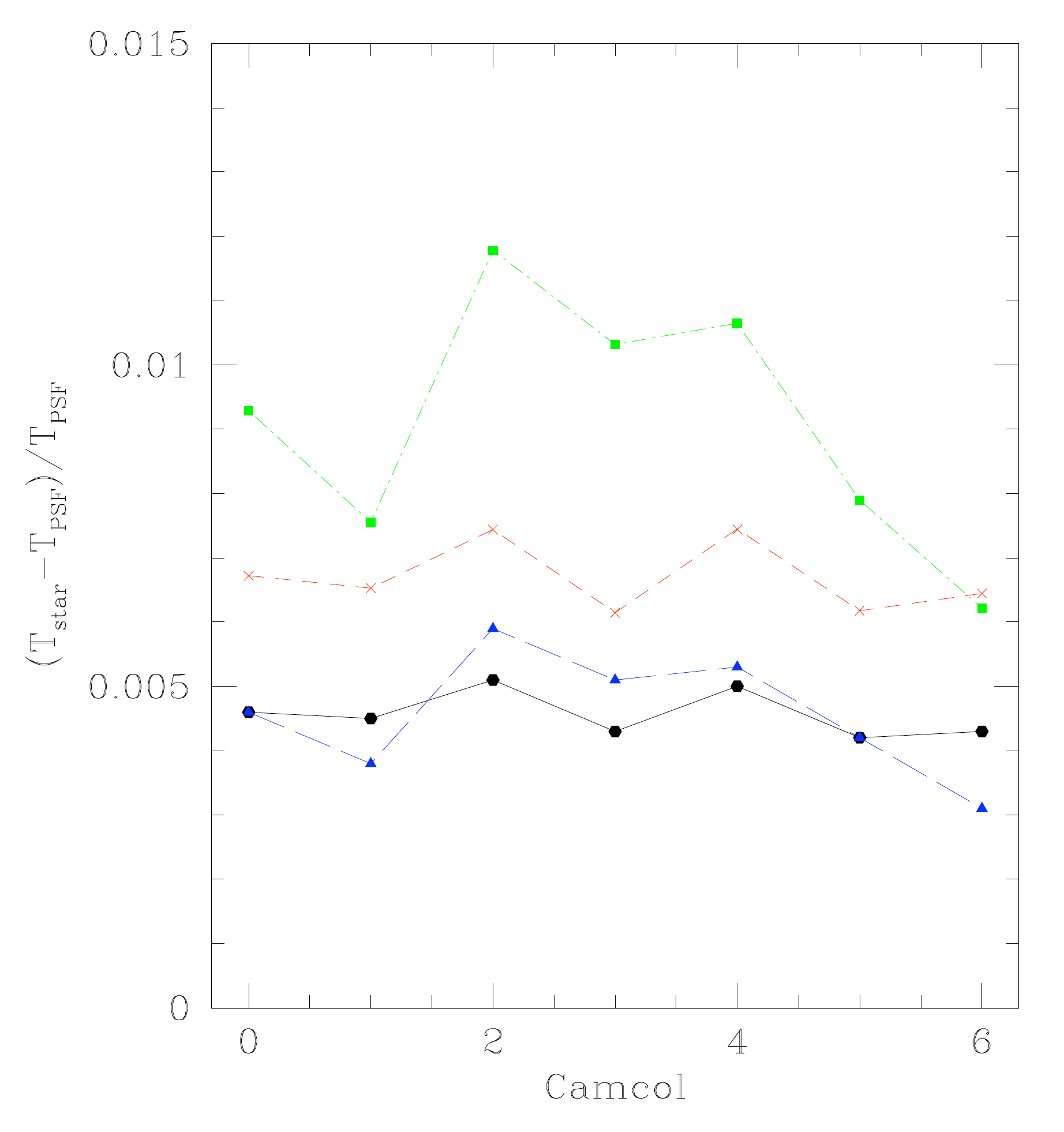} \\
\includegraphics[width=0.95\columnwidth,angle=0]{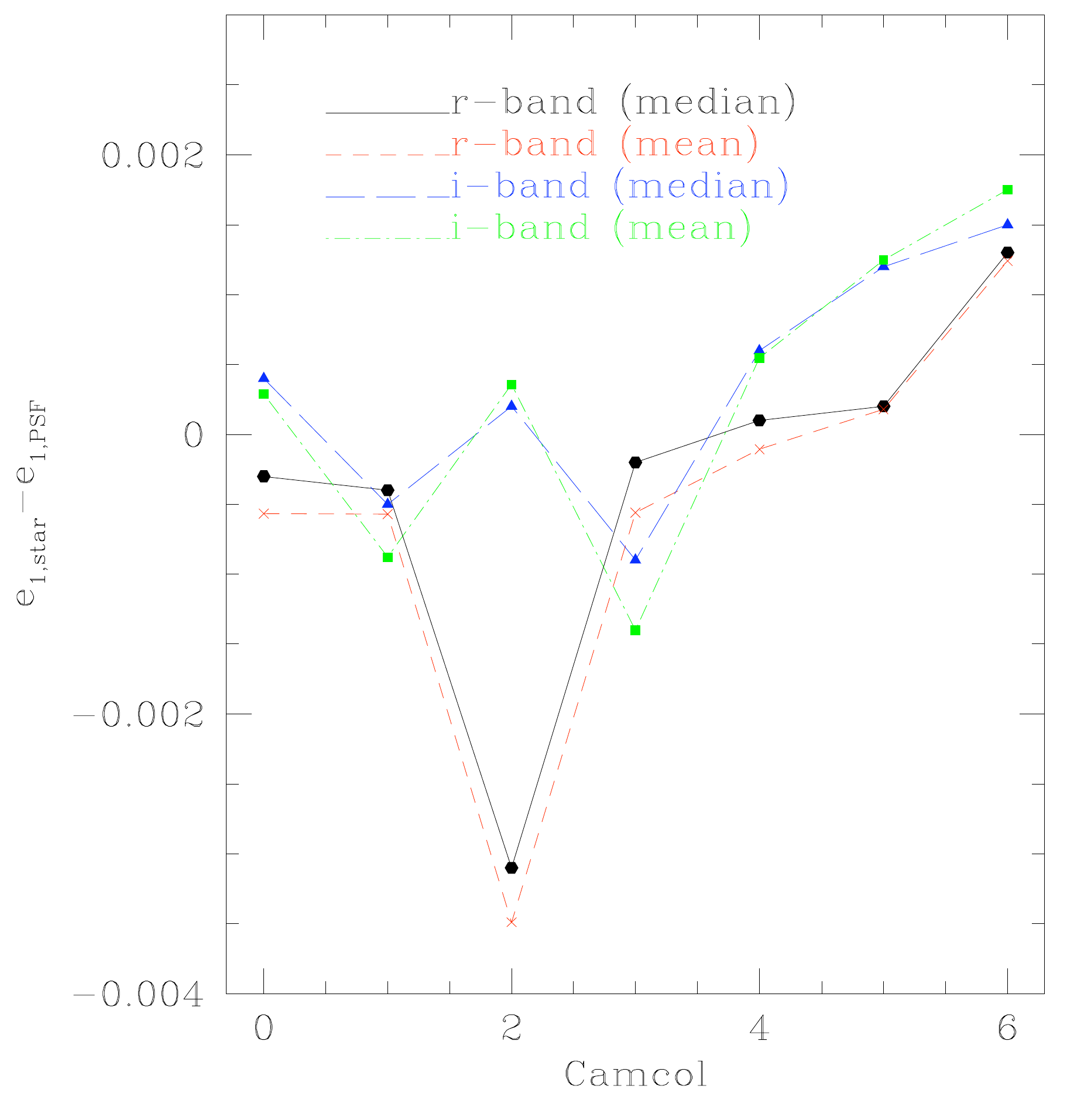} &
\includegraphics[width=0.95\columnwidth,angle=0]{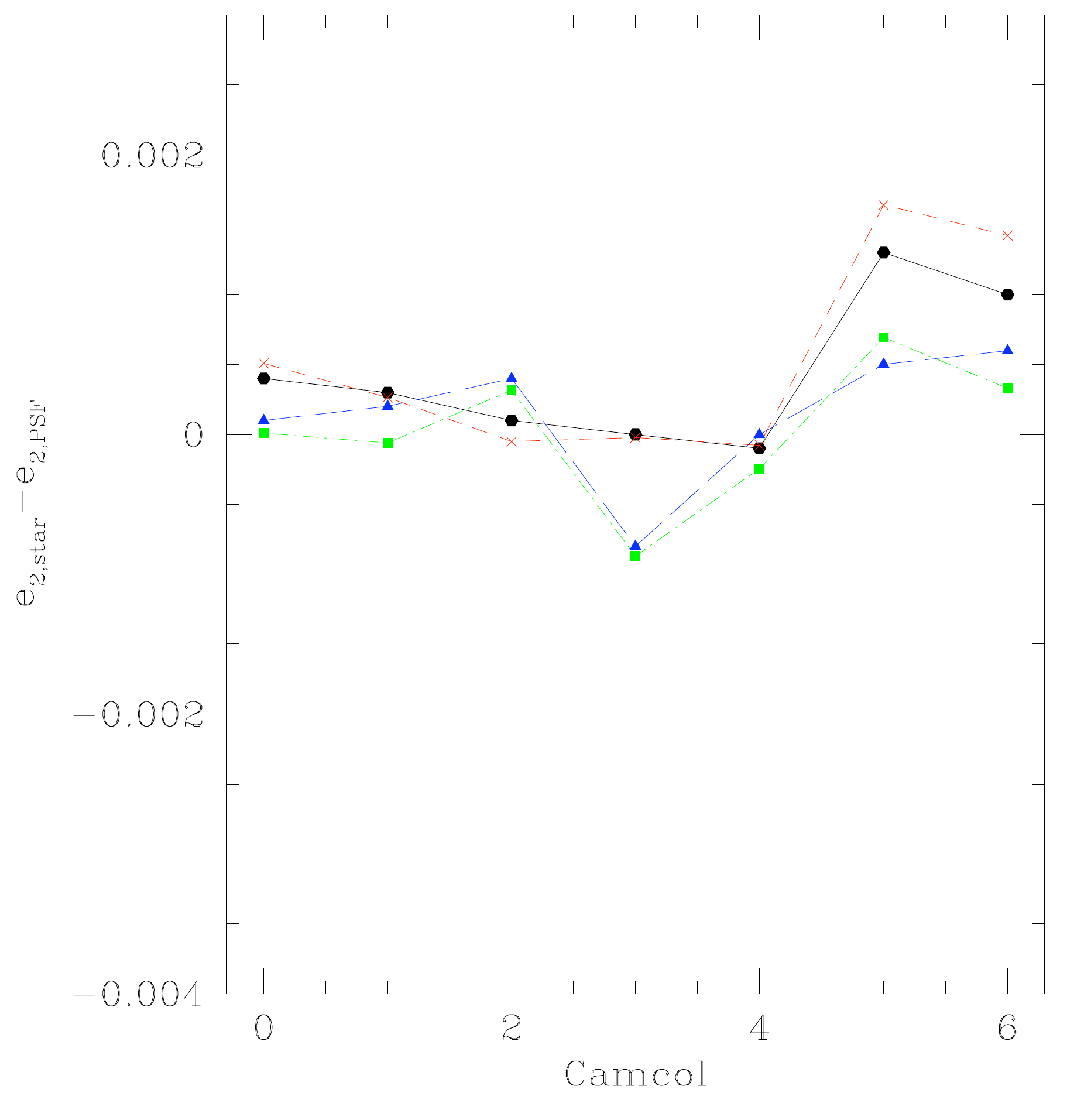} \\
\end{array}$
\caption{\label{fig:psfmodel} Plots relating to PSF modeling
  failures.  {\em Top left:} Histograms of $r$-band $\Delta e_1$, $\Delta e_2$,
and $\Delta \ln{T}$ for a sample of $\sim 5\times 10^5$ stars in
randomly selected fields in the source catalogue, where the quantities
that are plotted are defined in Eq.~(\ref{eq:deltaq}) and associated text. 
{\em Top right:} The typical value of $\Delta \ln{T}$ as a function of
camcol, where the value at camcol$=0$ is averaged over all 6 camcols.
The size of the statistical errors are similar to the size of the
points themselves, so they have been omitted.  The four lines, as
labeled in the lower left panel, show
results for the two bands ($r$ and $i$) and show the median and mean
trend, which differ because the distribution (shown in the top left
panel) is skewed. {\em Bottom left:} Same as top right, for $\Delta
e_1$.  {\em Bottom right:} Same as previous, for $\Delta e_2$.
}
\end{center}
\end{figure*}

The other panels of Fig.~\ref{fig:psfmodel} show trends in the
quantities from Eq.~\ref{eq:deltaq} with camcol, where the results
in camcol$=0$ show the results for all camcols averaged together.  As
shown in the bottom panels, while the median and mean deviations in
the PSF model ellipticities are of order $5\times 10^{-4}$ (with the
median and means agreeing quite well, given that the distributions in
the upper left panel are well behaved), there is an interesting trend
with camcol for the $r$-band, with the $r$-band PSF model ellipticity
in camcol 2 being on average wrong by $\sim 3\times 10^{-3}$ (about $20$ times
worse than for the data in all camcols, overall).  This trend was
originally noted in Huff et al. (2011, {\em in prep.}), and it means
that if we average the galaxy shapes in camcol 2, they have a small
systematic deviation from zero as well, because the wrong PSF ellipticity
was used in the PSF correction process.  Possible explanations include
improper nonlinearity corrections on the $r$-band camcol 2 CCD,
because the stars used to create the PSF model have significant
nonlinearity corrections, but the stars used for this test and the
galaxies typically used for the shape catalogue do not.  Fortunately,
while this does impact large-scale systematic additive shears, it does
not affect the shear calibration since the PSF model size is not
nearly as affected as its ellipticity.

The upper right panel shows statistics of $\Delta \ln{T}$ in both $r$
and $i$ band. The median and mean values differ because the
distribution is, as shown in the top left panel, noticeably skewed.
There is no drastic trend with camcol (as shown in the upper right panel).  For the value of $\Delta
\ln{T}$ that goes into our estimate of shear calibration bias due to
improper PSF modeling, we will use a lower limit of zero, because the
trends in this plot could in principle be caused by contamination of
the ``star'' sample by some number of very poorly resolved galaxies.
Our upper limit will be the upper line on the upper right panel,
which gives $0.01$ (1 per cent error in the PSF trace).  Note that our
lower and upper limits are {\em not} driven by the spread in the
histogram in the upper left panel, because we assume that shear
calibration biases from deviations in the PSF model due to noise will
simply average out.  So, we only worry about how well we can constrain
the typical value of $\Delta \ln{T}$.

Then we use  
\beq 
\frac{\delta
  \gamma}{\gamma} = -(R_2^{-1}-1) \frac{\delta T^{(P)}}{T^{(P)}}
\eeq
from \citet{2004MNRAS.353..529H} (accounting for the difference in the sign convention in the definition of $\Delta \ln{T}$ from Eq.~\ref{eq:deltaq}).
This equation says that the shear calibration bias due to incorrect PSF size
estimation is more important for poorly resolved galaxies than for
well-resolved galaxies, as one would expect.  
The sign of the effect is such that if the PSF size estimate is biased low,
the shape measurements would also be biased low.
our corrections are not large enough.  This means that the shape measurements are too low.
When we include
the fact that the average value of $\langle R_2^{-1}-1\rangle$ for the
catalogue is $0.70$ (using $\zphot>\zlens$ for the lens
redshifts used in this work), and calculating a weighted mean using the
lensing weights from Eq.~\ref{eq:wls}, we estimate an uncertainty for
the shear calibration bias due to systematic PSF reconstruction errors
to be in the range $[0.0, 0.007]$.  In practice, we assume a mean bias
in the shear due to PSF model errors of $-0.4\pm 0.4$ per cent.

\subsubsection{Shear responsivity error}

As shown in Sec.~\ref{subsec:lens_calc}, our shear estimator relies on
the fact that the shape definition in Eq.~(\ref{eq:shapedef}) responds to shear
in a particular way: $\delta e_+ /\delta \gamma_+ \approx 2(1-\erms^2)
\equiv 2\ssh$, where the shear responsivity depends on the intrinsic
RMS ellipticity per component for the source sample.  As described in
Sec.~\ref{subsubsec:erms}, accurate estimation of \ssh\ requires us to
be able to subtract the shape measurement error to get the `true' RMS
ellipticity of the population.  We have already shown there that our
shape measurement errors are underestimated, leading to an
overestimation of \erms, underestimation of \ssh, and overestimation
of the shear.  As we concluded there, the true value of \erms\ should
be in the range $0.35$--$0.37$, giving \ssh$=
1-\erms^2=0.863$--$0.878$.  When computing the signal using the shape
measurement errors in the catalogue, we had estimated \ssh$=0.848$.
This means that our mean calibration bias for which we must correct is
$+2.5$ per cent, and its $1\sigma$ uncertainty is, conservatively, $\pm 1$
per cent.

\subsubsection{PSF dilution}

PSF dilution is the rounding of the galaxy shape due to the PSF.  A
key purpose of PSF-correction is to apply a correction for the PSF
dilution.  The accuracy of the re-Gaussianization dilution correction,
as a function of the galaxy resolution factor and S\'{e}rsic index,
was shown by \cite{2003MNRAS.343..459H} and \cite{2005MNRAS.361.1287M}
using noiseless idealised simulations; more recently, \cite{2012MNRAS.420.1518M} did
the same test but using a realistic distribution of galaxy
morphologies (including small-scale structure, starting from COSMOS
galaxy images) and SDSS PSFs.

The simulations in \cite{2012MNRAS.420.1518M} specifically mimicked the SDSS imaging
conditions in the COSMOS field, which have slightly better seeing and
somewhat higher sky noise than typical SDSS data.  From the results
shown for the noisy simulations there, we can directly derive the
calibration bias due to PSF dilution and noise rectification bias
together, rather than deriving separate estimates for each one. 

\subsubsection{Noise rectification}

Noise rectification bias is bias in shear estimates due to the finite
$S/N$ of the galaxy images.  Because the estimation of the galaxy
shape is a nonlinear process involving measurement of galaxy moments,
the noise in the original image means we cannot necessarily estimate
galaxy shapes that are unbiased in the presence of noise.
\cite{2004MNRAS.353..529H} and M05 gave analytic approximations for
the noise rectification bias for methods such as re-Gaussianization
that rely on using adaptive moments of the image to estimate the
shape.  As shown there, the magnitude and sign of the effect depends
on how well resolved the galaxy is, but it becomes large and positive
for poorly resolved galaxies, which is the motivation for our
imposition of a cut on resolution factor at $R_2=1/3$.  In this work, we rely on the simulated SDSS data described
above, from \cite{2012MNRAS.420.1518M}.  The idea is that simulating
data without any noise allows us to derive a calibration bias due to
incorrect PSF dilution corrections, and when we add noise to the
simulations, we are measuring a combination of the calibration biases
from both the incorrect dilution corrections and the noise
rectification bias.

The resulting bias and uncertainty on that bias are $-6\pm 2$ per cent
(the $-4\pm 2$ per cent from \citealt{2012MNRAS.420.1518M} included,
as discussed there, a $+2$ per cent bias due to shear responsivity
errors, which we want to treat separately for the purpose of this
paper).  However, there are additional effects that must be taken into
account in any realistic analysis, which we account for here.  

First, we must consider the fact that in \cite{2012MNRAS.420.1518M} we
applied only a crude galaxy selection to the simulated galaxies:
rather than processing the simulations with {\sc Photo} so that we
could cut on the model magnitudes, we simply
applied an estimated $S/N$ cut (using the estimated $\sigma_\gamma$
based on the sky noise, ellipticity, and resolution factor).  The
other cuts were similar to those employed here, on the resolution
factor and total ellipticity.  The first question we face is whether
the galaxy population that results is similar in $S/N$ and resolution
factor to the observed one in the real catalogue.  To avoid issues of
sample variance, we ask this question specifically using the real
source catalogue in the COSMOS region.  There, we find that there are
3~695 galaxies with COSMOS galaxy postage stamps used for the
simulation that pass all cuts in the real data, as compared with 3~680
in the simulations.  These numbers are statistically consistent once
we account for the fact that different random noise fields were added
to the simulations than to the real data. This finding suggests that
the cuts in the simulations cannot be effectively too different from
the model magnitude cuts in the real data.   Moreover, the
two-dimensional distributions of resolution factor and $S/N$ were
nearly consistent, with the effective mean values the same in the
simulations as in the real data to within $3$ per cent.  Given that
the shear calibration was only found to be a weak function of those
properties in \cite{2012MNRAS.420.1518M}, we assume that the effect of
the crude cuts imposed on the simulations leads to a negligible change
in the errors due to PSF dilution corrections and noise rectification
bias with respect to those in the real data.

The
next issue is the fact that the shear calibration bias depends on the
galaxy population, which (as shown previously in
Sec.~\ref{subsec:imagecond}) depends on the observing conditions.
Therefore, we must simulate SDSS data with other observing conditions
besides that in the COSMOS field.  To carry out this test, we chose 8
random locations within the footprint of the source catalogue, and
simulated the COSMOS field as it would have looked at those positions,
including the full PSF model and sky noise. We then checked how much
the shear calibration depends on the conditions.  While detailed
results will be shown in Mandelbaum et al. (2011, {\em in prep.}), we
find that the shear calibration bias (in contrast to the bias due to
\photoz, see N11) is not demonstrably a function of
observing conditions when we allow them to vary in a reasonable way as
sampled by these random points, including variations in PSF
ellipticity, PSF size, PSF kurtosis, and sky noise.  For the eight random positions, we
find a shear calibration bias due to PSF dilution of $-4.8 \pm 2.5$
per cent, which is statistically consistent with the $-6\pm 2$ per cent
from \citep{2012MNRAS.420.1518M} within our claimed 2 per cent
($1\sigma$) uncertainty.  

In addition, we must include in the simulation such realistic effects
as the selection of galaxies with $\zphot>\zlens$ and the galaxy weighting
by $1/\Sigma_c^2$ (Eq.~\ref{eq:wls}), rather than taking unweighted
averages of all galaxies passing basic shape cuts as in
\cite{2012MNRAS.420.1518M}.  We have carried out several tests of this
issue; the basic tests have included first cuts on \photoz\ and then
the realistic weighting, with two possible \photoz.  The first are the
COSMOS \photoz\ \citep{2009ApJ...690.1236I}, which have an RMS
uncertainty $<0.01$ for the apparent magnitudes considered here.  The
second are simulated SDSS \photoz\ that start with the COSMOS \photoz\
and put in the error model estimated by N11 for the SDSS ZEBRA
\photoz.\footnote{We cannot include \photoz\ as part of our simulation
  of SDSS data directly, because it would require us to (a) simulate
  more than just $i$ band and (b) measure galaxy colours using the
  same method as for real data, including processing the simulations
  with the SDSS {\sc Photo} pipeline.  However, as demonstrated in
  Sec.~\ref{subsec:imagecond}, the \photoz\ errors and ellipticity
  errors are not correlated between different observations, so it is
  fair to treat them separately rather than within one self-consistent
  framework.} These tests reveal that the average calibration bias
due to PSF dilution and noise rectification bias changes by $-0.5$ per
cent (i.e., significantly less than our quoted uncertainty of 2 per
cent from the simulations in \citealt{2012MNRAS.420.1518M}).  We
therefore correct for a systematic bias due to these two effects of
$-5.3$ per cent--- the $-4.8$ per cent averaged over random positions
within SDSS with a $-0.5$ per cent correction due to weighting
effects.  Our estimated $1\sigma$ uncertainty is $\pm 2$ per cent.

\subsubsection{Selection bias}

The final shear calibration bias that we consider is selection bias.
As described in \cite{2004MNRAS.353..529H}, the dominant selection
bias for our catalogue is due to the responsivity cut of $R_2>1/3$ in
both bands.  For a galaxy of a given area, those that are more
elongated will have a larger $R_2$ than those that are
round.\footnote{One could imagine redefining the galaxy resolution in a
  way that does not lead to such a selection bias.}  This means that
a shear will effectively increase the $R_2$ value of any given galaxy.
For a galaxy near the resolution limit, this means that if its
intrinsic (pre-lensing) shape is aligned with the weak lensing shear,
then its ellipticity and therefore $R_2$ will be increased and it may
be included in our sample, whereas those that are anti-aligned with
the shear will be made more round and therefore have a reduced $R_2$,
so they might fall out of our sample.  Effectively, this means that
the assumptions behind any weak lensing analysis, that the mean
tangential galaxy ellipticities should go to zero in the absence of
lensing, will be violated.  The overall effect of this selection bias
is to enhance the estimated shear, unless there is also an upper limit
on resolution which will give an effect of opposite sign (the relative
magnitude of the effects due to lower and upper limits on resolution
depend on the resolution factor distribution).

Unfortunately, the simulations based on COSMOS data from
\cite{2012MNRAS.420.1518M} that are used to estimate the magnitude of
shear biases due to incorrect PSF dilution corrections and noise
rectification biases do {\em not} include this particular calibration
bias.  The reason for this is that those simulations were generated
using pairs of galaxies that were identical but for a 90 degree
rotation.  When the galaxy pairs are sheared, one has its ellipticity
and $R_2$ increased, and the opposite occurs to the other, so
requiring $R_2>1/3$ for both will not cause the same selection bias as
in the real data, where we require $R_2>1/3$ for two noise
realisations with the same orientation (i.e., the $r$ and $i$ band
data).

To estimate this selection effect, we rely on the arguments in M05
that, in the Gaussian approximation, we can use 
\beq 
\frac{\delta
  \gamma}{\gamma} = \frac{R_{2,\mathrm{min}}
  (1-R_{2,\mathrm{min}})}{\ssh} \erms^2 n(R_{2,\mathrm{min}}).  
\eeq
Here, $n(R_{2,\mathrm{min}})$ comes from the histogram of $R_2$ values
derived from using, for each galaxy, the minimum value of $R_2$ in $r$
or $i$ band, and evaluating the histogram at the lower bound of $1/3$.
The reason for this is that our selection imposes $R_2>1/3$ in {\em
  both} bands, and therefore the lower of the two determines whether a
galaxy makes it into the sample.  We have plotted this histogram in
Fig.~\ref{fig:basic_hist}; however, for the sake of this calculation,
we must use a weighted histogram that includes (a) the \photoz\ cuts
for $\zphot > \zlens$ and (b) the weighting that goes into calculation
of the lensing signal.  When we do this, we estimate $\delta
\gamma/\gamma \sim 5$ per cent.  Given the approximations going into
this calculation, we assign a $1\sigma$ uncertainty to this
calibration bias equal to half its value, or $\pm 2.5$ per cent.  

We have attempted a more precise empirical estimate using the
simulations from \cite{2012MNRAS.420.1518M}.  To do so, we impose our
selection criteria on galaxy pairs from two independent noise maps of
the galaxy with the same orientation (to mimic our real SDSS selection
in $r$ and $i$).  In this case, we expect the selection bias to
operate in a similar way as in the real data.  We can do this again using
galaxy pairs from two independent noise maps of the 90 degree rotated
orientation, which will have the same selection bias, and then average
those results with the first set of results to beat down the noise.
In this case, however, the shape noise cancellation is no longer exact
(because there are many boundary cases where the original or rotated
orientation does not get included, either due to noise or the
selection bias) and therefore the results are more noisy.  However,
the preliminary estimate of this effect is half as large as the one
from the analytic estimate, or $2.5$ per cent, and is likely more
reliable given that it has far fewer assumptions about the galaxy
models and the way the resolution factors respond to shear for those
models.  For the purpose of this work, we therefore assume a
calibration bias due to selection bias of $2.5\pm 2.5$ per cent, and
defer a more precise estimate from simulations to future work,
Mandelbaum et al. (2011, {\em in prep}).

Note that there is an additional possible selection bias, due to our
requirement that $e_\mathrm{tot}<2$.  However, any bias in the shear
due to this cut will have been implicitly included in the estimates of
shear calibration bias from the COSMOS-based simulations used to
estimate PSF dilution corrections and noise rectification bias, so we
do not consider it separately.

\subsubsection{The total systematic error budget}\label{subsubsec:totalerr}

We find that the calibration bias due to calculating the
lensing signal estimated using the procedure in
Sec.~\ref{subsec:lens_calc}, after combining all effects from the rest
of this section, is $+0.5$ per cent.  Therefore, for figures in this
paper, we multiply the
signal by a factor of 0.995 (accumulating all factors from this
subsection).  In practice, when performing fits to the signal, we apply
the inverse of this calibration factor to the theoretical signals
before comparing with the data, rather than doing anything to the data.

Then, as described at the start of Sec.~\ref{subsec:shear_calib}, we
first add the uncertainties for the last three types of bias linearly,
because they are all related.  This gives a $1\sigma$ calibration
uncertainty due to these three effects of $4.5$ per cent.  These are
then added in quadrature with the first four effects, to give a total
calibration uncertainty of 5 per cent.  We defer work to lower this
systematic error budget, which is dominated by shear calibration
effects, to future work.

As a basic test of our understanding of shear systematics, we also
estimate the {\em relative} calibration bias when computing lensing
signals using the source sample split into $r<21$ and $r>21$ samples
containing 60 and 40 per cent of the galaxies, respectively.
Given the nearly identical range of $z_\mathrm{src}$ covered by these
samples, we expect theoretically that the lensing signals should be
the same, and as in M05, we exploit this to test for shear calibration
biases that might differ for the different galaxy populations.  
Carrying through the same calculations as in the previous subsection,
but for $r<21$ and $>21$ sources separately, we estimate calibration
biases of $-0.3$ and $-3.3$ per cent respectively.  We will test in Sec.~\ref{subsubsec:sys_calibbias} whether
the observed signal ratio for the two samples is consistent with our
understanding of the shear calibration after correcting for those
factors (which provides a basic test of our combined understanding of
all the above systematic errors).

\subsection{Scale-dependent shear systematics}
\label{subsec:scaledep_sys}

In M05 and several subsequent works (most notably
\citealt{2006MNRAS.370.1008M,2006MNRAS.372..758M})
issues were raised regarding systematic errors in the shape catalogue
that would cause scale-dependent systematic errors in the lensing
signal.  We discuss several such observational effects in turn.  Note that
scale-dependent issues relating to theoretical uncertainties can be
resolved at the stage of modeling, so they 
are not discussed here.  These include the impact of
lensing magnification (previously studied in
\citealt{2006MNRAS.372..758M}, negligible here),  dust
extinction (for dust associated with the lens galaxies on large
scales, as in \citealt{2010MNRAS.405.1025M}), and the
possible need to model the contribution to the mass profile from the
stellar component of the lens galaxies are
discussed later, in Sec.~\ref{subsec:density_profiles}.

\subsubsection{Sky subtraction}
\label{subsubsec:skysub}

A point that was raised in M05, and further quantified in
\citet{2006MNRAS.372..758M}, is that the version of the SDSS {\sc
  Photo} pipeline used up through DR7 has difficulty properly
estimating the sky around large and bright galaxies, such as those
that are typically used as lens galaxies for galaxy-galaxy lensing.
As described in \cite{2011ApJS..193...29A}, the new version of {\sc
  Photo} used for DR8 ({\sc v5\_6}) has an improved sky subtraction
algorithm which alleviates some of the original problem, but not all.
Since galaxy-galaxy lensing measurements require the robust detection and
measurement of apparently faint galaxies (sources) nearby bright ones
(lenses), the residual sky subtraction problems can affect the lensing
signal computation up to $\sim 100$\arcsec\ which corresponds to 41
kpc at $z=0.02$ or 185 kpc at $z=0.1$ (these redshifts bracket the typical
redshifts of the lens sample used for this work).

The sky subtraction error can have two potential impacts on
galaxy-galaxy lensing measurements.  First, we may fail to detect some
faint galaxies due to the sky being overestimated. Since we normalise by
the number of galaxies around random points (in order to 
remove the effects of physically-associated sources), we would end
up multiplying our signal by a normalising factor that is too low, and
therefore underestimating the galaxy-galaxy lensing signal.  Second,
unsubtracted sky gradients could in principle impart some tangential
shear to the source galaxies.  Thus, we must check for both errors in the
number counts of faint galaxies near bright ones, and systematic
errors in the shears.

For this test, we rely on the apparent magnitude histogram of the
parent disk lens sample, in R11, which is dominated by galaxies
fainter than $r\sim 17$, along with figure 4 in
\cite{2011ApJS..193...29A}.  The lower right panel of that figure
shows that for $17<r<17.5$ lenses, the number density of detected
sources will be suppressed by $\sim 2.5$ per cent for
$20\lesssim\theta\lesssim 80$\arcsec.  This suggests that our boost
factors and therefore measured lensing signals must be underestimated
by about this amount on those scales ($<33$ kpc for lenses at
$z=0.02$, and $<150$ kpc for lenses at $z=0.1$).

In addition, we rely on some additional tests--- not shown in
\cite{2011ApJS..193...29A}--- using stars as ``lenses'' and looking
for a tangential shear signal using the source catalogue.  The idea is
that there should not be any signal, so we can assume that observed
signals are due to incorrect sky estimation leading to gradients that
turn into tangential or radial shears.   While we find strong evidence for
such shear signals using the M05 catalogue to $\theta=25$\arcsec, they
are reduced with the new catalogue such that they are only evident for
$\theta<10$\arcsec\ (at the $4\sigma$ level).  Since we do not use
such scales for science, we do not have to worry about this.

\subsubsection{Systematic shear}\label{subsubsec:sysshear}

As was clearly demonstrated in \cite{2006MNRAS.370.1008M}, there is a
slight ``systematic shear'' in the catalogue--- a coherent smearing of
the PSFs and the PSF-corrected galaxy shapes along the scan direction,
due to the fact that the re-Gaussianization PSF correction method as
implemented for this work allows $\sim 5 \times 10^{-3}$ of the PSF
ellipticity to leak into the estimated shears \citep{2012MNRAS.420.1518M}.  Given the
factor of shear responsivity and the typical PSF ellipticity of $\sim
0.05$ in SDSS, this corresponds to an additive systematic ellipticity of
$|e_\mathrm{sys}| \sim 3\times 10^{-4}$ per galaxy.

As a rule, this systematic shear is not a problem for galaxy-galaxy
lensing, since the azimuthal averaging of the tangential galaxy shears
at a given transverse separation $R$ removes it from the measured
lensing signal.  However, this is no longer true once we are at large
enough $R$ that survey edge effects are important.  To lowest order,
we can then use the lensing signal around random points to remove this
systematic shear from the real galaxy-galaxy lensing signal, provided
that any correlations between the lens density and the systematic
errors that determine the size of the systematic shear are also obeyed
by the random catalogue.  In Sec.~\ref{subsubsec:sys_random}, we will
show the galaxy-galaxy lensing signal around random points obeying the same
redshift distribution and area coverage as our real lenses, and
demonstrate that for this work, we are still on sufficiently small
scales that the systematic shear is irrelevant.  We defer a discussion
of larger scales, where it may be important, to future work
(Mandelbaum et al. 2011, {\em in prep.}).

%% file: Tables/catalogue_info.tex
\begin{table*}
\begin{tabular}{lcccc}
\hline
\multicolumn{5}{c}{Overall statistics} \\
\hline
\multicolumn{4}{l}{Galaxies passing all cuts} & 39~267~029 \\
\multicolumn{4}{l}{Fraction that passed shape cuts} & 0.71 \\
\multicolumn{4}{l}{Total area, deg$^2$} & 9~243 \\
\multicolumn{4}{l}{Total area of lens sample for this work, deg$^2$} & 7~131 \\
\multicolumn{4}{l}{Mean number density (gal$/$arcmin$^2$)} & 1.18\\
\multicolumn{4}{l}{RMS ellipticity per component} & $0.36$ \\
\hline
Property    &  \multicolumn{3}{c}{Percentiles} & Mean \\
  & 16 & 50 & 84 & \\
\hline
$r$-band model magnitude & 19.56 & 20.79 & 21.47 & 20.52 \\
$R_{2,r}$ & 0.46 & 0.63 & 0.80 & 0.63 \\
\photoz\ & 0.15 & 0.37 & 0.55 & 0.36 \\
$z$ (from N11) & 0.21 & 0.39 & 0.62 & 0.42 \\
$A_r$ & 0.042 & 0.079 & 0.136 & 0.087 \\
PSF FWHM ($r$, arcsec) & 1.03 & 1.21 & 1.42 & 1.23 \\
PSF FWHM ($i$, arcsec) & 0.95 & 1.14 & 1.35 & 1.15 \\
$10^{0.4A_i}\sigma_{\mathrm{sky},i}$ (nmgy) & 0.040 & 0.045 & 0.052 &
0.046 \\
\hline
\end{tabular}
\caption{Basic information about the source catalogue presented in
  this paper, including overall numbers and area; statistics of galaxy
properties; and statistics of the observing conditions.}
\label{tab:catalogue_info}
\end{table*}

%% file: Tables/rxy.tex
\begin{table*}
\begin{tabular}{lccccccc}
\hline
 & $\!\!\!\!r$-band PSF FWHM [arcsec]$\!\!\!\!$  & $R_{2,r}$ & $e_1$ & $e_2$ &
 $e_\mathrm{tot}$ & \zphot\ & $r$ \\
\hline
$r$-band PSF FWHM [arcsec]$\!\!\!\!$ & $1$ & $\mathbf{-0.329}$ & $0.000$ &
$0.000$ & $\mathbf{+0.031}$ & $\mathbf{-0.010}$ & $\mathbf{-0.032}$ \\ 
$R_{2,r}$ & & $1$ & $\mathbf{-0.015}$ & $\mathbf{-0.013}$ &
$\mathbf{-0.007}$ & $\mathbf{+0.010}$ & $\mathbf{-0.361}$ \\
$e_1$ & & & $1$ & $-0.001$ & $\mathbf{+0.043}$ & $-0.001$ &
$\mathbf{+0.006}$ \\
$e_2$ & & & & $1$ & $\mathbf{+0.023}$ & $+0.002$ & $+0.005$ \\
$e_\mathrm{tot}$ & & & & & $1$ & $-0.001$ & $+0.001$ \\
\zphot\ & & & & & & $1$ & $\mathbf{+0.080}$ \\
$r$ & & & & & & & $1$ \\
\end{tabular}
\caption{Pearson correlation coefficient $r_{xy}$ between pairs of
  observational conditions and/or galaxy properties, for a matched
  sample of galaxies observed with good and with poor seeing.  In all
  cases, given the large sample size, the statistical uncertainty
  $\Delta r_{xy}=0.002$.  $r_{xy}$ entries that are nonzero at
  $\ge3\sigma$ significance are shown in bold.}
\label{tab:rxy}
\end{table*}

%% file: systests_RM.tex
\subsection{Tests of systematics}
\label{subsec:lens_sys}

In this section, we present tests of systematics on the lensing signals calculated
using the procedure in Sec.~\ref{subsec:lens_calc} and corrected for the
calibration bias factors estimated in Sec.~\ref{subsec:shear_calib}.

\subsubsection{Random points test}
\label{subsubsec:sys_random}

\begin{figure}
\includegraphics[width=3in]{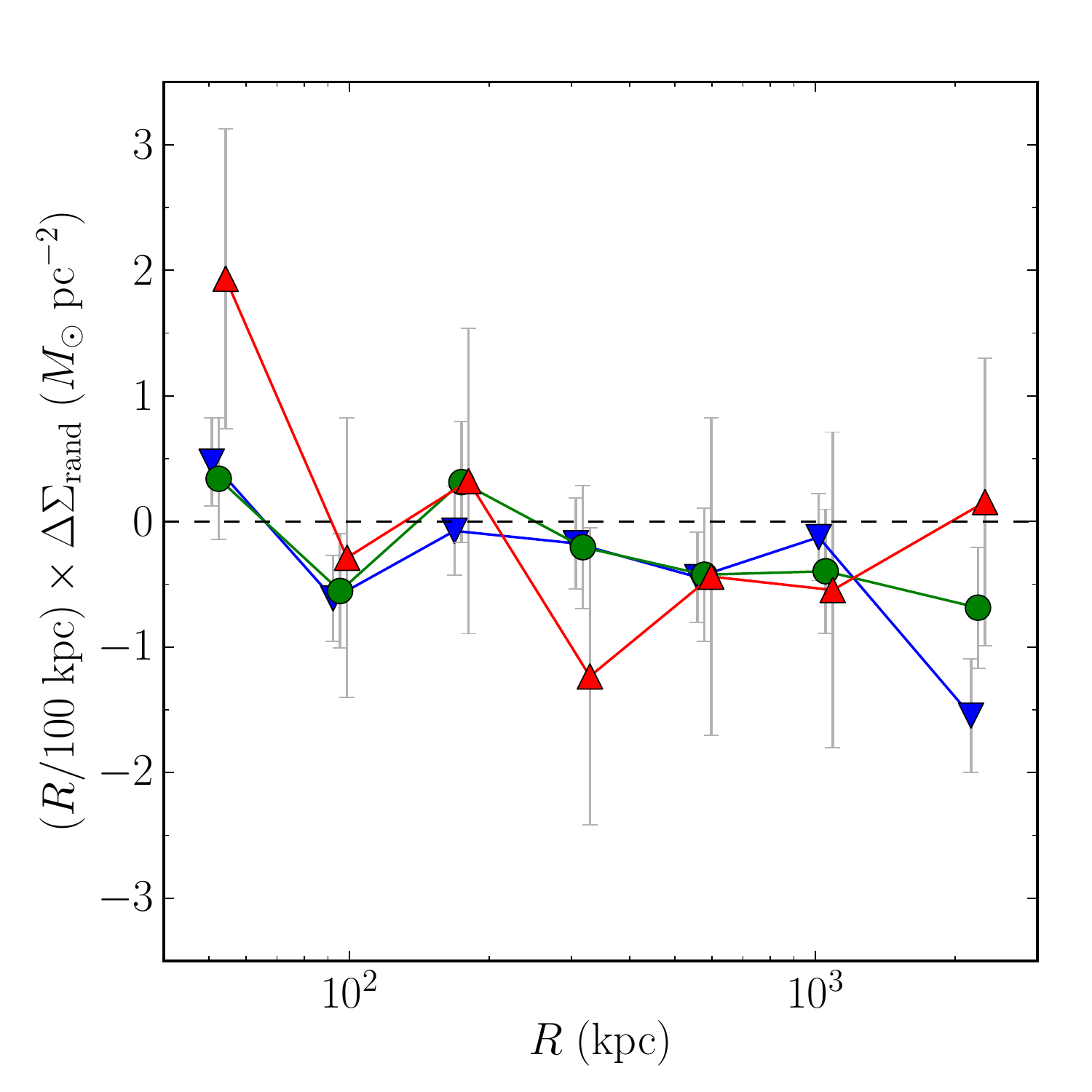}
\caption{Measured lensing signals $\ds_{\rm rand}(R)$ around stacked random galaxies for three $\mstr$ bins
with weighted mean stellar masses of 0.62, 2.68, and $6.52 \times 10^{10} M_\odot$
(blue inverted triangles, green circles, and red triangles, respectively).}
\label{fig:ds_random}
\end{figure}

Figure~\ref{fig:ds_random} shows the lensing signal $\ds_{\rm rand}(R)$ around stacked random galaxies for three 
$M_*$ bins with weighted mean stellar masses of 0.62, 2.68, and 6.52 $\times 10^{10} M_\odot$
(blue inverted triangles, green circles, and red triangles, respectively).

As shown, these signals are approximately consistent with zero, which
indicates that additive shear systematics do not significantly affect
the measured signals on these scales.  There is a slight tendency for
the signals to be negative, which is more pronounced (and
statistically significant) on far larger scales than are used in this
work (Mandelbaum et
al. 2011, {\em in prep}). This is most noticeable for the last data
point in the lowest stellar mass bin.  Using the full (non-rebinned)
data, the $\chi^2$ for a fit to zero signal is $49.5$, $23.5$, $23.4$
for the lowest to highest stellar mass bins.  When computing the
$p(>\chi^2)$, i.e. the probability of getting a $\chi^2$ value at
least as large as the observed one due to random chance, we use a
simulation described in \cite{2004MNRAS.353..529H} to account for the
noise in the bootstrap covariance matrices, which tends to
artificially increase the $\chi^2$ values so that they deviate from
the expected $\chi^2$ distribution.  Using that simulation, we find
$p(>\chi^2)=1$, $59$, and $59$ per cent, respectively.  The low $p$-value for the
lowest stellar mass bin is driven by the last data point.  However,
this slight sign of systematic shear does {\em not} affect our ability
to do science: because galaxy-galaxy lensing is a cross-correlation, we can
simply subtract the signal around random points from the signal around
real lenses.\footnote{More detailed investigations of the efficacy of
  the random points subtraction procedure will be presented in
  Mandelbaum et al. (2011, {\em in prep}).}

\subsubsection{45-degree test}
\label{subsubsec:sys_45}

\begin{figure}
\includegraphics[width=3in]{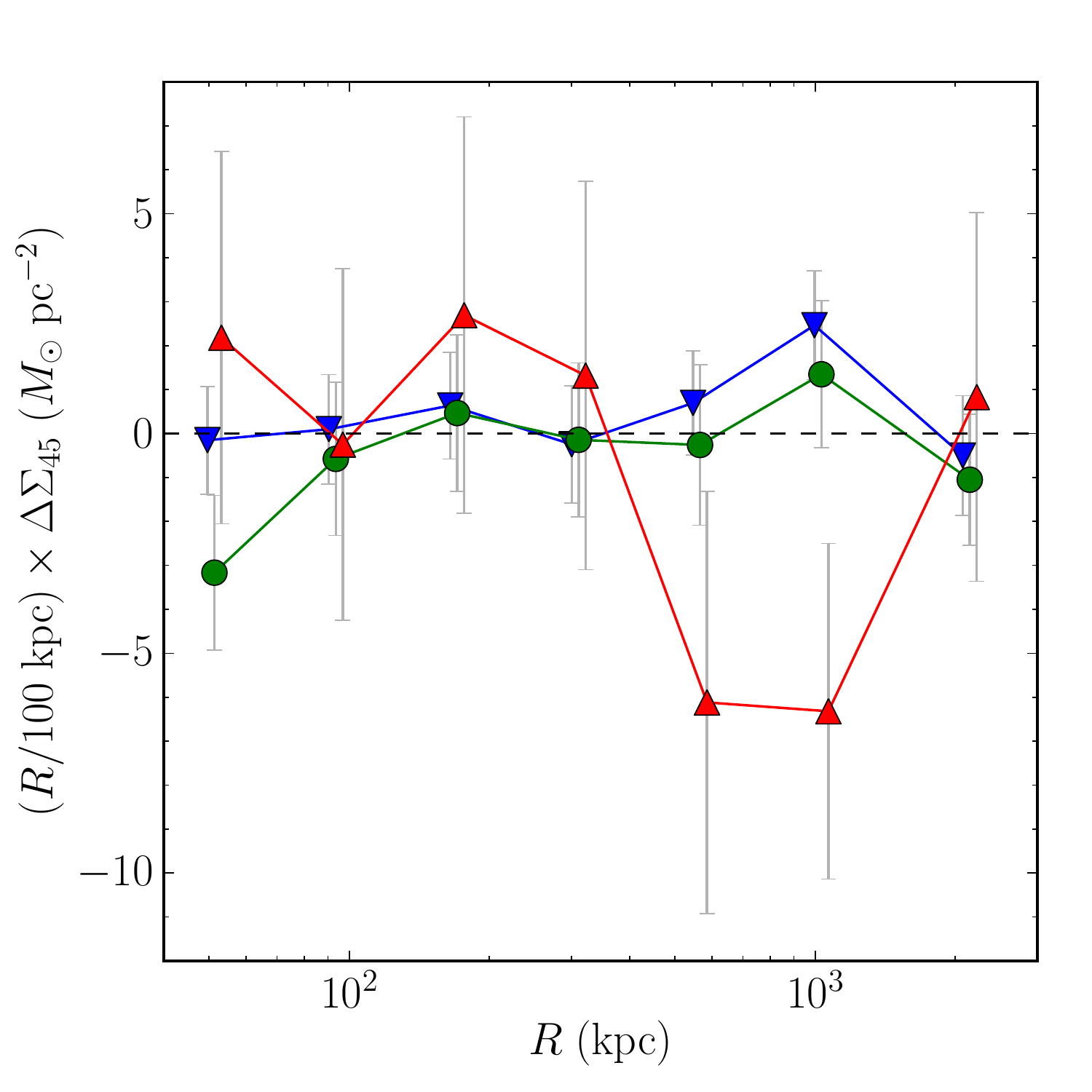}
\caption{Measured lensing signals $\ds_{45}(R)$ for three $\mstr$ bins
with weighted mean stellar masses of 0.62, 2.68, and $6.52 \times 10^{10} M_\odot$
(blue inverted triangles, green circles, and red triangles, respectively).}
\label{fig:ds_45deg}
\end{figure}

While gravitational lensing causes a coherent tangential shear effect
around the lens galaxies, it does not cause any average shape
distortion in the other (45 degree) ellipticity component.  However,
there are systematics that could cause such a nonzero
$\Delta\Sigma_{45}$, so we measure it (using the analogous equation to
Eq.~\ref{eq:dsestimator} but with the other shear component) as a systematics
test.  The result is shown in Fig.~\ref{fig:ds_45deg}; as shown, it is
consistent with zero for all 3 samples, with a $\chi^2$ for a fit to
zero of $16.4$, $15.5$, $36.3$ (23 degrees of freedom), giving a
$p(>\chi^2) = 90$, $93$, and $12$ per cent, for the lowest to highest stellar mass bins. 
Thus, there is no clear
evidence of systematic errors from the 45-degree test.

\subsubsection{Ratio test}
\label{subsubsec:sys_calibbias}

As described in Sec.~\ref{subsubsec:totalerr}, we have separately
calculated the 
lensing signals for the sources split into $r<21$ and $r>21$
separately, to ensure that we properly understand the shear
calibration bias as it depends on galaxy properties. 

We fit the lensing signals for the $r<21$ and $r>21$ source samples to the best-fit halo profile for our fiducial fit to the lensing signal for all sources (described in Sec.~\ref{subsec:fiducial_fits}), i.e., we fix the shape and only allow the normalization to vary. We find best-fitting scale factors of $0.99\pm 0.13$ and $1.04\pm 0.16$ for the $r<21$ and $r>21$ source samples, respectively, after taking the average over the three stellar mass bins (since the bins largely sample the same source populations, given that they are all at low redshift, below most of the sources). Note that the 13 and 16 per cent uncertainties do not represent our systematics floor; the systematics floor is set by the arguments in Section~4, and we carry out this test to confirm that there are no obvious observational reasons not to trust the arguments given there. We find no evidence for an offset between the lensing signals for the different source samples, and the calculated ratio, $0.95\pm 0.19$, is consistent with the predicted ratio of 1.03.



%% file: Tables/nfwfit_ratios.v15.tex
   9.792 & $  1.61\pm  0.13$ & $  0.15^{+  0.05}_{-  0.04}$ & $  1.26\pm  0.08$ \\
  10.428 & $  1.36\pm  0.09$ & $  0.26^{+  0.06}_{-  0.05}$ & $  1.39\pm  0.06$ \\
  10.814 & $  1.42\pm  0.13$ & $  0.23^{+  0.08}_{-  0.06}$ & $  1.27\pm  0.08$ \\

%% file: Tables/nfwfit_ratios.v18.tex
   9.772 & $  1.75\pm  0.14$ & $  0.10^{+  0.04}_{-  0.03}$ & $  1.12\pm  0.08$ \\
  10.431 & $  1.43\pm  0.13$ & $  0.22^{+  0.08}_{-  0.06}$ & $  1.31\pm  0.09$ \\
  10.812 & $  1.37\pm  0.20$ & $  0.25^{+  0.15}_{-  0.09}$ & $  1.31\pm  0.13$ \\

%% file: Tables/pdf_hsmratio.tex
   9.0 & $  1.27$ & $  1.74$ & $  2.09$ & $  2.33$ & $  2.59$ \\
   9.1 & $  1.28$ & $  1.71$ & $  2.03$ & $  2.26$ & $  2.51$ \\
   9.2 & $  1.29$ & $  1.69$ & $  1.98$ & $  2.18$ & $  2.43$ \\
   9.3 & $  1.28$ & $  1.66$ & $  1.93$ & $  2.11$ & $  2.34$ \\
   9.4 & $  1.28$ & $  1.63$ & $  1.88$ & $  2.05$ & $  2.25$ \\
   9.5 & $  1.28$ & $  1.59$ & $  1.83$ & $  1.98$ & $  2.15$ \\
   9.6 & $  1.29$ & $  1.57$ & $  1.77$ & $  1.92$ & $  2.06$ \\
   9.7 & $  1.30$ & $  1.55$ & $  1.72$ & $  1.85$ & $  1.98$ \\
   9.8 & $  1.31$ & $  1.52$ & $  1.67$ & $  1.79$ & $  1.90$ \\
   9.9 & $  1.30$ & $  1.49$ & $  1.62$ & $  1.72$ & $  1.81$ \\
  10.0 & $  1.28$ & $  1.46$ & $  1.57$ & $  1.66$ & $  1.74$ \\
  10.1 & $  1.28$ & $  1.43$ & $  1.52$ & $  1.60$ & $  1.67$ \\
  10.2 & $  1.28$ & $  1.39$ & $  1.47$ & $  1.55$ & $  1.61$ \\
  10.3 & $  1.23$ & $  1.35$ & $  1.43$ & $  1.50$ & $  1.57$ \\
  10.4 & $  1.17$ & $  1.31$ & $  1.39$ & $  1.46$ & $  1.53$ \\
  10.5 & $  1.14$ & $  1.26$ & $  1.35$ & $  1.43$ & $  1.51$ \\
  10.6 & $  1.06$ & $  1.23$ & $  1.32$ & $  1.41$ & $  1.50$ \\
  10.7 & $  1.02$ & $  1.19$ & $  1.31$ & $  1.41$ & $  1.51$ \\
  10.8 & $  0.96$ & $  1.18$ & $  1.33$ & $  1.44$ & $  1.52$ \\
  10.9 & $  1.02$ & $  1.23$ & $  1.42$ & $  1.55$ & $  1.75$ \\
  11.0 & $  1.15$ & $  1.37$ & $  1.61$ & $  1.92$ & $  2.17$ \\

%% file: Tables/pdf_ovvratio.tex
   9.0 & $-0.198$ & $-0.101$ & $-0.022$ & $ 0.094$ & $ 0.249$ \\
   9.1 & $-0.174$ & $-0.081$ & $-0.009$ & $ 0.099$ & $ 0.244$ \\
   9.2 & $-0.150$ & $-0.065$ & $ 0.003$ & $ 0.101$ & $ 0.238$ \\
   9.3 & $-0.126$ & $-0.046$ & $ 0.016$ & $ 0.106$ & $ 0.232$ \\
   9.4 & $-0.095$ & $-0.028$ & $ 0.028$ & $ 0.108$ & $ 0.226$ \\
   9.5 & $-0.072$ & $-0.013$ & $ 0.039$ & $ 0.115$ & $ 0.219$ \\
   9.6 & $-0.049$ & $ 0.003$ & $ 0.051$ & $ 0.118$ & $ 0.211$ \\
   9.7 & $-0.026$ & $ 0.018$ & $ 0.063$ & $ 0.121$ & $ 0.203$ \\
   9.8 & $-0.003$ & $ 0.036$ & $ 0.076$ & $ 0.124$ & $ 0.200$ \\
   9.9 & $ 0.021$ & $ 0.053$ & $ 0.085$ & $ 0.129$ & $ 0.193$ \\
  10.0 & $ 0.042$ & $ 0.067$ & $ 0.098$ & $ 0.134$ & $ 0.194$ \\
  10.1 & $ 0.057$ & $ 0.082$ & $ 0.109$ & $ 0.139$ & $ 0.189$ \\
  10.2 & $ 0.071$ & $ 0.093$ & $ 0.119$ & $ 0.144$ & $ 0.187$ \\
  10.3 & $ 0.081$ & $ 0.104$ & $ 0.129$ & $ 0.154$ & $ 0.193$ \\
  10.4 & $ 0.087$ & $ 0.112$ & $ 0.136$ & $ 0.163$ & $ 0.209$ \\
  10.5 & $ 0.088$ & $ 0.116$ & $ 0.143$ & $ 0.172$ & $ 0.212$ \\
  10.6 & $ 0.082$ & $ 0.116$ & $ 0.147$ & $ 0.178$ & $ 0.230$ \\
  10.7 & $ 0.075$ & $ 0.112$ & $ 0.144$ & $ 0.186$ & $ 0.244$ \\
  10.8 & $ 0.064$ & $ 0.096$ & $ 0.133$ & $ 0.183$ & $ 0.251$ \\
  10.9 & $-0.008$ & $ 0.053$ & $ 0.099$ & $ 0.160$ & $ 0.227$ \\
  11.0 & $-0.158$ & $-0.075$ & $ 0.028$ & $ 0.111$ & $ 0.180$ \\

%% file: appA_RM.tex
\section{Source catalogue generation procedure}\label{S:generation}

The generation of the new catalogue begins with the explicit selection of
what data to use.  We first select runs and then within them, portions
of runs, based on the following requirements: 
\begin{enumerate}
\item The files required for all the steps of the catalogue reduction
  procedure exist in their proper format (including psField, fpAtlas,
  fpObjc, astrometry, and photometric calibration).
\item The IMAGE\_STATUS flag must be $\le 4$.  Any higher order bits
  being set would indicate fundamental issues with the data quality
  (e.g., bad focus).  If it is set to $2$ or $4$, this may indicate
  issues with photometricity; however, we allow these data to be used
  provided that the photometric calibration procedure itself did not
  raise any flags.
\item PHOTO\_STATUS$==0$, indicating that the data were able to be
  processed by the {\sc Photo} pipeline.
\item The PSP\_STATUS flag must be $0$, which means that the PSF was able
  to be interpolated across the field using the standard second-order
  quadratics. 
\item The field must be classified as photometric according to the
  uber-calibration procedure \citep{2008ApJ...674.1217P}, i.e., we
  require CALIB\_STATUS==1.
\item The $r$-band PSF FWHM at the center of the field must be $<1.8$\arcsec.
\item The $r$-band extinction (calculated as $2.751 \times E(B-V)$, using the dust maps
  from \citealt{1998ApJ...500..525S} to obtain $E(B-V)$ and the
  extinction-to-reddening ratios from \citealt{2002AJ....123..485S} to
  convert them into the $r$-band extinctions) must be $<0.22$ magnitudes at
  the center of the field. (We later require $<0.2$ on a per-galaxy
  basis; the less stringent cut on the extinction at the field center
  simply eliminates fields for which no galaxies will pass the later
  cut.)
\end{enumerate}

Our goal is to select a galaxy sample that has reliable shape
measurements using the single-epoch images alone; no attempt is made
to combine multiple measurements for the same galaxy.\footnote{In part,
this choice is meant to avoid selection biases that can arise if the
galaxy detection has significantly different resolution or apparent
flux in the two runs, as can happen due to sky noise fluctuations for
galaxies near the detection limit.  Moreover, it means that while our
source number density is a function of the imaging conditions, it is
not a very strong function of the number of runs overlapping a given
position.} 
However, our initial run selection makes no attempt to
avoid overlapping areas, as we will process all the galaxies and then
reconcile multiple detections later.

Beginning from the list of reliable run/camcol/field combinations, we
then ran a set of scripts to loop over those combinations and measure
the galaxy shapes within a field.  In practice, this process ran 
on anywhere from 10--100 processors simultaneously, enabling us to
process 777 SDSS runs 
in around 4 weeks.  

Within a given field, we first obtain all necessary information from
the {\sc Photo v5\_6} outputs: astrometry including colour-dependent
terms, PSFs, photometric calibration including flat-fielding
corrections, catalogue of selected objects passing the $S/N>5$ {\sc
  Photo} object detection threshold, and information needed for the
noise model (the gain, dark variance, and sky level). We then impose
some preliminary, loose galaxy selection criteria:
\begin{enumerate}
\item The object must be classified as a galaxy (OBJC\_TYPE$==$3).
  {\sc Photo} carries out star/galaxy separation by comparing two
  measures of photometry, the cmodel and PSF magnitudes.  The cmodel
  magnitudes are obtained from the best-fitting non-negative linear
  combination of two profiles: the best-fitting de Vaucouleurs model
  and the best-fitting exponential model, each determined via separate
  fits to the object light profile.  The PSF magnitudes are simply
  determined by fitting the object light profile to a PSF, only
  allowing the amplitude to vary.  Objects with psfMag $-$ cmodelMag
  $>0.145$ are classified as galaxies.  While this has been shown
  \citep{2005MNRAS.361.1287M,2008MNRAS.386..781M} to be somewhat
  inaccurate at $r\gtrsim 21$, and probabilistic methods
  \citep{2002ApJ...579...48S} based on additional criteria do a better
  job. We find that (a) stars that are accidentally classified as
  galaxies do not end up in our shape catalogue because they fail our
  resolution cut (we quantify this statement in
  Sec.~\ref{subsec:shear_calib}) applied at a later stage of
  processing, and (b) galaxies that are accidentally classified as
  stars would have failed our resolution cut the vast majority of the
  time anyhow.  So, the use of OBJC\_TYPE in this case does not lead
  to significant stellar contamination or loss of useful galaxies.
\item SDSS fields within a given run are defined such that they
  overlap by 24.5\arcsec\ with the next field.  We discard objects
  located in the overlap region in one of the two fields in which they
  appear, to avoid duplicate detections. 
\item We apply cuts to the $r$ and $i$ band model magnitudes ($r<22$,
  $i<21.6$) {\em before} correcting for galactic extinction. We will
  later impose additional cuts to obtain a catalogue with a flux cut
  that is constant in extinction-corrected magnitudes.  The model
  magnitudes are defined using the better of an exponential or de
  Vaucouleurs fit to the object light profile in the $r$ band; fits in
  the other bands simply rescale the 2d $r$-band model, allowing for a
  stable determination of galaxy colours needed for \photoz\
  estimation. 
\item We require a robust detection of the galaxy at $S/N\ge 5$ in the
  unbinned data in $r$ and $i$ bands (the BINNED1 flag should be set
  in $r$, $i$, and overall).  
\item We exclude galaxies that have the following flags set: SATURATED,
  SATURATED\_CENTER, EDGE, LOCAL\_EDGE, MAYBE\_CR, MAYBE\_EGHOST,
  SUBTRACTED, BRIGHT, TOO\_LARGE, BADSKY.
\item We reject galaxies that have the BLENDED flag set, unless
  NODEBLEND is also set.  This avoids the measurement of shapes for
  deblended parents and their children.  Note that the SDSS deblender
  will set this flag for galaxy pairs that are nearby (it does not mean that the light profiles
  actually strongly overlap with that of another object).  The result
  is that $\sim 15$--20 per cent of all detected objects in SDSS are
  deblended.\footnote{\texttt{http://www.sdss.org/DR7/algorithms/deblend.html}}
\item There is a flag cut that depends on the galaxy apparent
  magnitude: those that are fainter than $r=19.2$ are rejected if the
  INTERP and CR flags are set in $r$ or $i$; those that are brighter
  are rejected if INTERP\_CENTER is set in $r$ or $i$.  This different
  treatment is done because the brighter galaxies tend to take up
  sufficient area that they might overlap with a bad pixel requiring
  interpolation simply by chance, so we are more permissive in
  allowing for interpolation, as long as it is not too close to the
  object centroid.  Note that $r<19.2$ galaxies are $11$ per
  cent of the catalogue, but their relative weight in any lensing
  analysis is low because they are quite low redshift, and therefore
  are either in front of many lens samples, or receive a low weight
  due to the small lens-source separation.
\item There is a very preliminary and loose resolution cut on the
  galaxy resolution (where the quantity used to impose this cut is defined below).
\end{enumerate}

 For all of the galaxies passing the above 
cuts, 
we first obtained the full PSF estimate from the {\sc PSP}
pipeline \citep{2001ASPC..238..269L}.  This PSF estimate is obtained
via a Karhunen-Lo\'eve (KL) transform, which uses a set of bright
stars to determine basis functions and then to fit their coordinates
to spatially varying (quadratic) functions.  It can be reconstructed
for a given SDSS run, camcol, field, and filter as a function of
position on the CCD using the publicly available {\sc read\_psf} C
code\footnote{\tt
http://www.astro.princeton.edu/\~{}rhl/readAtlasImages.tar.gz} that
reconstructs the basis functions and the variation of the coefficients
across the field from the SDSS psField files.

Then, we ran the re-Gaussianization PSF correction
software \citep{2003MNRAS.343..459H} on the $r$ and $i$ band Atlas
images (which are postage stamp images with the sky level subtracted
and pixels belonging to other objects masked out, that can be read
using the {\sc read\_atlas\_image} code that is part of the same code
package as {\sc read\_psf}).  This code measures PSF-corrected galaxy
ellipticities according to the shape definition in
Eq.~(\ref{eq:shapedef}), where ellipticities are derived using the
``adaptive moments.''  

 In general, the definition of moments requires performing sums over
 the image that are the discrete approximation to the following
 integrals: 
\beqa \nonumber
M_{ij}^{\rm (method)} &=& \int I({\bmath x}) \ w_{\rm method}({\bmath x}) \\ \label{eq:xxmom}
&&\times \ ({\bmath x}-{\bmath x_0})_i ({\bmath x}-{\bmath x_0})_j \rmd{\bmath x}.
\eeqa
The adaptive moments are the results of  minimizing the integral
\beq \nonumber
E = \frac{1}{2} \int \left| I({\bmath x})
- A\exp\left[-\frac{1}{2} ({\bmath x}-{\bmath x}_0)^T {\bf M}^{-1}
({\bmath x}-{\bmath x}_0) \right] \right|^2 \rmd^2{\bmath x} 
\eeq
over the quantities $(A,{\bmath x}_0,{\bf M})$.  This procedure
amounts to weighting by a weight function $w^\mathrm{(adapt)}({\bmath
  x})$ corresponding to the best-fitting {\em elliptical} Gaussian
that represents the image itself, which in practice is determined
iteratively.  Given the moment matrix ${\bf M}$, we can define
ellipticity via
\beqa
e_1 &=& \frac{M_{xx}-M_{yy}}{M_{xx}+M_{yy}} \nonumber \\
e_2 &=& \frac{2M_{xy}}{M_{xx}+M_{yy}}.
\eeqa

Given the adaptive moments of the observed galaxy images (${\bf M_I}$)
and the PSF model interpolated to the position of that galaxy (${\bf
  M_P}$), we can make a  simplest definition of a resolution factor
\beq
R_{2,\mathrm{simple}} = 1-\frac{T_P}{T_I}
\eeq
in terms of the traces of those moment matrices.  This $R_2$ tends to
1 for well-resolved galaxies and 0 for those that are completely
unresolved.  The initial, 
loose resolution factor cut used to select galaxies for shape
measurement is $R_2>1/4$ in either $r$ or $i$ band; after carrying out
the shape measurement we will impose a more stringent cut to be
described below.

In the case of a Gaussian PSF and galaxy, the PSF-correction could be
trivially carried out via subtraction of the moment matrices.  The
re-Gaussianization method is specifically designed to address both the
deviation of the galaxy and the PSF from Gaussianity to some order.

We begin by correcting for the non-Gaussianity of the PSF in a way
that is exact to first order in PSF non-Gaussianity.  To do so, the
code finds the best-fitting Gaussian approximation to the PSF (which
is generally more extended than a Gaussian), and uses the fit residual
to construct an image $I'$ of the galaxy as it would have appeared
with a Gaussian PSF.\footnote{For more detail, see
  \cite{2003MNRAS.343..459H}.}  Then, using the re-Gaussianized
image, a PSF correction is carried out on the galaxy and PSF moments,
using the procedure from \cite{2002AJ....123..583B} to correct for the
non-Gaussianity of the galaxy to first order.  In the course of
carrying out this procedure, we define a new resolution factor
\beq\label{eq:r2}
R_2 \equiv 1-\frac{T_P}{T_{I'}}
\eeq
using the moment matrix of the re-Gaussianized galaxy image.  This
resolution factor definition is used for all subsequent cuts on
resolution factor.

Finally, we must define the shape measurement error $\sigma_e$ per
component.  To do so, we use a simple estimator from
\cite{2002AJ....123..583B} which is equivalent to
$\sigma_\gamma=2/\nu$ ($\nu$ is the significance of the detection),
where $\sigma_e=2\sigma_\gamma$.  In terms of the actual quantities
that we actually measure, we first define a sky variance as
\beq
\sigma_\mathrm{sky}^2 = \frac{\mathrm{sky}}{\mathrm{gain}} +
\sigma_\mathrm{dark}^2, 
\eeq
where the first term results from the
Poisson noise due to the photons in the sky, and the second is due to
the dark current (current that builds up due to heat even in the
absence of photons).  The sky level is high enough that the noise is
effectively Gaussian, and it is uncorrelated from pixel to pixel.
Moreover, the sky noise dominates over the noise from the galaxy flux
for the large majority of galaxies in the catalogue, for $r\ge 20$.
Then, we determine
\beq\label{eq:sigmaedef}
\sigma_e = \frac{\sqrt{4\pi}\sigma_\mathrm{sky}\sigma_I}{F\,R_2}
\eeq
where $\sigma_I^4 = \mathrm{det} \,{\bf M_I}$ and $F$ is the total
galaxy flux.  
We present tests of these $\sigma_e$ values in Sec.~\ref{subsubsec:erms}.

After all fields were processed, we ran a reconciliation procedure to
decide between multiple detections, and impose our final (more
stringent) set of galaxy selection criteria.  To do so, we first
eliminated all galaxies at positions with $r$-band extinction
$A_r>0.2$.  For galaxies passing this cut, we collected all detections
of any single galaxy from all the
runs that were processed through the pipeline (using a tolerance of
1\arcsec\ to define multiple detections).   Then, we chose the
detection in the observation with the smallest PSF FWHM as our primary
detection of that galaxy.   Finally, for the full list of galaxies
(now using only the primary detection of each one) we required that
the extinction-corrected $r$-band model magnitude satisfy $r<21.8$.

For each galaxy, we have both $r$- and $i$-band shape measurements.
We combine the measurements in the two bands as follows: we define the
galaxy $S/N$ in each band $\alpha$ as being $f_\alpha /
\sigma_{f,\alpha}$ where $f$ is the model flux.  We then weight the
galaxy shape measurements $(e_{1,\alpha}, e_{2,\alpha})$ by
$(S/N)_{\alpha}^2$, and take the weighted average.  For the purpose of
our science analyses, we require shape measurements in both $r$ and
$i$ bands, with $R_2\ge 1/3$ in each.  Additionally, we require
$e_\mathrm{tot} = \sqrt{e_1^2
+ e_2^2} < 2$ (where $e_1$ and $e_2$ are the band-averaged
ellipticities).  This cut helps avoid shape measurements that are
excessively dominated by noise, while at the same time avoiding
selection biases
that can be incurred by imposing the apparently more physical cut of
$e_\mathrm{tot}<1$ (since noise can result in observations with
$e_\mathrm{tot}>1$, which means that imposing a cut at that value
results in cutting off part of the error distribution, biasing the mean).   The resulting catalogue has 43~378~516 unique galaxy
detections in 9~493 deg$^2$.  

The final step was to run the template-based Zurich Extragalactic
Bayesian Redshift Analyzer (ZEBRA, \citealt{2006MNRAS.372..565F}) for
all galaxies in the catalogue.  The procedure used in detail is
described in N11.  In brief, we used a set of templates from
\cite{1980ApJS...43..393C} observed across a long wavelength baseline
in the local universe, supplemented by synthetic starburst spectra by
\cite{1996ApJ...467...38K}, and then interpolated to create a full set
of 31 templates in total.  We ran ZEBRA in the maximum-likelihood (ML)
mode, then selected the \photoz\ based on the peak likelihood
marginalized over template, using a $z<1.5$ prior which is reasonable
for single-epoch SDSS photometry.  We did not use any of the following
ZEBRA options: photometry self-calibration, template optimisation, or
Lyman-$\alpha$ IGM absorption.  As described in N11, which quantifies
the effect of the \photoz\ bias and scatter on galaxy-galaxy lensing
measurements, we imposed additional galaxy cuts based on the ZEBRA
outputs, requiring that (a) the resulting \photoz\ not be one of the
boundary values ($0$ or $1.5$), and (b) the template not be one of the
two starburst templates (or an interpolated one in that range), since
the galaxies that are classified as starburst typically have unusually
large \photoz\ errors (the spectra of star-burst galaxies are
sufficiently featureless in our range of wavelengths that \photoz\
estimation is very difficult).  After imposition of those cuts,
eliminating $\sim 10$ per cent of the sample, the catalogue contains
39~267~029 galaxies.

%% file: appB_RM.tex
\section{Differences from M05}\label{S:differences}

In terms of area coverage, the catalogue from M05 included imaging data
acquired until 2004 June 15 (imaging run 4682), whereas this catalogue
includes all publicly available imaging data from SDSS. As a
consequence, after all quality cuts were imposed, the resulting area
increased from 7~002 to 9~493 deg$^2$. For the science
results presented in this paper, we use a subset of that area covering
the portion of the DR7 spectroscopic sample area covering our
extinction cut, 7~131 deg$^2$.  Note that the
7~002 deg$^2$ of the original source catalog includes some imaging
area without spectroscopy; it is not strictly a subset of the DR7
spectroscopic sample.  Use of the original source catalogue would
require us to eliminate $\sim 20$ per cent of the DR7 spectroscopic
lens sample.

The M05 catalog relied on {\sc Photo v5\_4} outputs.
\cite{2011ApJS..193...29A} detailed
the differences between {\sc v5\_4} and {\sc v5\_6} used for the new
catalog; in brief, the primary difference that is relevant for our
purposes is that the new version of {\sc Photo} has an improved (but
not fully corrected) sky subtraction algorithm that corrects some of
the deficiencies in faint galaxy detections near bright objects first
noted in M05 and subsequently documented in SDSS data release papers
(DR4, \citealt{2006ApJS..162...38A}, and others).  We discuss the impact of the residual sky errors for
this catalog in Sec.~\ref{subsec:scaledep_sys}.

An additional difference between the catalogs is the updated
photometric calibration.  The old catalog used calibrations based on
the 0.5-m Photometric Telescope (PT, \citealt{2006AN....327..821T}),
which became defunct partway through the survey.  As a consequence,
the photometric calibration was wrong for $\sim 4$ per cent of the
area of the old catalog, since the PT calibration files contained
incorrect results due to their not being maintained.  This error in 4
per cent of the area, and in general, the worse photometric calibration
performance of the PT calibrations with respect to ubercalibration,
affected both object selection and also the \photoz\ performance.  In
the new catalog, we use the uber-calibration procedure
\citep{2008ApJ...674.1217P}, which provides a stable photometric
calibration at the 1 per cent level for $griz$, and 2 per cent for
$u$, yielding greater uniformity in galaxy selection and \photoz.

The catalogue from M05 employed two methods of estimating redshifts
for the source galaxies.  For those galaxies at $r<21$
(extinction-corrected model magnitude), the {\sc kphotoz} ({\sc v3\_2}, \citealt{2003AJ....125.2348B}) was used.  Unfortunately this code tended to fail
for galaxies at fainter magnitudes, so for those at $r\ge 21$, we
simply utilized a source redshift distribution $\rmd N/\rmd z$
motivated by early data from the DEEP2 survey data in the Extended
Groth strip \citep{2003SPIE.4834..161D}.  Significant later work \citep{2008MNRAS.386..781M} was necessary to
quantify more precisely the biases in the lensing signal due to these
two separate methods of redshift estimation.  For the new catalogue,
as described in Sec.~\ref{S:generation}, we have utilized a single \photoz\ code to calculate photometric
redshifts for all sources regardless of their apparent magnitude, 
ZEBRA.  Extensive tests of the
impact of using the ZEBRA \photoz\ on the calibration of the
galaxy-galaxy lensing signal in SDSS were presented by N11.  The
use of a single method of redshift estimation simplifies the process
of using the catalog to calculate lensing signals, since we no longer
have to calculate them separately for $r<21$ and $r>21$, correct for
the different biases, and then combine them.

An additional small difference from the M05 catalogue results from the
correction of two bugs in the process of determining the PSF to use
for PSF correction.  The first bug resulted from incorrect usage of
the {\sc read\_PSF} software to extract the KL PSF from the psField
files: in particular, the CCD row and column were swapped.\footnote{The
  SDSS fields are not square, so this bug resulted in the code
  requesting that {\sc read\_PSF} return PSFs outside of the range of
  row and column that should be included in the field.  However, {\sc
    read\_PSF} simply extrapolated the PSF models off the edge of the
  field without complaint.}  In practice, this amounts to noise in
the PSF model used for PSF correction, since the PSF does not vary
systematically when we flip the CCD along the diagonal.  The noise in
the PSF model becomes noise in the estimated
shear. 

The second bug was that the PSF returned by the {\sc read\_PSF} software
included a ``soft bias'' of 1000 counts per pixel, but this soft bias
was not subtracted off before using the PSF for PSF-correction.  While
this sounds alarming in principle, in fact the PSF model itself has a
very high flux normalization (peak flux of $3\times 10^4$ counts) so
this additive constant is not very noticeable.  However, it does make
the PSF seem slightly more extended than it actually
is, an effect that is quantified in Sec.~\ref{subsec:syst_cat}.   
As shown there, a comparison of shapes
in the old vs. in the new catalog can demonstrate the impact of these
bugs, which turns out to be quite minor ($\sim 1$--$2$ per cent).